\documentclass[a4paper,fleqn,usenatbib]{mnras}

\usepackage[T1]{fontenc}
\usepackage{ae,aecompl}

\usepackage{graphicx}	
\usepackage{amsmath}	
\usepackage{amssymb}	
\usepackage{subfig}
\usepackage{pdflscape}
\usepackage{threeparttable}
\usepackage{caption3}
\usepackage[all]{hypcap}        
\usepackage[capitalise]{cleveref}
\crefname{paragraph}{}{}
\crefformat{equation}{Eq.~#2#1#3}
\usepackage{array}

\graphicspath{{figures/}}

\newcommand{\Ha}{\text{H}\ensuremath{\alpha}}
\newcommand{\Hb}{\text{H}\ensuremath{\beta}}
\newcommand{\Nii}{[\text{N {\sc ii}}]}
\newcommand{\Oiii}{[\text{O {\sc iii}}]}
\newcommand{\Sii}{[\text{S {\sc ii}}]}
\newcommand{\kms}{km~s$^{-1}$}
\newcommand{\ebv}[1]{E($B-V$)$^{\textrm{#1}}$}
\newcommand{\msun}{M$_\odot$}
\newcommand{\Ni}{$^{56}$Ni}
\newcommand{\Hii}{\ion{H}{ii}}

\newcolumntype{L}[1]{>{\raggedright\let\newline\\\arraybackslash\hspace{0pt}}m{#1}}
\defcitealias{dopita16}{D16}
\defcitealias{marino13}{M13}

\title[SN~Iax environments]{Investigating the diversity of supernovae type Iax: A MUSE and NOT spectroscopic study of their environments}

\author[J. D. Lyman et al.]{
\parbox{\textwidth}{J.~D.~Lyman,$^1$\thanks{E-mail: J.D.Lyman@warwick.ac.uk}
F.~Taddia,$^2$
M.~D.~Stritzinger,$^3$
L.~Galbany,$^4$
G.~Leloudas,$^{5,6}$
J.~P.~Anderson,$^7$
J.~J.~Eldridge,$^8$
P.~A.~James,$^9$
T. Kr\"uhler,$^{10}$
A.~J.~Levan$^1$
G.~Pignata$^{11,12}$
and E.~R.~Stanway$^1$
}
\\\\
$^1$Department of Physics, University of Warwick, Coventry CV4 7AL, UK\\
$^2$The Oskar Klein Centre, Department of Astronomy, Stockholm University, AlbaNova, 10691, Stockholm, Sweden\\
$^3$Department of Physics and Astronomy, Aarhus University, Ny Munkegade 120, 8000, Aarhus C, Denmark\\
$^4$PITT PACC, Department of Physics and Astronomy, University of Pittsburgh, Pittsburgh, PA 15260, USA\\
$^5$Department of Particle Physics and Astrophysics, Weizmann Institute of Science, Rehovot 7610001, Israel\\
$^6$Dark Cosmology Centre, Niels Bohr Institute, University of Copenhagen, Juliane Maries vej 30, 2100 Copenhagen, Denmark\\
$^7$European Southern Observatory, Alonso de C\'ordova 3107, Casilla 19, Santiago, Chile\\
$^8$Department of Physics, University of Auckland, Private Bag 92019, Auckland, New Zealand\\
$^9$Astrophysics Research Institute, Liverpool John Moores University, IC2, Liverpool Science Park, 146 Brownlow Hill, Liverpool L3 5RF, UK\\
$^{10}$Max-Planck-Institut f\"ur extraterrestrische Physik, Giessenbachstra{\ss}e, 85748 Garching, Germany\\
$^{11}$Departamento de Ciencias Fisicas, Universidad Andres Bello, Avda. Republica 252, Santiago, 8320000, Chile\\
$^{12}$Millennium Institute of Astrophysics (MAS), Nuncio Monse\~nor S\'{o}tero Sanz 100, Providencia, Santiago, Chile
}

\date{Accepted XXX. Received YYY; in original form ZZZ}

\pubyear{2017}

\begin{document}
\label{firstpage}
\pagerange{\pageref{firstpage}--\pageref{lastpage}}
\maketitle

\begin{abstract}
SN~2002cx-like Type Ia supernovae (also known as SNe~Iax) represent one of the most numerous peculiar SN classes. They differ from normal SNe~Ia by having fainter peak magnitudes, faster decline rates and lower photospheric velocities, displaying a wide diversity in these properties. We present both integral-field and long-slit visual-wavelength spectroscopy of the host galaxies and explosion sites of SNe~Iax to provide constraints on their progenitor formation scenarios.  
The SN~Iax explosion site metallicity distribution is similar to that of core-collapse (CC) SNe and metal-poor compared to normal SNe~Ia. Fainter members, speculated to form distinctly from brighter SN~Iax, are found at a range of metallicities, extending to very metal-poor environments. Although the SN~Iax explosion sites' ages and star-formation rates are comparatively older and less intense than the distribution of star forming regions across their host galaxies, we confirm the presence of young stellar populations (SP) at explosion environments for most SNe~Iax, expanded here to a larger sample. Ages of the young SP (several~$\times 10^{7}$ to $10^8$~yrs) are consistent with predictions for young thermonuclear and electron-capture SN progenitors. The lack of extremely young SP at the explosion sites disfavours very massive progenitors such as Wolf-Rayet explosions with significant fall-back. We find weak ionised gas in the only SN~Iax host without obvious signs of star-formation. The source of the ionisation remains ambiguous but appears unlikely to be mainly due to young, massive stars.
\\\\
\end{abstract}


\section{Introduction}

The relative homogeneity of Type Ia supernovae (SN~Ia)  has allowed them to serve as precise 
extragalactic distance indicators  after the application of empirically derived relations for light curve shape \citep{phillips93} and color \citep{tripp98}. The similarity among the vast majority of SNe~Ia suggests a commonality among their progenitor stars and their explosion physics.
SNe~Ia are generally thought to be the complete disruption of Chandrasekhar-mass white dwarfs (WDs)  which undergo a thermonuclear runaway process of explosive carbon and oxygen burning.
Despite a large population of `normal' SNe~Ia, increasing numbers of apparently similar events, but with spectroscopic and light curve peculiarities have been discovered. It is not clear whether such under- \citep[e.g.][]{filippenko92b,leibundgut93, turatto96} and over-luminous \citep{ruizlapuente92,phillips92,filippenko92a, howell06} events represent extreme extensions of the SN Ia progenitor model or are distinct in their progenitor stars and explosion mechanisms.
One particular event, SN~2002cx, also appeared to fall into the category of SN~Ia, following the traditional SN classification scheme \citep{filippenko97}, but garnered the title of the `most peculiar known' SN~Ia \citep{li03}. 
With early spectral similarity to the over-luminous SN~1991T-like events -- consisting of conspicuous high-ionization spectral features associated with \ion{Fe}{ii} and \ion{Fe}{iii} -- but being relatively faint at peak ($M_R \sim -17.7$~mag), SN~2002cx prompted the classification of other `SN~2002cx-like events' \citep[e.g.,][]{jha06, narayan11}. A comprehensive study of all objects designated as SN~2002cx-like events (hereafter SNe~Iax) was presented by \citet{foley13}. The sample of
has since expanded and a recent review by \citet{jha17} discusses the current status of SN~Iax observations and theory.

The primary distinctions of SNe~Iax from normal SNe~Ia are fainter peak magnitudes, lower ejecta velocities, no secondary near-infrared maxima (with light curves peaking in optical before near-infrared, unlike normal SNe~Ia), and late time spectra that do not become fully nebular \citep[see][for more in depth discussion]{jha06, foley13, foley16}.
However even within the events classified as members of this group there appears a large degree of diversity.
The low ejecta velocities of SNe~Iax ($2000-8000$ cf. $10000-15000$~\kms{} for normal SNe~Ia around peak) implies that their explosions are of significantly lower energies, and their faint peak magnitudes indicate production of smaller amounts of radioactive \Ni{} ($0.003-0.27$ cf. $\sim 0.4-1.0$~\msun{} for normal SN Ia, e.g. \citealt{stritzinger06, childress15}). One of the leading scenarios to explain SNe~Iax is that they are deflagrations of Chandrasekhar mass WDs \citep[e.g.][]{branch04, phillips07, jordan12, kromer13} that have accreted helium-rich material from a companion star \citep{foley09, liuzw15a}. Due to the extension of SNe~Iax to low explosion energies and inferred ejecta masses, there is indirect evidence that at least some fraction of these SNe could leave bound remnants. 

Support for the helium-accreting WD scenario has come from direct progenitor constraints and environmental constraints. \citet{mccully14} present the detection of a blue source at the location of SN~2012Z that is similar to the Galactic helium nova V445 Pup and limits for SN~2014dt provided by \citet{foley15b} are also consistent with a similar progenitor system. \citet{liuzw15b} suggest a hybrid CONe WD + He-star as the most favourable progenitor system for SN~2012Z. The delay time to explosion for WD+He-star systems has been investigated by \citet{wang13, wang14} and  \citet{liuzw15a} who found a timescale of $\sim$100~Myr, in good agreement with observational constraints on their ages. \citet{lyman13} found that SN~Iax host galaxy types closely match one drawn from the star-formation (SF) rate density of the local universe and the environments of SNe~Iax trace the SF of their hosts at a similar level to core-collapse SNe~II with expected delay times of tens-hundred Myr. This statistical estimate of their environment ages was backed up through analyses of the local stellar populations (SP) of SNe 2008ha \citep{foley14} and 2012Z \citep{mccully14}, which found environments harbouring young SP with ages of 10--80~Myr. 

Consistent with their young environments, it has also been suggested that some SNe~Iax may be due to the collapse of massive stars. One particular example that has drawn attention is SN~2008ha. This was an extremely weak SN, even in comparison to other SN~Iax, producing 0.003~\msun{} of \Ni{} and possibly ejecting only a few tenths~\msun{} of material \citep{valenti09,foley09}, although it is not alone in this regard \citep[e.g., SN~2010ae;][]{stritzinger14}. Massive star models explored to explain these very low-energy and -luminosity events include the collapse of very massive Wolf-Rayet (WR) stars with significant fall-back onto a nascent black hole \citep{valenti09, moriya10} and stripped-envelope electron-capture (EC) SNe \citep{pumo09, moriya16}. The scenario of very low-luminosity SNe accompanying massive stellar collapse may be appealing to explain apparently SN-less gamma ray bursts \citep[GRBs, e.g.][]{fynbo06, dellavalle06, michalowski16}, if indeed they are related to other long-duration GRBs (LGRBs), and thus a result of the collapse of massive stars \citep[see, e.g.,][]{gehrels06,galyam06}.

Adding further diversity to the sample we note that for one event, SN~2008ge, no evidence for a young SP at the explosion site, or within the S0 host galaxy, was found by  \citet{foley10b}. Furthermore, for two events, SNe~2004cs and 2007J, helium (and potentially hydrogen) was detected in their spectra. Doubt as to their validity as SNe~Iax members has been raised by \citet{white15} who classify them instead as core-collapse Type IIb SNe, although this was refuted by \citet{foley16}.

It is apparent that the group of SNe that have been labelled SN~Iax are a heterogeneous group, and multiple progenitor systems and explosion mechanisms could be responsible \citep{foley13, jha17}.
The sample has now grown to a size where statistical studies can elucidate further information on the progenitor systems, beyond characterising individual examples. Environmental diagnostics have played a central role in the understanding of other well-known transient types \citep[see][and references therein]{anderson15} and, in particular, the Multi-Unit Spectroscopic Explorer \citep[MUSE;][]{bacon10} instrument mounted at the Very Large Telescope (VLT) is revolutionising the way such studies are performed. MUSE is an integral field spectrograph (IFS) providing spatial and spectral information at the transient explosion site and across the host galaxy system in a single pointing \citep[e.g.,][]{galbany16, prieto16, kruehler17, izzo17}.

In this paper we perform spectroscopic environmental measurements for a large number of SNe Iax using both IFS and long-slit data from VLT/MUSE and Nordic Optical Telescope/Andalucia Faint Object Spectrograph and Camera (NOT/ALFOSC), thereby providing additional constraints on the nature of the progenitor systems and investigating whether their wide-ranging explosion characteristics are matched by large spreads in environmental measures. In \cref{sec:obs} we present our observations. Our methods for the MUSE-observed sample are given in \cref{sec:musemethods} and in \cref{sec:notmethods} we highlight similarities and differences for our NOT-observed sample's analysis. Results are presented and discussed in \cref{sec:results,sec:discussion}, respectively.  Throughout the paper we use $\log(\textrm{O}/\textrm{H})+12 = 8.69$ dex as the solar abundance \citep{asplund09} and adopt the cosmological parameters $H_0 = 73.24$~\kms{}~Mpc$^{-1}$ \citep{riess16} and $\Omega_\text{m} = 0.3$.

\section{Observations and Data reduction}
\label{sec:obs}

The current number of SN Iax discovered is around 50 events
\citep{jha17}.
Our sample consists of 37 SNe Iax host galaxies observed with VLT/MUSE or NOT/ALFOSC. 
 The hosts span a redshift range of 0.004 to 0.08. Membership of our sample as SN Iax up to SN~2012Z is presented in \citet{foley13}. Additionally we add SN~2013dh \citep{jha13}, SN~2013en \citep{liuzw15c}, SN~2013gr \citep{hsiao13}, SN~2014ey\footnote{\url{https://wis-tns.weizmann.ac.il/object/2014ey}}
 ({\em Carnegie Supernova Project}, in prep), SN~2014ck \citep{tomasella16}, SN~2014dt \citep{ochner14}, SN~2015H \citep{magee16}, PS~15aic \citep{pan15}, PS~15csd \citep{harmanen15}, SN~2015ce \citep{balam15} and SN~2016ado \citep{tonry16}.  
 Here we detail the observations taken for our sample on the two instruments.

\subsection{VLT/MUSE}
\label{sec:museobs}

Observations of the host galaxies were carried out in a MUSE programme running between September 2015 and March 2016 apart from the host of SN~2002bp, which was observed as part of an earlier programme (as this galaxy has also hosted other SNe) in April 2015.
The time lag between the SN and our observations is more than two years for all but two events, SNe 2014dt and 2015H. In both these cases it is possible to identify broad forbidden line emission arising from the SN ejecta (see \citealp{foley16} for an overview of SN~Iax late-time spectra). These are analysed further in \cref{app:14dt15H} where the late time SN spectra and any impact on our environmental analyses are presented. In essence, we do not consider the presence of faint underlying signal from these hydrogen-poor, non-interacting events to significantly impact our emission-line based environmental analyses.

The strategies for sky-subtraction with MUSE depended on the angular size of each host galaxy. For galaxies that did not cover the field of view of MUSE, blank sky spaxels within the on-source exposures could be used to create the sky-spectrum to subtract. Otherwise, we included two off-source shorter exposures of nearby blank sky interspersed among the on-source exposures. Four on-source exposures, rotated 90~degrees from each other, were used in each case to account for detector artefacts. The details of the exposures taken are given in \cref{tab:musesample}. The seeing values were determined by the FWHM of a point source in the (spectrally) flattened data cubes, in the absence of a suitable source we used the observatory's estimate of the conditions provided in the headers. All data were reduced and combined using version 1.6.2 of the ESO MUSE pipeline via {\sc reflex} \citep{freudling13} with default parameters. The ESO MUSE pipeline employs an empirical method for the removal of the sky signal, however this can leave significant residuals, particularly in the redder part of the spectrum where sky lines are prominent. To combat this, Zurich Atmospheric Purge (ZAP\footnote{\url{http://muse-vlt.eu/science/tools}}; \citealt{soto16}) has been developed. We additionally applied this method to our already reduced data cubes to correct these residuals. Where we took off-source sky exposures, we applied ZAP to the reduced off-source blank-sky exposures, before applying the results to the on-source combined data cubes. For those hosts that did not fill the field of view of MUSE, the correction to be applied by ZAP was calculated using blank regions of the on-source data cubes. Although some residuals were still present, they were significantly reduced compared to the standard sky-subtraction method.

\begin{table*}
\begin{threeparttable}
\caption{Observations of SN Iax host galaxies with VLT/MUSE.}
\begin{tabular}{llcrlcc}
  \hline
  SN name   & Host galaxy   & $z$   & $R_{25}$ &  Date Obs.            & Exp. time               & Seeing   \\
            &               &       & (arcsec) &                       & (s)                     & (arcsec) \\  
  \hline
 1991bj & IC~344        & 0.018 & 14.4     & Sept 2015             & $4 \times 700$\tnote{b} & 1.7      \\
 2002bp & UGC~6332      & 0.021 & 36.9     & Apr 2015              & $4 \times 555$          & 1.2      \\    
2002cx & CGCG~044-035  & 0.024 & 20.8     & Jan 2016              & $4 \times 700$\tnote{b} & 1.7 \\
2004cs & UGC~11001     & 0.014 & 38.7     & Mar 2016              & $4 \times 555$          & 1.1      \\    
2005P  & NGC~5468      & 0.009 & 78.9     & Jan 2016              & $4 \times 555$          & 1.0      \\   
2005hk & UGC~272       & 0.013 & 43.4     & Oct 2015              & $4 \times 555$          & 1.4      \\   
2008ae & IC~577        & 0.030 & 15.8     & Nov 2015              & $4 \times 700$\tnote{b} & 0.9      \\
2008ge & NGC~1527      & 0.004 & 111.5    & Sept 2015             & $4 \times 555$          & 0.6      \\   
2008ha & UGC~12682     & 0.005 & 40.5     & Nov 2015              & $4 \times 555$          & 1.4      \\   
2009J  & IC~2160       & 0.016 & 61.3     & Sept 2015             & $4 \times 555$          & 2.0      \\    2010ae & ESO~162-017   & 0.004 & 61.3     & Sept 2015            & $4 \times 700$\tnote{b} & 1.8      \\
2010el & NGC~1566      & 0.005 & 249.6    & Sept 2015             & $4 \times 520$          & 1.5      \\   
2011ce & NGC~6708      & 0.009 & 34.5     & Sept 2015             & $4 \times 555$          & 0.9      \\    
2012Z  & NGC~1309      & 0.007 & 65.7     & Oct 2015              & $4 \times 555$          & 0.5      \\   
2013gr & ESO~114-007   & 0.007 & 54.6     & Sept 2015             & $4 \times 555$          & 1.8      \\   
2014dt & NGC~4303      & 0.005 & 193.7    & Jan 2016              & $4 \times 520$          & 1.9      \\   
2014ey & CGCG~048-099 & 0.032 & 19.8& Feb/Mar 2016\tnote{a} & $5 \times 700$\tnote{b} & 0.7 \\
2015H  & NGC~3464      & 0.012 & 77.1     & Dec 2015              & $4 \times 555$          & 1.1      \\   
  \hline
\end{tabular}
\label{tab:musesample}
\begin{tablenotes}
 \item [a] {The best five exposures from two incomplete observing attempts were combined.}
 \item [b] {Sky subtraction was done using blank regions of on-source exposures.}
\end{tablenotes}
\end{threeparttable}
\end{table*}

\subsection{NOT/ALFOSC}
\label{sec:notobs}

We observed 21 host galaxies of SNe~Iax at the NOT using ALFOSC over two campaigns during March, September 2016 and April 2017. The data were reduced largely following the same procedures described in \citet{taddia13, taddia15b}, which we additionally describe here. 

We obtained long-exposure ($\gtrsim$~1800~s), long-slit spectra of the \Hii{} regions within the host galaxies, by placing the slit at the SN position and choosing a position angle such that the galaxy centre or other bright \Hii{} regions near the SN location were captured. In most cases the slit contained a few \Hii{} regions bright enough for our analysis. For all but SN~2008A the slit was positioned to go through the host galaxy nucleus. The instrumental setup chosen was the same as adopted for the study presented in \citet{taddia13,taddia15}, i.e. ALFOSC with grism \#4 (wide wavelength range $\sim$3500$-$9000~\AA) and a 1 arcsec-wide slit, resulting in a FWHM (full width half maximum) spectral resolution of $\sim$16--17~\AA. Details of the observations and exposure times adopted for each host galaxy observation are listed in \cref{tab:notsample}.

First, the 2-D (dimensional) spectra were bias subtracted and flat-field corrected in a standard way. When available, multiple exposures were then median-combined to remove any spikes produced by cosmic rays. Otherwise, we removed them using the L.A.Cosmic removal algorithm \citep{vandokkum01}.  

We extracted the trace of the brightest object in the 2-D spectrum (either the galaxy nucleus, a bright star, or a \Hii{} region with a bright continuum) and fitted 
with a low-order polynomial. The precision of this trace was checked by plotting it over the 2-D spectrum. We then shifted the same trace in the spatial direction to match the position of each \Hii{} region visible in the 2-D spectrum, and then extracted a 1-D spectrum for each \Hii{} region. The extraction regions were chosen by looking at the \Ha{} flux profile. The extracted spectra were wavelength and flux calibrated using an arc-lamp spectrum and a spectrophotometric standard star, observed the same night or (in March 2016) during the same week, respectively. Following \citet{stanishev07}, we removed the second order 
contamination, which characterises the spectra obtained with grism~\#4, from each spectrum. In this study, we included all the 
spectra showing at least \Ha{} and \Nii{}~$\lambda$6583 emission lines.

\begin{table*}
\begin{threeparttable}
\caption{Observations of SN Iax host galaxies with NOT/ALFOSC.}
\begin{tabular}{llcrlcc}
  \hline
  SN name             & Host galaxy                    & $z$            & $R_{25}$ & Date Obs. & Exp. time    & Seeing     \\
                      &                                &                & (arcsec) &          & (s)          & (arcsec)    \\
  \hline
1999ax             & SDSS J140358.27+155101.2       & 0.023\tnote{a} & \ldots  & Mar 2016  & $1200 \times 3$ & 0.8      \\
2002cx             & CGCG~044-035                   & 0.024          & 20.8    & Mar 2016  & $1200 \times 3$ & 1.1      \\
2003gq             & NGC 7407                       & 0.021          & 59.9    & Sep 2016  & $1800 \times 3$ & 0.8      \\
2004gw             & CGCG 283-003                   & 0.017          & 38.8    & Mar 2016  & $1200 \times 3$ & 1.1      \\
2005cc             & NGC 5383                       & 0.008          & 94.9    & Mar 2016  & $1800 \times 1$ & 1.2      \\
2006hn             & UGC 6154                       & 0.017          & 29.3    & Mar 2016  & $1800 \times 2$ & 1.7      \\
2007J              & UGC 1778                       & 0.017          & 36.1    & Sep 2016  & $1800 \times 3$ & 1.3      \\
2007qd             & SDSS J020932.73-005959.8       & 0.043          & 25.1    & Sep 2016  & $1800 \times 3$ & 1.0      \\
2008A              & NGC 634                        & 0.016          & 62.7    & Sep 2016  & $1800 \times 3$ & 0.7      \\
2009ku             & APMUKS(BJ) B032747.73-281526.1 & 0.079          & \ldots  & Sep 2016  & $1800 \times 3$ & 1.7      \\
2011ay             & NGC 2315                       & 0.021          & 40.5    & Mar 2016  & $1200 \times 3$ & 1.0      \\
PS1-12bwh             & CGCG 205-021                   & 0.023          & 23.8    & Apr 2017  & $1800 \times 3$ & 1.1      \\
2013dh             & NGC 5936                       & 0.013          & 43.4    & Mar 2016  & $1200 \times 3$ & 1.6      \\
2013en             & UGC 11369                      & 0.012          & 34.5    & Mar 2016  & $1200 \times 2$ & 1.5      \\
2014ck             & UGC 12182                      & 0.005          & 38.8     & Mar 2016  & $1800 \times 1$ & 1.7      \\
2014ek             & UGC 12850                      & 0.023          & 28.7    & Sep 2016  & $1800 \times 3$ & 1.0      \\
2015H              & NGC 3464                       & 0.012          & 77.1    & Mar 2016  & $1200 \times 3$ & 1.3      \\
2015ce             & UGC 12156                      & 0.017          & 28.7    & Sep 2016  & $1800 \times 3$ & 1.0      \\
PS~15aic              & SDSS J133047.95+380645.0       & 0.056          & 22.3    & Mar 2016  & $1800 \times 2$ & 1.2      \\
PS~15csd              & SDSS J020455.52+184815.0       & 0.044\tnote{b} & \ldots  & Sep 2016  & $1800 \times 3$ & 0.7      \\
2016ado            & SDSS J020305.81-035024.5       & 0.043          & \ldots  & Sep 2016  & $1800 \times 3$ & 1.3      \\
  \hline
\end{tabular}
\label{tab:notsample}
\begin{tablenotes}
 \item [a] {An adjacent galaxy is present at $z = 0.059$ (as measured from \Ha{} in the NOT spectrum).}
 \item [b] {Redshift from \Ha{} emission of host galaxy in PESSTO classification spectrum.}
\end{tablenotes}
\end{threeparttable}
\end{table*}

\section{MUSE data analysis and methods}
\label{sec:musemethods}

To analyse the data cubes, we used {\sc ifuanal}\footnote{\url{https://github.com/Lyalpha/ifuanal}}, a package developed in Python to perform spaxel binning, stellar continuum and emission line fitting of IFU data following work done by \citet{stanishev12} and \citet{galbany14}. Further documentation is available for the package but we also detail the relevant analysis procedures in the following subsections.

The effects of Galactic reddening were removed for each cube using the extinction maps of \citet{schlafly11} and adopting a $R= 3.1$ \citet{cardelli89} reddening law before correcting to rest-frame using redshifts for each host given by the NASA/IPAC Extragalactic Database (NED)\footnote{\url{http://ned.ipac.caltech.edu/}}.

\begin{figure*}
\centering
\subfloat{\includegraphics[width=1\linewidth]{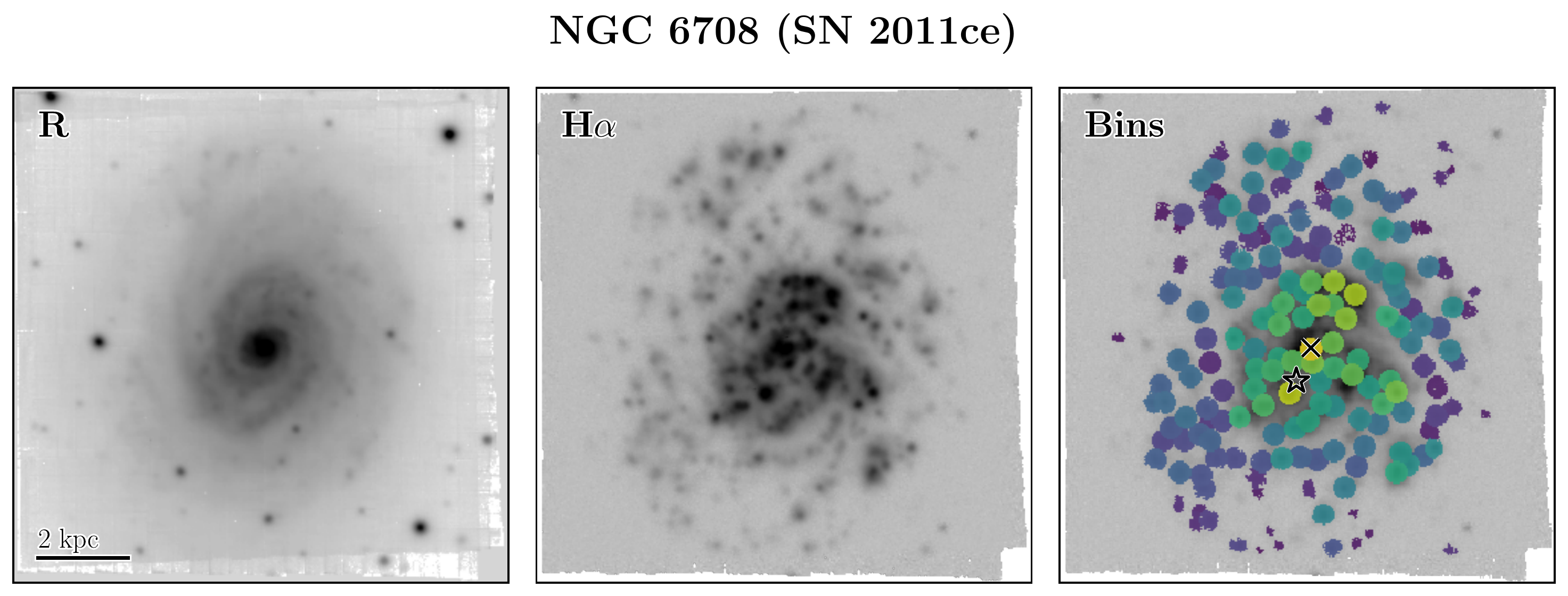}}\\
\subfloat{\includegraphics[width=0.333\linewidth]{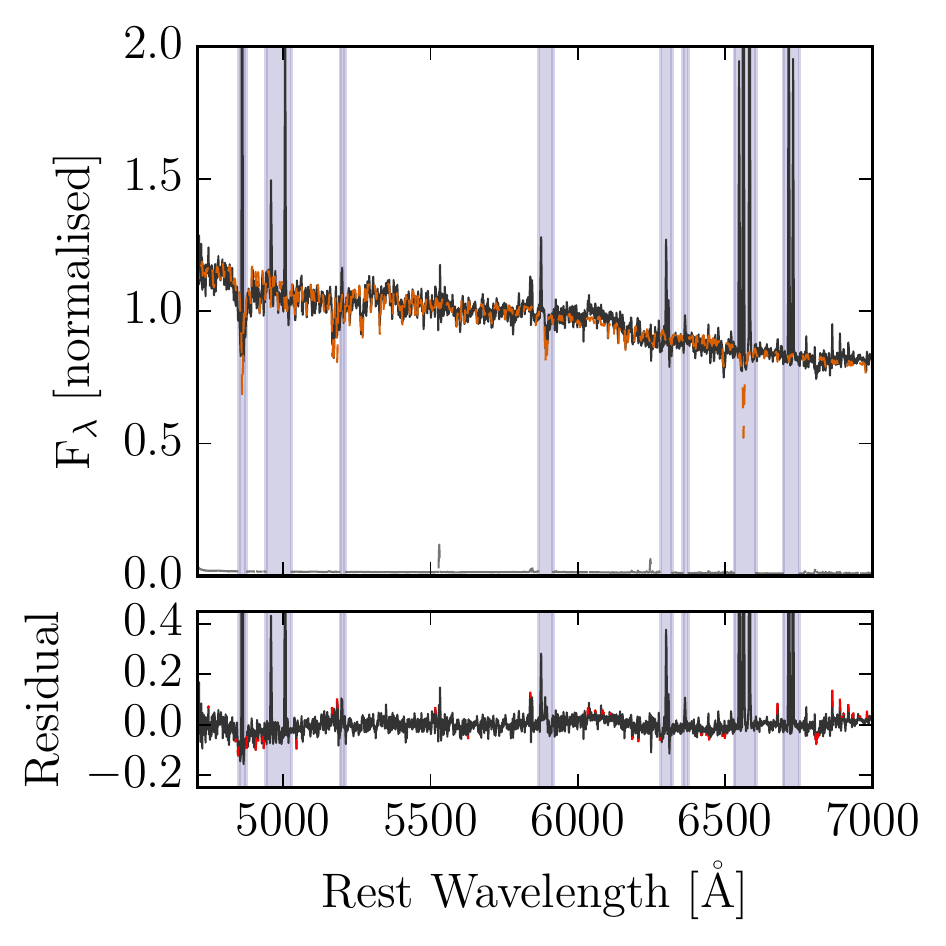}}
\subfloat{\includegraphics[width=0.666\linewidth]{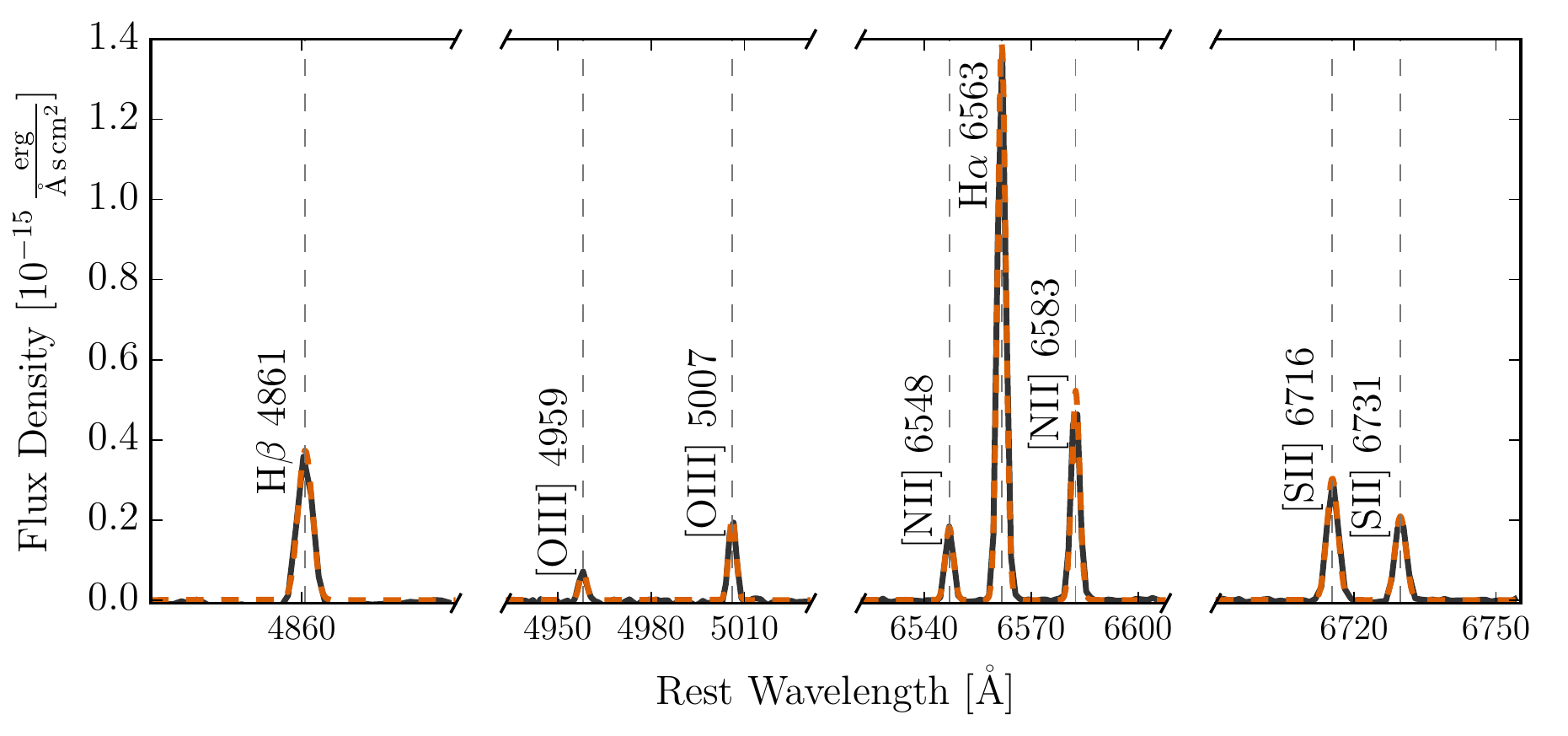}}
 \caption{An example of the binning and fitting routines performed on each data cube described in \cref{sec:binning,sec:contfitting,sec:emisfitting}, as shown for NGC~6708 (the host of SN 2011ce). {\em Top:} The spectral axis has been convolved with the transmission profile of the appropriate filter to produce an $R$-band {\em (left)} and \Ha{} {\em (middle)} image of the MUSE cube; North is up, East is left. The bins that were produced by the binning algorithm (\cref{sec:hiibinning}) are shown (arbitrarily coloured based on their \Ha{} brightness) overlaid by the location of the nucleus and the SN explosion site, given by black $\times$ and `star' symbols, respectively {\em (right)}. {\em Bottom:} The continuum fitting of an example spectrum {\em (left)}. Black shows the observed spectrum and orange the {\sc starlight} fit. Regions around strong emission lines were masked and these are shown as vertical shaded regions. Red elements in the residual plot are those flagged as bad pixels or clipped by {\sc starlight}. The continuum fits were subtracted from the spectra before emission line fitting {\em (right)}. Orange dashed line shows the multiple gaussian fit over the observed residual emission spectrum -- line identifications for the main lines used in our analysis are shown.} 
 \label{fig:museanalysis}
\end{figure*}

\subsection{Spaxel binning}
\label{sec:binning}

As the spaxels in MUSE data are typically smaller than the seeing and typical sizes of objects of interest (i.e. \Hii{} regions), we utilised binning methods in order to combine the signal of adjacent physically-related spaxels, as well as improve the SNR for faint regions. For every galaxy except NGC~1527 (the host of SN~2008ge) we adopted a binning method aimed at segmenting \Hii{} regions. For NGC~1527 we used the Voronoi binning method based on achieving a target SNR of the bins in the continuum. Additional to these algorithmically created bins, we manually added two custom bins, a fibre-like 2~arcsec bin on the host nucleus and a bin at the explosion site of the SN using the same radius limit as for our \Hii{} region binning (\cref{sec:hiibinning}). For any cases where it was evident that foreground stars were affecting the binning algorithm, these were removed using a circular mask before repeating the binning. A single weighted-average spectrum was used to represent each bin, which was forwarded for stellar continuum and emission line fitting. An overview of the binning performed for each data cube is given in \cref{tab:bins}.

\subsubsection{\Hii{} region binning}
\label{sec:hiibinning}

In order to separate \Hii{} regions into bins we use a method based on the {\sc Hiiexplorer}\footnote{\url{http://www.caha.es/sanchez/HII_explorer/}} algorithm \citep{sanchez12b, galbany16}. A narrow 30~\AA{} filter was simulated in the data cubes, centred on the \Ha{} emission line, and from this we subtracted a continuum level determined by interpolation of the flux in two simulated neighbouring continuum filters. From the \Ha{} map a region of blank sky was used to determine the background median and standard deviation. Seeds of potential bins were found as all peaks in the flux distribution that were $>8\sigma$ above the background median. The \Ha{} map was smoothed with a narrow gaussian before the peak detection algorithm was run to avoid finding spurious peaks due to noise fluctuations. Seeds were ordered in descending \Ha{} flux and, starting with the brightest seed, every pixel in the \Ha{} map that satisfied the following conditions was added to the candidate bin:

\begin{enumerate}
 \item within 250~pc (or 1/2 $\times$ FWHM of the MUSE cube if this was larger) of the seed pixel; 
 \item a flux at least 10~per~cent of the seed pixel flux;
 \item $>5\sigma$ above the background median.
\end{enumerate}

If the pixels satisfying these conditions were a contiguous group of 8 or more then the bin was stored, otherwise the bin was rejected. The next seed (that had not already been assigned a bin) was then used. This process was repeated until all seeds were used. 
 
The nature of the {\sc Hiiexplorer} algorithm means that bins in bright areas of the galaxy are grown to a similar size (i.e. the maximum radius specified). Our limit is conservative in comparison to observed sizes of \Hii{} regions, however our choice was driven by the approximate mode of \Hii{} region sizes \citep[e.g.][]{oey03}. Thus, although we will crop any giant regions that can have diameters up to $\sim$1~kpc, our limit means the majority of bins are representative of \Hii{} region sizes. As shown in the top panels of \cref{fig:museanalysis}, the algorithm captures each of the individual regions above our flux limit, although some of the diffuse emission and the outer extents of very large complexes may be missed.

The typical minimum flux in the \Ha{} narrow band image we consider as a bin seed was \mbox{a few~$\times 10^{-18}$~erg~sec$^{-1}$~cm$^{-2}$~\AA{}$^{-1}$}. The number of bins found by the algorithm for each galaxy is shown in \cref{tab:bins}.

\subsubsection{Continuum binning}
\label{sec:contbinning}

NGC~1527 is a smooth-profiled, early-type S0 galaxy \citep{foley10b} and as such we chose another method of binning the spaxels. Specifically, we adopted the Voronoi binning method of \citet{cappellari03}, designed to attain a minimum signal-to-noise in the continuum ($5590-5680$~\AA{}) for each bin whilst maximising spatial resolution. Individual spaxels were required to have a signal-to-noise ratio (SNR) of 3 in the continuum window and these were binned by the algorithm to a target of SNR = 40 per bin. In practice, individual spaxels in the central $\sim 2$~kpc of the galaxy had SNR $\gtrsim$~40 and so these were not binned. Even with the reasonably high SNR target, bin sizes only reached radii of $\sim 100-150$~pc in the outskirts of the field of view.

\subsubsection{Host nucleus bins}
\label{sec:surveybin}

For each MUSE cube where the central nucleus of the host galaxy was in the field of view we simulated a representative survey fibre by creating a 2~arcsec diameter bin centred on the nucleus. The results pertaining to these bins will be referred to as ``nucleus'' values.

\subsubsection{SN explosion site bins}
\label{sec:expsites}

We manually added a bin at the location of the SN explosion site in each host. The explosion site bins were circular with a diameter equal to that used for the \Hii{} region binning (\cref{sec:hiibinning}). The location was found via astrometric alignment of a SN discovery image or, if no image could be obtained, we used offsets from the host centre given in discovery IAU circulars (the method used for each is given in \cref{tab:bins}). Where a suitable image was found, the data cube was convolved with a filter transmission matching the SN image filter to create a MUSE image. The MUSE field of view is comparatively small at $\sim 1$~arcmin and so, generally, only a small number of common sources with which to tie the astrometric alignment were present. We therefore first registered the SN and MUSE images with their WCS information before fitting a transformation that allowed a shift only. 

Two of our MUSE images detected their respective SN at late times - SN~2014dt and SN~2015H, and we used these as a test of our alignment procedures. With 3 sources to tie the fit, our transformed location of SN~2015H was 0.6 pixels (0.12 arcsec) offset from the centroid location of the SN in the MUSE image itself. For SN 2014dt the MUSE image is quite poor seeing and we are limited to using diffuse \Hii{} regions as common sources; nevertheless we find our transformed location to be only 1 pixel offset from the centroid of the SN. When compared to the typical sizes of the explosion site bins, uncertainties in the alignment of $\sim 1$ pixel do not significantly impact our results.

\begin{table*}
\begin{threeparttable}
\caption{Details of spatial binning for the data cubes.}
\begin{tabular}{lccll}
  \hline
  SN name   & Binning              &  $N_\text{bins}$ & Survey fibre &  Explosion site        \\
            & algorithm\tnote{a}   &                  & bin?         &  astrometry\tnote{b}   \\
  \hline                           &                  &              &                        \\
  1991bj & \Hii{} region        &  31              &  Y           &  NTT/EMMI (R)        \\
  2002bp & \Hii{} region        &  67              &  Y           &  IAUC offsets          \\
  2002cx & \Hii{} region        &  17              &  Y           &  IAUC offsets          \\
  2004cs & \Hii{} region        &  90              &  Y           &  IAUC offsets          \\
  2005P  & \Hii{} region        &  127             &  N           &  Swope/SITe3 ($r$)     \\
  2005hk & \Hii{} region        &  87              &  Y           &  VLT/FORS1 (R)       \\
  2008ae & \Hii{} region        &  58              &  Y           &  Swope/SITe3 ($r$)     \\
  2008ge & continuum            &  10237           &  Y           &  Gemini/GMOS-S ($r$)   \\
  2008ha & \Hii{} region        &  50              &  Y           &  Swope/SITe3 ($r$)     \\
  2009J  & \Hii{} region        &  61              &  Y           &  Swope/SITe3 ($r$)     \\
  2010ae & \Hii{} region        &  39              &  Y           &  NTT/EFOSC (R)       \\
  2010el & \Hii{} region        &  88              &  Y           &  NTT/EFOSC (R)       \\
  2011ce & \Hii{} region        &  182             &  Y           &  NTT/EFOSC (R)       \\
  2012Z  & \Hii{} region        &  311             &  Y           &  Swope/SITe3 ($r$)     \\
  2013gr & \Hii{} region        &  48              &  Y           &  NTT/EFOSC (V)       \\
  2014dt & \Hii{} region        &  109             &  N           &  SOAR/Goodman (clear)  \\
  2014ey & \Hii{} region        &  95              &  Y           &  Swope/SITe3 ($r$)     \\
  2015H  & \Hii{} region        &  148             &  Y           &  NTT/EFOSC (V)       \\
  \hline
\end{tabular}
\label{tab:bins}
\begin{tablenotes}
 \item [a] {The \Hii{} region and continuum binning are described in \cref{sec:hiibinning,sec:contbinning}, respectively.}
 \item [b] {If an image of the SN was used to astrometrically align the MUSE cube, the Telescope/instrument (filter) is given. Otherwise offsets from the host given in the discovery IAU Circular were used.}
\end{tablenotes}
\end{threeparttable}
\end{table*}

\subsection{Stellar continuum fitting}
\label{sec:contfitting}

Although our analysis of the underlying stellar continuum is restricted as our regions of interest are emission dominated, and generally without sufficient signal in the continuum to permit a robust analysis from the fitting of absorption indices, we nevertheless must account for the continuum in order to determine accurate emission line results. In order to fit the stellar continuum we used the spectral synthesis package {\sc starlight} \citep{cidfernandes05}. A minimisation of the (model $-$ data) residuals is performed by summing contributions of single stellar populations (SSPs) after masking regions dominated by emission lines and assuming a dust screen following the same reddening law we assume for Galactic extinction. SSP base models were taken from \citet{bc03}\footnote{Using the 2016 update at \url{http://www.bruzual.org/bc03/}.} for the MILES spectral library \citep{sanchezblazquez06} using a \citet{chabrier03} stellar initial mass function (IMF) over the range 0.1--100~\msun{}. 
The components for the base models comprised 16 ages from 1 Myr to 13 Gyr for each of 4 metallicities (Z~$= 0.004, 0.008, 0.02, 0.05$). We note that tests with somewhat different base model sets and choice of IMF did not significantly affect continuum subtraction and so do not affect the emission line results on which our analyses rely. Our procedure largely mirrors that performed on MUSE data of supernova hosts elsewhere \citep[e.g.,][]{galbany16}.

\subsection{Emission line fitting}
\label{sec:emisfitting}

Emission line fitting was performed on the spectra after subtracting the best-fit continuum, as determined in \cref{sec:contfitting}. Prior to fitting, the continuum-subtracted spectra were median filtered with a width of 120~\AA{}\ -- this width was chosen in order to remove any broad residuals left by the imperfect continuum fitting, whilst leaving the narrow emission lines unaffected. The emission line model was a composite model of single gaussians initially centred at the wavelengths of \Hb{}, \Oiii{} $\lambda\lambda$4959,5007, \Ha{}, \Nii{}~$\lambda\lambda$6548,6583 and \Sii{} $\lambda\lambda$6716,6731 (the two exceptions to this set up, SNe 2002bp and 2008ge, are detailed later). The fitting was performed via the weighted Levenberg-Marquardt least-squares algorithm. Such fitting is however susceptible to finding local minima solutions that are dependent on the initial guesses for the gaussian parameters. In order to circumvent the dependency on initial guesses, and to restrict to physically plausible solutions we added the following caveats and steps to our fitting routine:

\begin{enumerate}
 \item The velocity offset of the gaussian means were restricted to within 500~\kms{} relative to the galaxy's rest-frame (i.e., after accounting for the redshift of the galaxy nucleus). This comfortably encloses the limits of rotational velocities seen in late-type galaxies \citep{sofue01}  whilst preventing misidentification of lines.
 \item The velocity offset of the Balmer and forbidden lines were each tied, i.e a single velocity offset for \Ha{} and \Hb{} was fit and a separate offset was fit for the other lines.
 \item The standard deviations of the gaussians were restricted to $\leq 130$~\kms{} and pairs of lines for \Oiii{}, \Nii{} and \Sii{} had tied values for the standard deviation (in \kms{}).
 \item In order to semi-brute force locate the global minimum, a list of several initial guesses for each of the means, standard deviations and amplitudes to be fit were made. A grid was then formed of all possible combinations of these parameters to use as initial guesses for the Levenberg-Marquardt fitter, which found the local minimum for each initial guess. The residual between the data and model was used as an estimator of the goodness of fit for each grid position and used to identify the best overall fit. 
\end{enumerate}

Where the fitting process failed, we manually inspected the emission line spectrum. In most cases, low signal-to-noise or non-detection of the emission features was the cause and we did not consider these bins in our further analysis. Another source of failure was the nuclear bins of some hosts that harbour an active galactic nucleus (AGN). AGN can have a variety of spectral morphologies including broad emission line components. Their presence prevents much of our emission line analyses of the host nuclei bins since metallicity and age indicators are calibrated based on ionising radiation from recently-formed young stars, these cases were identified as described in \cref{sec:ionising}.

Observations of UGC~6332, the host of SN~2002bp, were taken under relatively poor sky conditions, and its redshift means that the \Sii{} $\lambda 6731$ lies in a region of telluric absorption. We attempted to correct this with {\sc molecfit} \citep{smette15} using the galaxy nucleus (as there are no bright stars in the MUSE field of view for this host). However we were not able to clean the spectrum sufficiently to recover the line satisfactorily. Instead we opted to set the flux of \Sii{} $\lambda 6731$ equal to $0.7 \times$~\Sii{} $\lambda 6716$ for each bin. This was chosen as the relation provided a good fit to the vast majority of all other bins in our MUSE sample. We assigned a factor of two uncertainty in the estimated flux, which encompasses the range of intensity ratios for this doublet for typical \Hii{} regions \citep{osterbrock06}.

As mentioned, the host of SN~2008ge is an S0 galaxy with the consequential expectation of weak or no emission lines from SF. We found after a initial run with the full emission line list detailed above, that the fitting routine was failing as it relies on finding a fit simultaneously for each line. Thus, although inspection of certain bin spectra showed conspicuous emission in stronger lines (e.g.\ \Ha{}, \Nii{}), the fits were failing due to non-detections of the weaker lines. We opted to fit this observation for \Hb{}, \Ha{} and N{\sc ii} $\lambda\lambda$6548,6583 only, which allows us to still estimate \ebv{gas} and determine N2 metallicities for the small number of bins where these were detected. This host galaxy is discussed further in \cref{sec:08ge}.

The fitted gaussian parameters were used to determine the signal-to-noise ratio, flux, equivalent width (EW), FWHM and velocity offset of the lines. Uncertainties on these quantities were found by propagating the statistical uncertainties of the fitted parameters and accounting for photon noise in the manner of \citet[][and references therein]{gomes16b}. The \Ha{} and \Hb{} fluxes were used to determine the Balmer decrement compared to the expected ratio of 2.86 (assuming Case B recombination, $T_e = 10^4$~K and $n_e = 10^2$~cm$^{-3}$, \citealt{osterbrock06}) and give the reddening of the gas component, \ebv{gas}, again adopting a $R_V= 3.1$ \citet{cardelli89} reddening law. This was used to correct the flux and EW values for the effects of dust extinction. Although assumptions for the physical parameters of the gas are inherent to this correction, they are representative of observed \Hii{} regions. Furthermore, our metallicity measurements, for example, rely on emission lines nearby in wavelength and as such are largely insensitive to reddening.

Luminosity measurements have their statistical uncertainties quoted, however the systematic uncertainty due to the distance of the hosts is a dominating source in most cases. For example, a 500~\kms{} peculiar galaxy velocity at the median redshift of our sample ($z$ = 0.0165) equates to an uncertainty of $\sim 0.084$ dex in $L$(H$\alpha$).

\subsubsection{Ionising source}
\label{sec:ionising}

We created a BPT diagram \citep{baldwin81} for each host, in order to discard bins where the emission lines are not driven by the radiation from young, hot stars, such as regions around AGN, and thus where measures of line strengths and ratios are not appropriate tracers of metallicity, SFR etc. We used the redshift-dependent classification criterion for regular \Hii{} regions being powered by SF of \citet{kewley13}. The theoretically-determined limit on the ionisation ratios from pure SF is also used \citep{kewley01}. We analyse any region below the theoretical SF limit as the region is still consistent with being driven by SF, although there may be smaller contributions from other sources of ionisation. Above this limit we consider the ionising source of the bin to no longer be dominated by SF and do not include these in our results.

\begin{figure*}
\centering
\includegraphics[width=1\linewidth]{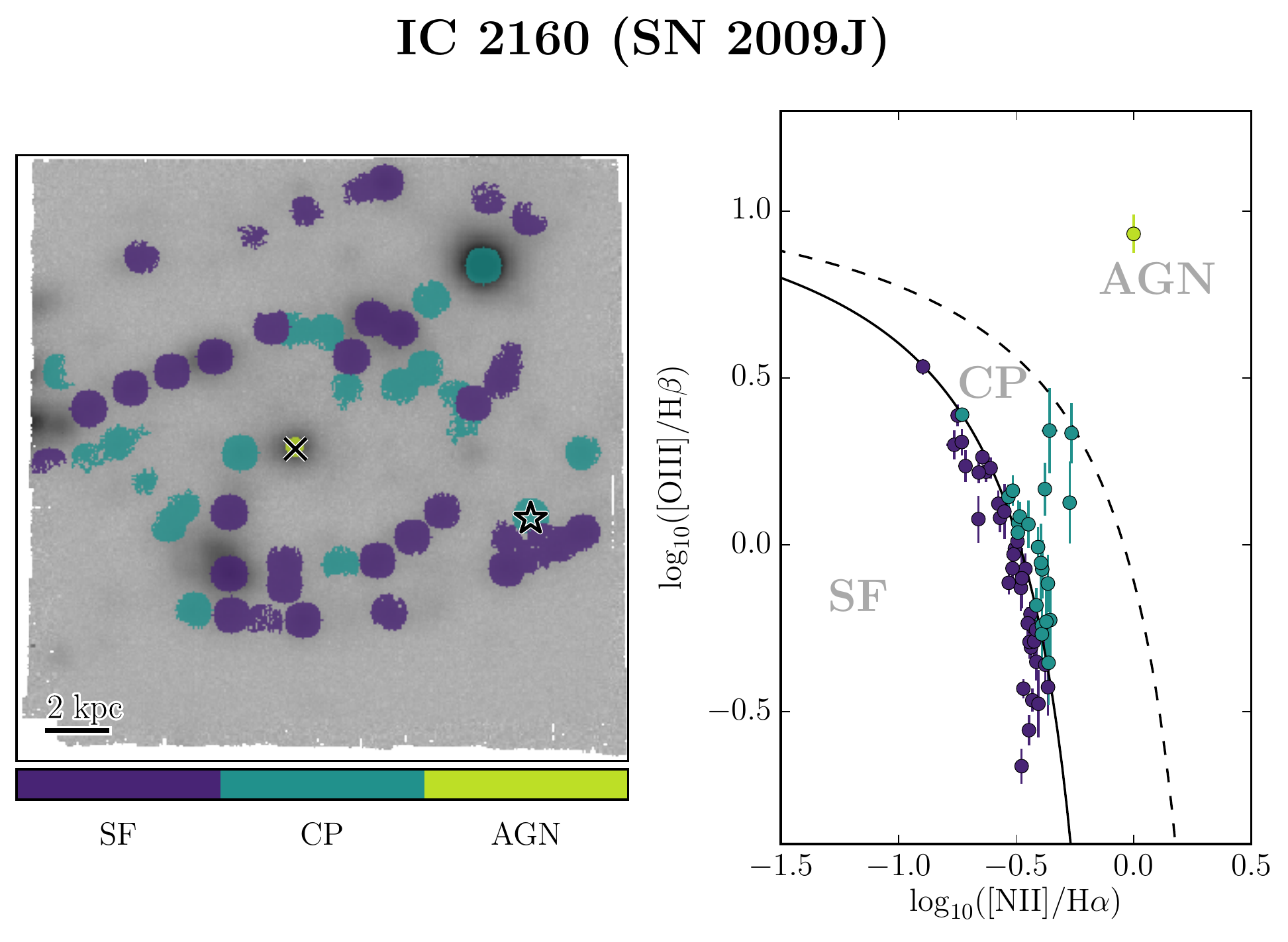}
 \caption{An example BPT map and diagram shown for IC~2160 (the host of SN 2009J). {\em Left:} Heatmap classification of the ionising source (SF: star-formation, CP: composite, AGN: active galactic nucleus) of each bin determined using the relations of \citet[][see text]{kewley01, kewley13}, overlaid on a \Ha{} map of the galaxy; North is up, East is left. The location of the host nucleus and the SN explosion site are indicated by the black $\times$ and `star' symbols, respectively. {\em Right:} The classification limits for each ionising source shown for \Nii{}/\Ha{} and \Oiii{}/\Hb{} line ratios with the location of each bin in this parameter space.}
 \label{fig:bptmap}
\end{figure*}

An example for the host of SN~2009J, which harbours an AGN and strongly star-forming regions, is shown in \cref{fig:bptmap}. Those bins that were found to be above the theoretical SF line (shown in this plot) were discarded from further analysis as they are powered primarily via other sources of ionisation.

\subsubsection{Metallicity}
\label{sec:methmetallicity}

Many emission line metallicity abundance indicators are present in the literature, with significant systematic differences in results between methods, in particular based on whether the method is empirical or theoretically motivated \citep[e.g.][]{kewley08}. The wavelength range of MUSE for low-redshift galaxies covers many emission lines used in various strong emission line methods for determining the gas-phase metallicity, although it does not extend sufficiently into the blue to cover the most commonly used temperature-sensitive lines -- a thorough discussion of metallicity determinations with a view to MUSE data of low redshift galaxies is given in \citet{kruehler17}. 
We concentrate here on determining metallicities on the scale of relative oxygen abundance and use the theoretically-motivated calibration based on photo-ionization models presented by \citet[][hereafter \citetalias{dopita16}]{dopita16}. This indicator uses the \Sii{} $\lambda\lambda6717, 6731$ lines as well as \Ha{} and \Nii{} $\lambda6583$ and is thus well-suited to MUSE observations of \Hii{} regions in the local Universe. The relation is given as:
\begin{equation}
 12 + \log_{10}(\text{O/H}) = 8.77 + y,
\label{eq:D16}
\end{equation}
where
\begin{equation}
 y = \log_{10}\frac{\Nii{} \lambda6583}{\Sii{} \lambda\lambda6717, 6731} + 0.264 \cdot \log_{10}\frac{\Nii{} \lambda6583}{\Ha}.
\end{equation}

\noindent This is a good description of the theoretical results over a wide range of metallicities from very sub- to super-solar. Although this indicator is robust to changing ionization parameter \citepalias{dopita16}, it was noted by \citet{kruehler17} that there exists an apparent $\sim 0.15$~dex systematic offset to lower abundances with this indicator compared to $T_e$-based values.

We also present values based on the O3N2 \mbox{($ = \log_{10}\left [\frac{\Oiii{} \lambda5007}{\Hb} \times \frac{\Ha}{\Nii{} \lambda6583}\right ]$)} calibration of \citet[][hereafter \citetalias{marino13}]{marino13}\footnote{Where we did not reliably detect \Oiii{} $\lambda$5007 in our cubes we use the N2 relation of \citetalias{marino13}. These cases are highlighted in tables of results.}, primarily to facilitate comparison with literature values for the environments of other transient types. There is a 1$\sigma$ uncertainty of 0.18~dex associated to measurements with this indicator due to the observed spread of $T_e$-based abundances about this relation. The indicator is given over the $Z$ range $\sim8.15-8.75$ dex. Where literature values were presented in the O3N2 calibration of \citet{pettini04}, we convert these to \citetalias{marino13}. We adopt a binning regime that means our effective spatial scale is similar to the size of \Hii{} regions, and we are therefore not as sensitive to significant variations of the ionisation parameter that can occur within individual regions on small spatial scales. For ionisation parameter-dependent abundance indicators, such as O3N2, this creates strong gradients within individual regions \citep{kruehler17}.

Although we correct our emission line fluxes for reddening (\ebv{gas} estimated by the Balmer decrement), we note that since the line ratios used in the above metallicity indicators are relatively nearby in wavelength, we are not strongly affected by uncertain reddening values.

We determined metallicity gradients for each host with a linear fit to the bin metallicities and deprojected galactocentric distances. 
Deprojected distances were found following the method of \citet{andersen13} to determine the position angle and inclination from the velocity map of the \Ha{} line and then normalised by the $R_{25}$ (25th $B$-band mag~arcsec$^{-2}$ surface brightness radius) value of the host. $R_{25}$ values were taken from NED. For SN~2008ge the velocity map used was that of the stellar continuum as determined by {\sc starlight}. For SNe where we do not directly detect emission lines at the location of the SN we use these metallicity gradients to estimate the metallicity of the region local to the SN. The uncertainty on this estimate is taken as the rms of the observed metallicity values about the linear fit.

\begin{figure*}
\centering
\includegraphics[width=1\linewidth]{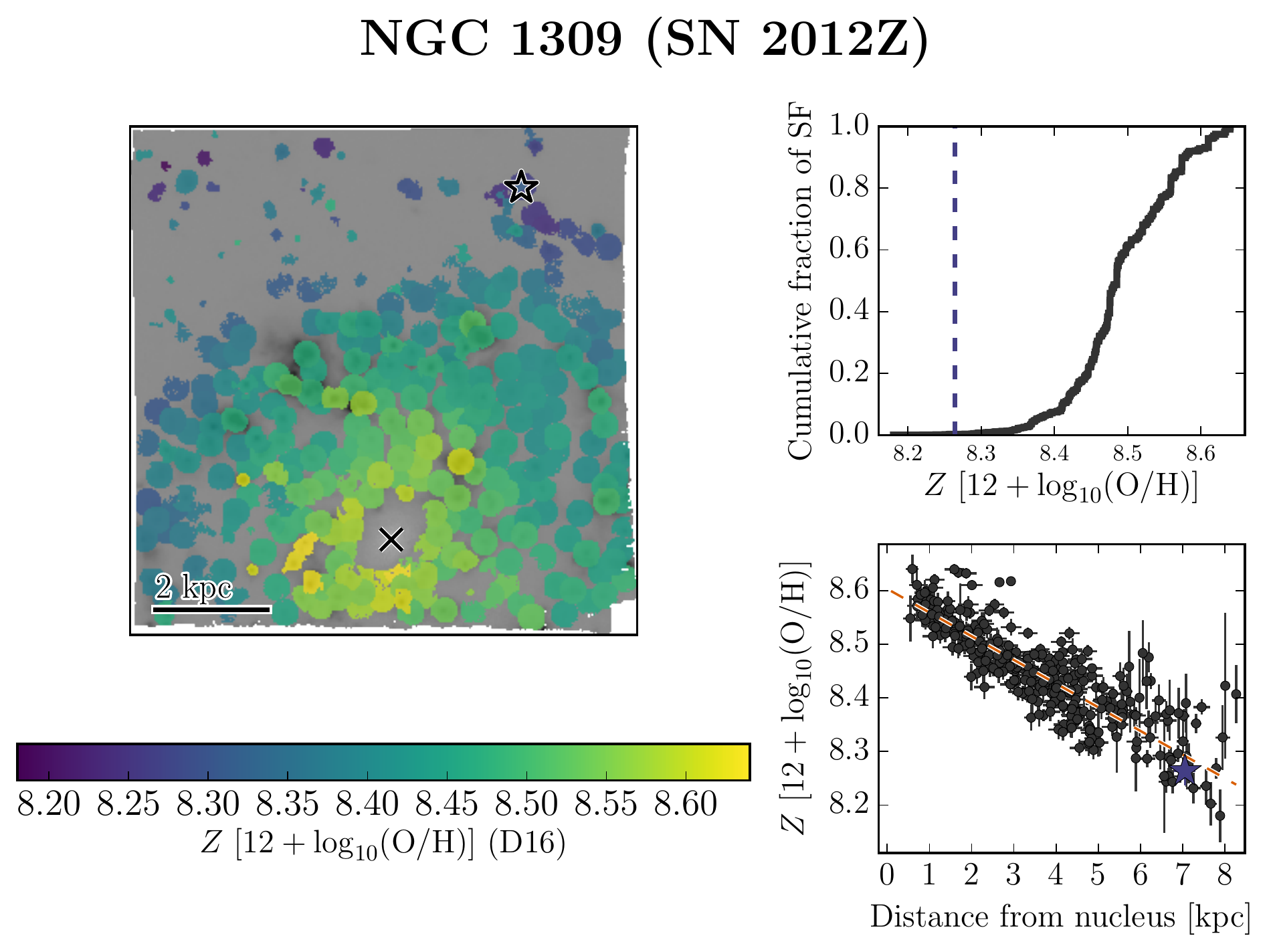}
 \caption{A gas-phase metallicity map for NGC~1309 (the host of SN~2012Z). Metallicities are given on the oxygen abundance scale using the relations of \citetalias{dopita16}. {\em Left:} A heatmap of the metallicity of each bin overlaid on a \Ha{} map of the galaxy; North is up, East is left. 
 The adopted locations of the nucleus and the SN explosion site are indicated by the black $\times$ and `star' symbols, respectively. {\em Right:} A cumulative distribution of the metallicity in the galaxy weighted by the SFR -- i.e. $L(\Ha{})$ -- of each bin {\em (top)} and the deprojected metallicity gradient of the galaxy {\em (bottom)}. The SN explosion site metallicity is highlighted in these panels also.} 
 \label{fig:d16map}
\end{figure*}

An example of a metallicity map and gradient is shown in \cref{fig:d16map} and stamps for all hosts (except SN~2008ge) are given in \cref{app:zstamps}.

\subsubsection{SF rates}
\label{sec:methsfr}

We use \citet{kennicutt98} to convert our extinction-corrected $L$(\Ha{}) values (determined as in \cref{sec:emisfitting}) into SFRs with the relation:

\begin{equation}
SFR~[M_\odot \text{yr}^{-1}] = 7.9 \times 10^{-42} L(\text{H}\alpha)~[\text{ergs s}^{-1}].
\end{equation}

\subsubsection{Ages}
\label{sec:methages}

\begin{figure}
 \centering
\subfloat{\includegraphics[width=\columnwidth]{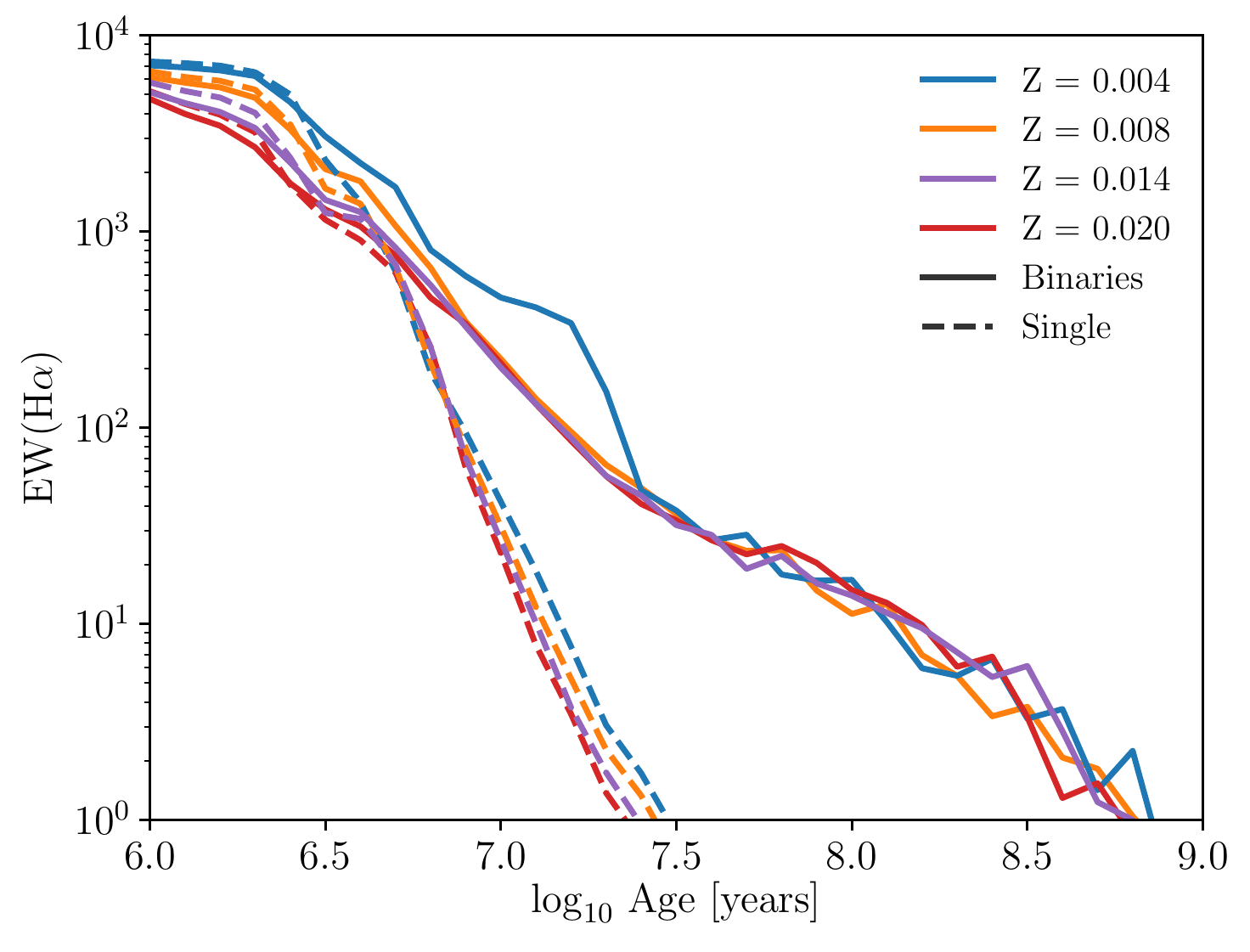}}
 \caption{The evolution of EW(\Ha{}) for an instantaneous SF episode calculated with {\sc bpass} and {\sc cloudy} (see text). The IMF used has a power-law slope of $\alpha = 2.35$ over the range 0.5 to 100 \msun{}. We stress here that this is for a single fiducial gas model of assumed size and density and that the physical properties of the gas distribution can affect results significantly \citep[see][Eldridge \& Stanway, in prep]{stanway14b}. Different metallicities are represented by colours with binaries and single populations as solid and dashed lines, respectively.} 
 \label{fig:bpassew}
\end{figure}

Ages for the youngest SP in galactic regions and at SN explosion sites are often made based on the EW of (primarily Balmer) emission lines in comparison to theoretical predictions. Such mappings between EW values and ages are subject to significant uncertainty due to potentially unaccounted-for effects (stellar multiplicity, SN feedback) and the large spread in physical characteristics of nebular gas in star forming regions (electron density, physical size).
With these caveats in mind, it is nevertheless widely accepted that the EW of nebular emission lines should have some inverse relation with the age of the youngest SP for an instantaneous or rapidly-declining SF history for the region. We used results from the Binary Population and Spectral Synthesis code, {\sc bpass} \citep[e.g.][]{eldridge09,eldridge12}, and processed the stellar continua with {\sc cloudy} \citep{ferland98}. This was done with {\sc bpass} version 2.1 \citep[][Eldridge \& Stanway, in prep]{stanway16} in the manner described in \citet{stanway14b} for a fiducial nebula gas model of 10$^2$~cm$^{-3}$ hydrogen gas density in a spherical distribution with inner radius of 10~pc. From this we obtained the evolution of EW measurements for \Ha{} with age for instantaneous SF episodes, which is shown in \cref{fig:bpassew}. These were compared with our measured EW values at the explosion sites to provide estimates of the age of the youngest SP at these locations. We note here the strong effect of including binary stellar evolution for these calculations, which act to strengthen the emission relative to the continuum, beginning after several~Myr. A binary fraction $< 1.0$ would result in evolution somewhere between the two cases shown in \cref{fig:bpassew}.
A pertinent study of the effect of including binary stellar evolution when determining age estimates of CCSN-hosting regions has found that regions are generally older than previously thought, thus revising progenitor initial mass estimates down \citep[][Xiao et al. in prep]{xiao17}

\subsubsection{Pixel statistics}
\label{sec:pixelstat}

The location of transients within the light distribution of their hosts has been used to infer further constraints on the nature of the progenitor systems \citep[see][for a review]{anderson15}. We use the Normalised Cumulative Rank (NCR) method presented in \citet{james06}: briefly, pixel values for an image of a transient host are sorted, cumulatively summed and then normalised by the total sum of the values. The location of the transient's explosion site in this cumulative sum provides the fraction of light in the host at a level lower than the intensity at the explosion site. When using an appropriate SF tracer, such as \Ha{}, the fraction of SF in the host below the level of the explosion site can be inferred; a distribution of \Ha{} NCR values describes the association to SF for a sample of transients. 

The NCR analysis has been presented for an initial sample of SNe~Iax in \citet{lyman13}. We used the same \Ha{} maps constructed from our MUSE data cubes that were used to create our spaxel bins (\cref{sec:hiibinning}) in order to calculate NCR values for a larger sample of SN~Iax. The maps were binned 3$\times$3 to circumvent astrometric uncertainties before applying the NCR method. Where we had an existing NCR value from \citet{lyman13} and our MUSE observation did not cover the extent of the host, we use the pre-existing value.

We further introduce a metallicity cumulative rank in \cref{sec:reszrank}, which we use to assess the presence of metallicity bias in the production of SNe~Iax within their host galaxies.

\section{NOT data analysis and methods}
\label{sec:notmethods}

\begin{figure*}
 \centering
\includegraphics[height=9cm,angle=90]{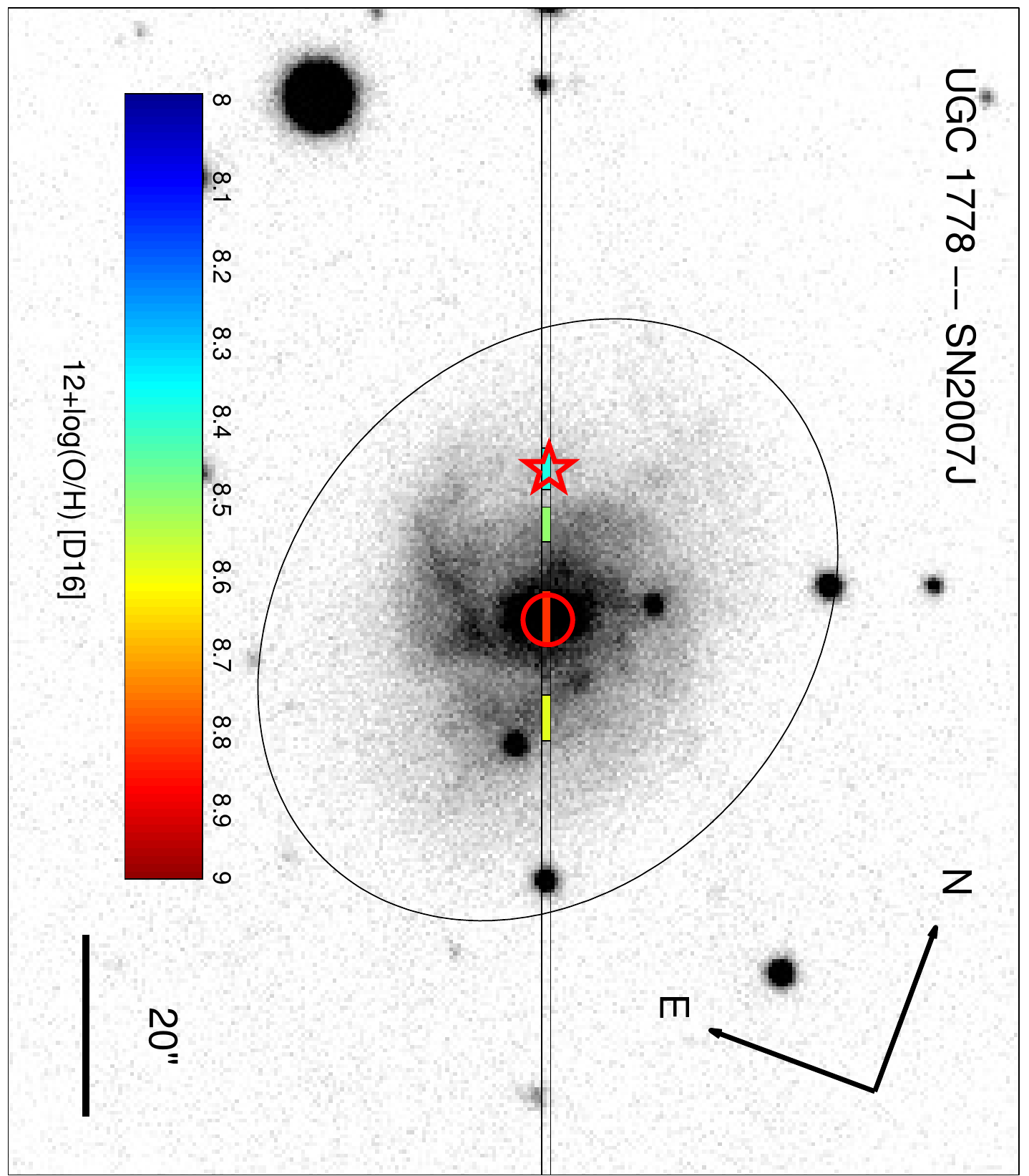}
\includegraphics[width=7.0cm,angle=0]{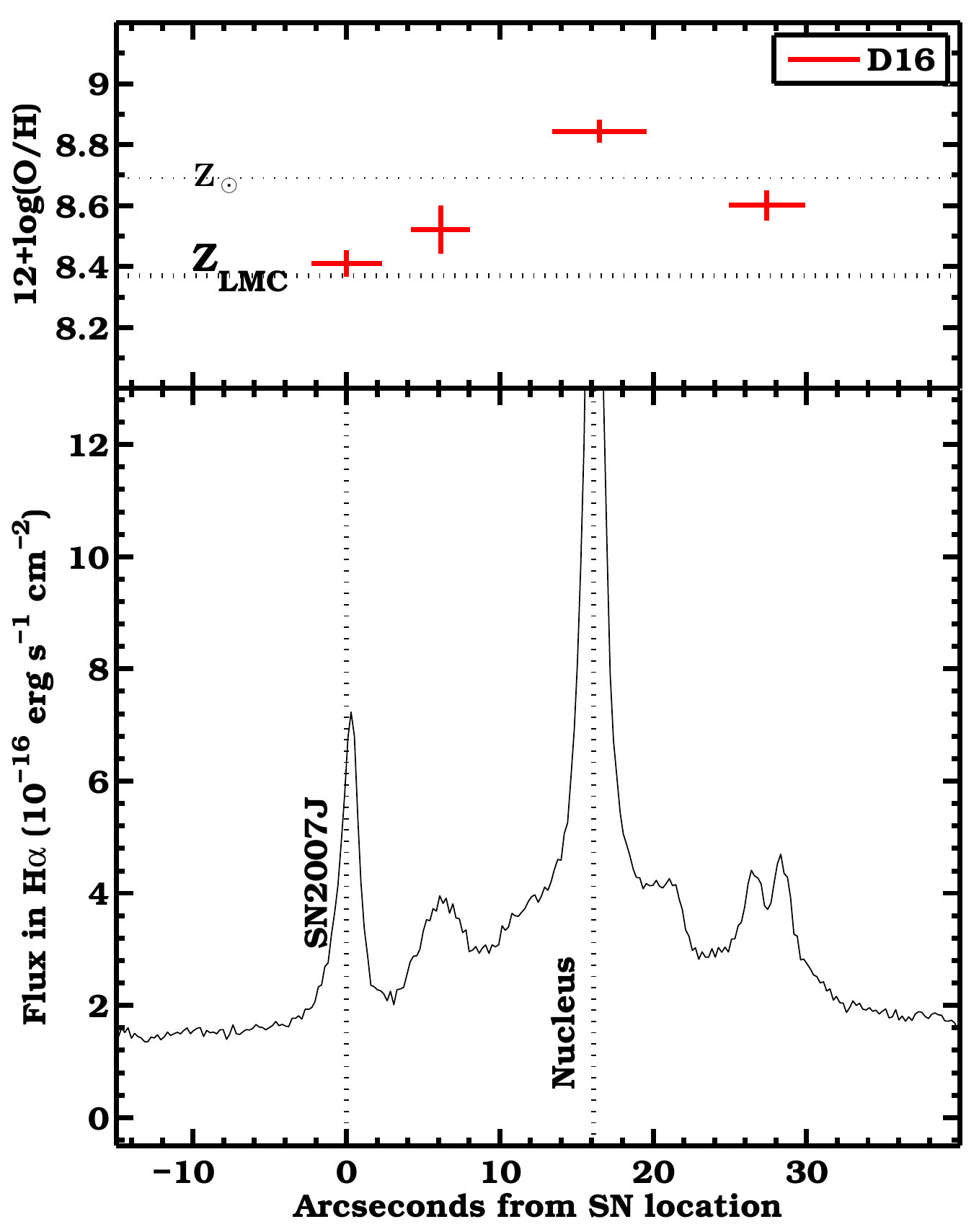}\\
\includegraphics[width=17cm,angle=0]{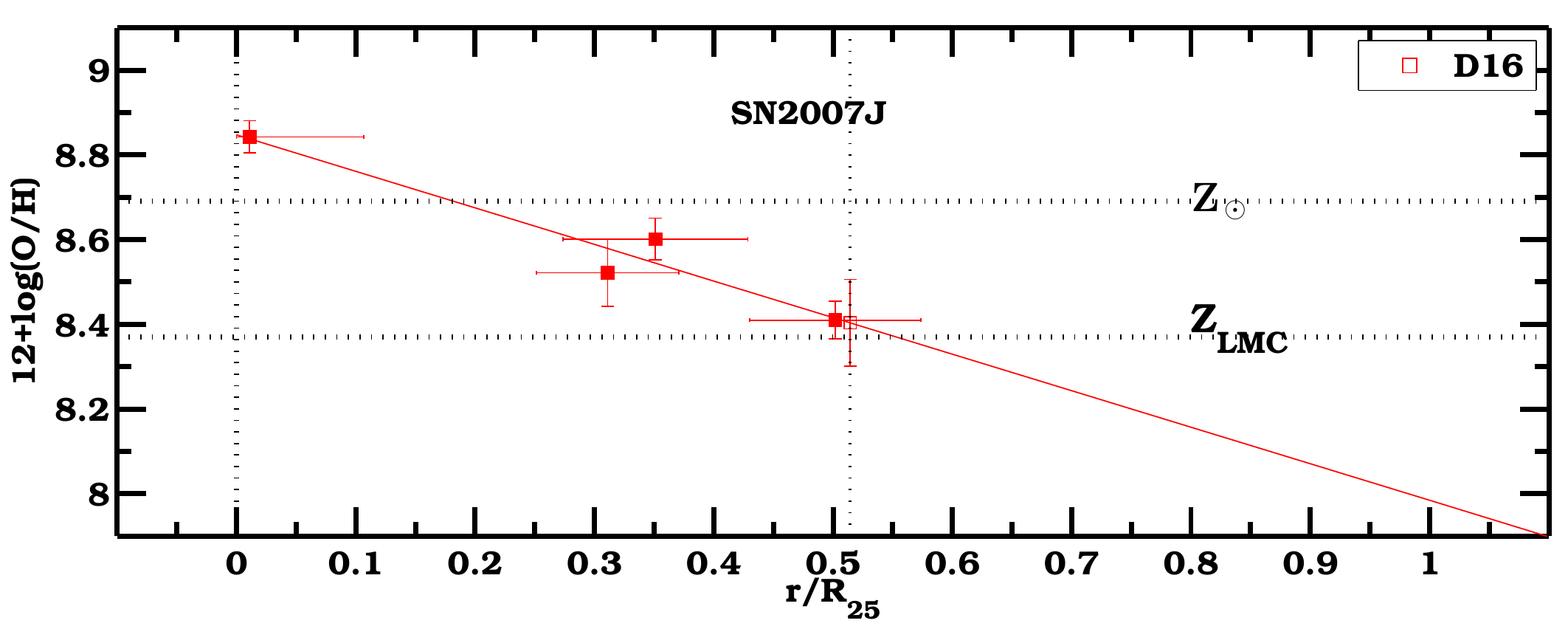}
  \caption{An example of the analysis performed on our NOT observations for UGC~1778 (the host of SN~2007J). {\em Top left:} The open filter acquisition image. Overlaid are the 25th $B$-band magnitude elliptic contour (black solid line), the position of SN~2007J (red star) and the centre of the galaxy (red circle). The slit position is shown, and a color code is used to show the \citetalias{dopita16} metallicity measurements at the position of each \Hii{} region that we inspected. {\em Top right:} The flux at \Ha{} along the slit, with the location of the SN and host nucleus marked. The \citetalias{dopita16} metallicity measurements are shown at the corresponding positions in the top sub-panel. {\em Bottom:} The metallicity gradient of UGC~1778. The linear fit on our D16 measurements is shown with a solid red line. The interpolated metallicity at the SN distance is marked with a empty red square and its uncertainty corresponds to the fit error. The positions of SN and nucleus are marked with vertical dotted lines. The solar metallicity and the LMC metallicity are indicated with horizontal dotted lines.}
  \label{fig:notanalysis}.
 \end{figure*} 

We here detail the analysis of our NOT observations, highlighting differences from the MUSE analysis. Further details of the analysis methods used can be found in \citet{taddia13,taddia15}. Prior to analysis, NOT spectra were corrected for Galactic extinction and dereddened in the same manner as the MUSE data cubes. The main differences from the MUSE analysis were the identification of \Hii{} regions versus spaxel binning routines and the fitting of emission lines due to a lower spectral resolution. Stellar continuum fitting, identification of the ionising source and our metallicity, SFR and age measurements followed those described in \cref{sec:musemethods}.

Intrinsic (deprojected) galactocentric offsets of SN explosion sites and other \Hii{} regions were calculated and normalised to $R_{25}$ following the method of \citet{hakobyan09,hakobyan12}. Host galaxy position angles and inclinations were obtained from NED, supplemented by data from HYPERLEDA\footnote{\url{http://atlas.obs-hp.fr/hyperleda/}}, since we were not able to determine a velocity map as was done for the MUSE data (\cref{sec:methmetallicity}). Again, these were normalised by the $R_{25}$ of the hosts. These deprojected distances were used to determine metallicity gradients for the host galaxies where more than two \Hii{} regions were extracted. For SNe where we could not extract a \Hii{} region underlying the explosion site we used these gradients to estimate the metallicity of the SN. Statistical uncertainties on these interpolated or extrapolated estimates were determined based on the uncertainties of our linear gradient fit (which generally dominate over the rms of \Hii{} regions about the relation, whereas the opposite is true for MUSE, \cref{sec:methmetallicity}). 

Since we only have long-slit spectra for these targets, we do not perform our pixel statistics, as done for the MUSE sample (\cref{sec:pixelstat}).

For two observations, those of UGC~12182 (SN~2014ck) and SDSS J020305.81-035024.5 (SN~2016ado) we were not able to perform our main emission line-based analysis at any location within the slit due to poor observing conditions. We remove these two events from our sample when discussing our spectroscopic results as they provide no meaningful measurements or limits (however we include the host-normalised offset of SN~2014ck, $r/R_{25} = 0.18$).

For the hosts of SN~1999ax and PS~15aic we were only able to extract emission at the host nuclei.\footnote{For SN~1999ax we additionally detect another nearby source of \Hii{} at an apparently higher redshift, this is discussed in \cref{sec:hetero}.} As we have no additional spatial information to be gained from the NOT long-slit spectra, we opted to use existing SDSS spectra which offer improved spectral resolution and depth on which to perform our analysis. 

For SN~2009ku and PS~15csd we take the host nucleus measurements as those of the SN explosion site owing to the small apparent size of the hosts and the very small offset of the SNe.

Within our NOT-observed sample we have two duplicates of MUSE-observed events: SNe~2002cx and 2015H. We found very good agreement (within 1$\sigma$) between our results obtained for the explosion site of SN~2002cx and host nuclei of each (the explosion site of SN~2015H was not detected in emission lines in the NOT data) for the two sets of data. We present our results for these SNe based on the MUSE data results since these are subject to smaller statistical uncertainties owing to greater signal-to-noise of the emission lines and increased spectral resolution, and also allowed for more robust determinations of the metallicity gradients as more \Hii{} regions could be analysed.

\subsection{H{\sc ii} region identification}

We performed an inspection of each acquisition image as well the \Ha{} line flux trace in order to identify \Hii{} regions from which to extract spectra (i.e. peaks in the flux trace). An example of this procedure is shown in the top of \cref{fig:notanalysis}, extraction widths were ad hoc in order to maximise the signal to noise ratio.

For our observation of SN 2008A the slit did not cover the host nucleus and thus we do not have a measurement of the nucleus for this host.

\subsection{Emission line fitting}

Emission line fitting followed broadly the same methodology as for the MUSE spectra (\cref{sec:emisfitting}), however, the lower resolution of the NOT spectra meant we had to use a different fitting routine as nearby lines are blended. Firstly, the \Ha{} and \Nii{} $\lambda\lambda$6548,6583 lines were simultaneously fit with the gaussians of fixed width (determined by the spectral resolution), and fixed centroid offsets using the known wavelengths of the lines. Furthermore, the flux of 
\Nii{} $\lambda$6548 was fixed as one third that of \Nii{} $\lambda$6583 \citep{osterbrock06} -- this assumption was required to allow for a proper fit of this faint line, although it is not used in our metallicity determinations, to remove its contamination from \Ha{}. Similarly, the \Sii{} $\lambda\lambda$6716,6731 doublet was fitted with two gaussians of fixed width and separation. We also attempted fits to \Hb{}, \Oiii{} $\lambda$5007 for each extracted spectrum, however, due to poorer sensitivity in the blue part of the spectrum we were were not able to measure these in several cases.
 
 \begin{table*}
\begin{threeparttable}
     \caption{Results for SN explosion sites and host nuclei.}
     \begin{tabular}{L{3cm}ccccc@{\hskip 0.4cm}clcl}
\hline
Galaxy          & SN                & Offset\tnote{a}             & $L$(H$\alpha$)              & EW(H$\alpha$)     &  NCR  & \multicolumn{2}{c}{$Z$ (SN site)\tnote{b}} & \multicolumn{2}{c}{$Z$ (Host nucleus)\tnote{c}} \\
                &                   & [${R_{25}}$]             & [$\log_{10}$ erg s$^{-1}$]  & [\AA{}]         &       & D16     & \multicolumn{1}{c}{M13} & D16     & \multicolumn{1}{c}{M13}    \\
\hline
\multicolumn{10}{c}{MUSE sample}\\
\hline
IC~344          & 1991bj          &  1.383 &   39.40$\pm$0.01    &  98.9$\pm$3.4  & 0.29           &  8.26$\pm$0.01  &  8.37$\pm$0.01  &  8.55$\pm$0.01  &  8.51$\pm$0.01  \\
UGC~6332        & 2002bp          &  0.920 &   37.13$\pm$0.36    &   1.8$\pm$1.5  & 0.00           &  \ldots         &  8.77$\pm$0.25\tnote{$\dagger$} &  \multicolumn{2}{c}{AGN}        \\
CGCG~044-035    & 2002cx          &  0.765 &   39.19$\pm$0.01    &  45.8$\pm$1.9  & 0.33           &  8.25$\pm$0.02  &  8.36$\pm$0.01  &  8.52$\pm$0.01  &  8.50$\pm$0.01  \\
UGC~11001       & 2004cs\tnote{\ddag}          &  0.519 &   39.23$\pm$0.01    &  78.6$\pm$2.2  & 0.72           &  8.42$\pm$0.02  &  8.43$\pm$0.01  &  8.68$\pm$0.01  &  8.56$\pm$0.02  \\
NGC~5468        & 2005P           &  0.443 &   37.50$\pm$0.02    &   9.2$\pm$0.9  & 0.06\tnote{d}  &  8.07$\pm$0.13  &  8.36$\pm$0.06  &   \multicolumn{2}{c}{not in FOV}  \\
UGC~272         & 2005hk          &  1.089 &   37.25$\pm$0.04    &  20.9$\pm$5.1  & 0.00           &  7.85$\pm$0.28  &  8.34$\pm$0.09  &  8.39$\pm$0.02  &  8.45$\pm$0.01  \\
IC~577          & 2008ae          &  0.881 &   39.06$\pm$0.01    &  58.6$\pm$5.4  & 0.07           &  8.66$\pm$0.03  &  8.50$\pm$0.02  &  8.99$\pm$0.04  &  8.53$\pm$0.01  \\
NGC~1527        & 2008ge          &  0.055 &   \ldots            &  \ldots        & 0.00           &  \ldots         &  \ldots         &  \ldots         &  \ldots         \\
UGC~12682       & 2008ha          &  0.263 &   37.91$\pm$0.01    &  24.9$\pm$1.0  & 0.41\tnote{d}  &  7.91$\pm$0.04  &  8.22$\pm$0.01  &  7.99$\pm$0.02  &  8.19$\pm$0.00  \\
IC~2160         & 2009J           &  0.551 &   38.81$\pm$0.01    &  34.3$\pm$1.7  & 0.00\tnote{d}  &  8.51$\pm$0.03  &  8.41$\pm$0.02  &   \multicolumn{2}{c}{AGN}         \\
ESO~162-017     & 2010ae          &  0.070 &   39.24$\pm$0.01    &  39.7$\pm$0.8  & 0.61           &  8.26$\pm$0.01  &  8.32$\pm$0.00  &  8.28$\pm$0.01  &  8.31$\pm$0.00  \\
NGC~1566        & 2010el          &  0.106 &   38.66$\pm$0.01    &  20.4$\pm$0.6  & 0.03           &  8.85$\pm$0.02  &  8.51$\pm$0.02  &   \multicolumn{2}{c}{AGN}         \\
NGC~6708        & 2011ce          &  0.136 &   39.94$\pm$0.01    &  86.4$\pm$1.9  & 0.59           &  8.76$\pm$0.01  &  8.59$\pm$0.01  &  8.88$\pm$0.01  &  8.60$\pm$0.02  \\
NGC~1309        & 2012Z           &  0.766 &   37.80$\pm$0.01    &  36.1$\pm$1.9  & 0.00\tnote{d}  &  8.26$\pm$0.03  &  8.41$\pm$0.01  &   \multicolumn{2}{c}{AGN}         \\
ESO~114-007     & 2013gr          &  0.266 &   37.78$\pm$0.01    &  31.0$\pm$1.5  & 0.06           &  7.84$\pm$0.05  &  8.27$\pm$0.01  &  7.96$\pm$0.03  &  8.27$\pm$0.01  \\
NGC~4303        & 2014dt          &  0.218 &   38.78$\pm$0.01    &  33.0$\pm$0.6  & 0.04           &  8.85$\pm$0.02  &  8.52$\pm$0.02  &   \multicolumn{2}{c}{not in FOV}  \\
CGCG~048-099    & 2014ey          &  0.616 &   38.87$\pm$0.01    &  47.8$\pm$3.6  & 0.22           &  8.36$\pm$0.04  &  8.43$\pm$0.02  &  8.67$\pm$0.01  &  8.57$\pm$0.02  \\
NGC~3464        & 2015H           &  0.426 &   37.97$\pm$0.02    &   7.6$\pm$0.4  & 0.01           &  8.66$\pm$0.07  &  8.50$\pm$0.10 &  8.97$\pm$0.05  &  8.58$\pm$0.01\tnote{$\dagger$} \\
\hline
\multicolumn{10}{c}{NOT sample}\\
\hline
SDSS J140358.27+155101.2        & 1999ax         & \ldots  & \ldots          & \ldots           & \ldots  &  \ldots           & \ldots          &  7.78$\pm$0.10  & 8.23$\pm$0.02  \\
NGC 7407                        & 2003gq         & 0.133   & 39.57$\pm$0.01  & 23.5$\pm$0.9     & \ldots  &  8.51$\pm$0.07    & 8.33$\pm$0.02\tnote{$\dagger$}   &  \multicolumn{2}{c}{AGN}     \\      
CGCG 283-003                    & 2004gw         & 0.568   & \ldots          & \ldots           & \ldots  &  \ldots           & \ldots          &  9.23$\pm$0.24     &  8.54$\pm$0.02     \\
NGC 5383                        & 2005cc         & 0.061   & 38.81$\pm$0.01  & 28.7$\pm$0.4     & \ldots  &  8.80$\pm$0.07    & 8.50$\pm$0.02\tnote{$\dagger$}   &  8.86$\pm$0.08     & 8.51$\pm$0.02\tnote{$\dagger$}     \\
UGC 6154                        & 2006hn         & 0.498   & 38.98$\pm$0.01  & 18.4$\pm$0.6     & \ldots  &  8.81$\pm$0.17    & 8.50$\pm$0.02   &  8.95$\pm$0.05     & 8.48$\pm$0.01     \\
UGC 1778                        & 2007J\tnote{\ddag}          & 0.514   & 39.35$\pm$0.01  & 116.9$\pm$9.4    & \ldots  &  8.41$\pm$0.04    & 8.38$\pm$0.02   &  8.84$\pm$0.04     & 8.45$\pm$0.01     \\
SDSS J020932.73-005959.8        & 2007qd         & 0.531   & \ldots          & \ldots           & \ldots  &  \ldots           & \ldots          &  8.54$\pm$0.03     & 8.46$\pm$0.01     \\
NGC 634                         & 2008A          & 0.789   & \ldots          & \ldots           & \ldots  &  \ldots           & \ldots          &  \multicolumn{2}{c}{not in FOV}        \\                                         
APMUKS(BJ) B032747.73-281526.1  & 2009ku\tnote{e}& \ldots     & 40.41$\pm$0.01          & 47.86$\pm$0.82        & \ldots     &  8.43$\pm$0.06    & 8.37$\pm$0.01   &  8.43$\pm$0.06  & 8.37$\pm$0.01  \\
NGC 2315                        & 2011ay         & 0.468   & \ldots          & \ldots           & \ldots  &  \ldots           & \ldots          &  \ldots     & 8.69$\pm$0.01     \\
CGCG 205-021                    & PS1-12bwh         & 0.318   & \ldots          & \ldots           & \ldots  &  \ldots           & \ldots          &  8.98$\pm$0.07  & 8.56$\pm$0.01  \\  
NGC 5936                        & 2013dh         & 0.219   & 39.14$\pm$0.01  & 34.3$\pm$3.7     & \ldots  &  8.77$\pm$0.05    & 8.58$\pm$0.01   &  9.18$\pm$0.03     & 8.63$\pm$0.01     \\
UGC 11369                       & 2013en         & 0.480   & 39.32$\pm$0.01  & 20.7$\pm$0.7     & \ldots  &  8.48$\pm$0.09    & 8.46$\pm$0.03\tnote{$\dagger$}   &  9.05$\pm$0.02     & 8.63$\pm$0.01\tnote{$\dagger$}     \\
UGC 12850                       & 2014ek         & 0.248   & 39.83$\pm$0.01  & 34.8$\pm$0.4     & \ldots  &  \ldots           & 8.50$\pm$0.02\tnote{$\dagger$}   &  \multicolumn{2}{c}{AGN}    \\
UGC 12156                       & 2015ce         & 0.394   & \ldots          & \ldots           & \ldots  &  \ldots           & \ldots          &  8.92$\pm$0.09     & 8.63$\pm$0.01     \\
SDSS J133047.95+380645.0        & PS~15aic          & 0.666   & 39.50$\pm$0.02  & 2.3$\pm$0.1      & \ldots  &  \ldots           & \ldots          &  \ldots         & 8.72$\pm$0.02\tnote{$\dagger$}  \\
SDSS J020455.52+184815.0        & PS~15csd\tnote{e} & \ldots  & 39.55$\pm$0.02  & 25.1$\pm$1.0     & \ldots  &  8.11$\pm$0.17    & 8.29$\pm$0.03   &  8.11$\pm$0.17  & 8.29$\pm$0.03  \\
\hline  
     \end{tabular}
\label{tab:results}
 \begin{tablenotes}
 \item[$\dagger$] The N2 relation of \citetalias{marino13} was used instead of O3N2.
 \item[\ddag] Helium detected in SN spectra.
 \item[a] The deprojected offset of the explosion site normalised by the $R_{25}$ value of the host.
 \item[b] The $12 + \log_{10}(\textrm{O/H})$ abundance measured in the scales of \citetalias{dopita16} and \citetalias{marino13} for the SN explosion site. Uncertainties are statistical only: \citetalias{marino13} have a systematic uncertainty of 0.18 dex; \citetalias{dopita16} do not have well quantified systematics but appear lower than $T_e$-based estimates \citep{kruehler17}.
 \item[c] As for $^\text{b}$ but for the host nucleus. `AGN' means the host was deemed to host an active galactic nucleus, `not in FOV' means the host nucleus was not captured in the IFU or slit.
 \item[d] NCR value taken from \citet{lyman13}.
 \item[e] Due to the SN being overlaid on the nucleus of an unresolved host at very small offset, we use the host nucleus results as those of the SN site also.

 \end{tablenotes}
\end{threeparttable}
\end{table*}

\begin{landscape}
\begin{table}
\begin{center}
\begin{threeparttable}
     \caption{Metallicity gradients for SN Iax host galaxies and SN explosion site metallicities. The metallicity gradients for SN Iax host galaxies are parameterised as $Z = m/R_{25} \times D + c$, where $D$ is the deprojected offset normalised by the host $R_{25}$, in the scales of \citetalias{dopita16} and \citetalias{marino13}. $Z_\textrm{SN, grad}$ are estimates of $Z$ based on these metallicity gradients at the offset of the SN, and $Z_\textrm{SN, meas}$ are our directly measured values for the \Hii{} underlying the SN position.}
     \begin{tabular}{ll@{\hskip 0.75cm}rccc@{\hskip 0.75cm}rccc}
\hline
Galaxy          & SN name         & \multicolumn{4}{c}{D16}                            & \multicolumn{4}{c}{M13}                            \\
                &                 & $m/R_{25}$        & $c$               & $Z_\textrm{SN, grad}$ & $Z_\textrm{SN, meas}$ & $m/R_{25}$        & $c$               & $Z_\textrm{SN, grad}$ & $Z_\textrm{SN, meas}$ \\
\hline
\multicolumn{10}{c}{MUSE sample}\\
\hline
IC~344 & 1991bj        & $-0.27\pm0.08$ & $ 8.59\pm0.08$ & $ 8.22\pm0.05 $ & $ 8.26\pm0.01 $ & $-0.13\pm0.07$ & $ 8.51\pm0.06$ & $ 8.33\pm0.04 $ & $ 8.37\pm0.01 $ \\
UGC~6332\tnote{$\dagger$} &2002bp        & $ 0.07\pm0.19$ & $ 8.82\pm0.12$ & $ 8.88\pm0.08 $ & \ldots & $ 0.01\pm0.06$ & $ 8.56\pm0.04$ & $ 8.57\pm0.02 $ & $ 8.77\pm0.25 $ \\
CGCG~044-035 & 2002cx        & $-0.43\pm0.11$ & $ 8.54\pm0.05$ & $ 8.21\pm0.03 $ & $ 8.25\pm0.02 $ & $-0.18\pm0.08$ & $ 8.49\pm0.03$ & $ 8.35\pm0.02 $ & $ 8.36\pm0.01 $ \\
UGC~11001 & 2004cs        & $-0.28\pm0.13$ & $ 8.62\pm0.07$ & $ 8.47\pm0.05 $ & $ 8.42\pm0.02 $ & $-0.22\pm0.07$ & $ 8.56\pm0.04$ & $ 8.45\pm0.02 $ & $ 8.43\pm0.01 $ \\
NGC~5468 & 2005P         & $-0.68\pm0.15$ & $ 8.70\pm0.08$ & $ 8.40\pm0.05 $ & $ 8.07\pm0.13 $ & $-0.38\pm0.10$ & $ 8.57\pm0.05$ & $ 8.40\pm0.04 $ & $ 8.36\pm0.06 $ \\
UGC~272 & 2005hk        & $-0.38\pm0.10$ & $ 8.30\pm0.06$ & $ 7.88\pm0.05 $ & $ 7.85\pm0.28 $ & $-0.18\pm0.08$ & $ 8.41\pm0.05$ & $ 8.22\pm0.04 $ & $ 8.34\pm0.09 $ \\
IC~577 & 2008ae        & $-0.23\pm0.11$ & $ 8.88\pm0.07$ & $ 8.68\pm0.04 $ & $ 8.66\pm0.03 $ & $-0.17\pm0.05$ & $ 8.64\pm0.04$ & $ 8.49\pm0.03 $ & $ 8.50\pm0.02 $ \\
UGC~12682 & 2008ha        & $-0.14\pm0.13$ & $ 7.97\pm0.07$ & $ 7.93\pm0.07 $ & $ 7.91\pm0.04 $ & $-0.06\pm0.12$ & $ 8.25\pm0.06$ & $ 8.24\pm0.05 $ & $ 8.22\pm0.01 $ \\
IC~2160 & 2009J         & $-0.79\pm0.23$ & $ 8.97\pm0.10$ & $ 8.53\pm0.06 $ & $ 8.51\pm0.03 $ & $-0.50\pm0.20$ & $ 8.66\pm0.10$ & $ 8.38\pm0.05 $ & $ 8.41\pm0.02 $ \\
ESO~162-017 & 2010ae        & $-0.24\pm0.17$ & $ 8.26\pm0.07$ & $ 8.24\pm0.06 $ & $ 8.26\pm0.01 $ & $-0.08\pm0.16$ & $ 8.35\pm0.06$ & $ 8.35\pm0.05 $ & $ 8.32\pm0.01 $ \\
NGC~1566 & 2010el        & $-0.64\pm0.75$ & $ 8.91\pm0.10$ & $ 8.84\pm0.05 $ & $ 8.85\pm0.02 $ & $ 0.13\pm0.44$ & $ 8.55\pm0.07$ & $ 8.57\pm0.03 $ & $ 8.51\pm0.02 $ \\
NGC~6708 & 2011ce        & $-0.54\pm0.10$ & $ 8.85\pm0.05$ & $ 8.78\pm0.04 $ & $ 8.76\pm0.01 $ & $-0.30\pm0.09$ & $ 8.62\pm0.04$ & $ 8.58\pm0.04 $ & $ 8.59\pm0.01 $ \\
NGC~1309 & 2012Z         & $-0.41\pm0.14$ & $ 8.60\pm0.05$ & $ 8.29\pm0.05 $ & $ 8.26\pm0.03 $ & $-0.24\pm0.10$ & $ 8.55\pm0.04$ & $ 8.37\pm0.03 $ & $ 8.41\pm0.01 $ \\
ESO~114-007 & 2013gr        & $-0.23\pm0.14$ & $ 7.99\pm0.06$ & $ 7.93\pm0.06 $ & $ 7.84\pm0.05 $ & $-0.16\pm0.18$ & $ 8.29\pm0.08$ & $ 8.24\pm0.07 $ & $ 8.27\pm0.01 $ \\
NGC~4303 & 2014dt        & $-0.27\pm0.30$ & $ 8.88\pm0.07$ & $ 8.82\pm0.05 $ & $ 8.85\pm0.02 $ & $ 0.07\pm0.28$ & $ 8.55\pm0.07$ & $ 8.57\pm0.03 $ & $ 8.52\pm0.02 $ \\
CGCG~048-099 & 2014ey        & $-0.36\pm0.09$ & $ 8.65\pm0.06$ & $ 8.42\pm0.06 $ & $ 8.36\pm0.04 $ & $-0.17\pm0.07$ & $ 8.51\pm0.06$ & $ 8.40\pm0.05 $ & $ 8.43\pm0.02 $ \\
NGC~3464 & 2015H         & $-0.51\pm0.20$ & $ 8.92\pm0.08$ & $ 8.70\pm0.05 $ & $ 8.66\pm0.07 $ & $-0.52\pm0.11$ & $ 8.77\pm0.05$ & $ 8.55\pm0.03 $ & $8.50\pm0.10$ \\
\hline
\multicolumn{10}{c}{NOT sample}\\
\hline
NGC 7407\tnote{$\dagger$}                   &  SN~2003gq   & $-0.04\pm0.43$                  & $8.61\pm0.13$                 & $8.60\pm0.14$          & $8.51\pm0.07$         & $ 0.30\pm0.32$                  & $8.38\pm0.10$                 & $8.42\pm0.11$          & $8.33\pm0.02$         \\
CGCG 283-003\tnote{$\dagger$}               &  SN~2004gw   & $-1.21\pm0.21$                  & $9.18\pm0.07$                 & $8.03\pm0.21$          & \ldots                & $-0.12\pm0.08$                  & $8.51\pm0.02$                 & $8.39\pm0.08$          & \ldots                \\ 
NGC 5383\tnote{$\dagger$}                   &  SN~2005cc   & $-0.97\pm0.21$                  & $8.86\pm0.01$                 & $8.80\pm0.02$          & $8.80\pm0.07$         & $-0.30\pm0.30$                  & $8.51\pm0.01$                 & $8.49\pm0.02$          & $8.50\pm0.02$         \\ 
UGC 6154                                    &  SN~2006hn   & $-0.14\pm0.03$                  & $8.94\pm0.02$                 & $8.82\pm0.03$          & $8.81\pm0.17$         & $ 0.06\pm0.10$                  & $8.49\pm0.06$                 & $8.55\pm0.11$          & $8.50\pm0.02$         \\
UGC 1778                                    &  SN~2007J    & $-0.86\pm0.16$                  & $8.85\pm0.06$                 & $8.40\pm0.10$          & $8.41\pm0.04$         & $-0.13\pm0.02$                  & $8.45\pm0.01$                 & $8.38\pm0.01$          & $8.38\pm0.02$         \\ 
SDSS J020932.73-005959.8\tnote{$\dagger$}   &  SN~2007qd   & $0.07$                          & $8.54$                        & $8.60\pm0.15$\tnote{a} & \ldots                & $0.04$                          & $8.46$                        & $8.50\pm0.15$\tnote{a} & \ldots                \\ 
NGC 2315\tnote{$\dagger$}                   &  SN~2011ay   &  \ldots                         & \ldots                        & \ldots                 & \ldots                & $-0.42 $                        & $8.69$                        & $8.50\pm0.15$\tnote{a} & \ldots                \\
CGCG 205-021\tnote{$\dagger$}               &  PS1-12bwh   & $-0.42$                         & $8.98$                        & $8.84\pm0.15$\tnote{a} & \ldots                & $ 0.00\pm0.05$                  & $8.51\pm0.04$                 & $8.51\pm0.04$          & \ldots                \\
NGC 5936                                    &  SN~2013dh   & $-1.92\pm0.55$                  & $9.17\pm0.10$                 & $8.75\pm0.16$          & $8.77\pm0.05$         & $-0.21\pm0.02$                  & $8.63\pm0.00$                 & $8.59\pm0.00$          & $8.58\pm0.01$         \\
UGC 11369\tnote{$\dagger$}                  &  SN~2013en   & $-0.96\pm0.37$                  & $9.07\pm0.14$                 & $8.61\pm0.24$          & $8.48\pm0.09$         & $-0.28\pm0.11$                  & $8.63\pm0.04$                 & $8.50\pm0.07$          & $8.46\pm0.03$         \\
UGC 12850\tnote{$\dagger$}                  &  SN~2014ek   &  \ldots                         & \ldots                        & \ldots                 & \ldots                & $-0.00\pm0.03$                  & $8.48\pm0.02$                 & $8.48\pm0.02$          & $8.50\pm0.02$         \\ 
UGC 12156\tnote{$\dagger$}                  &  SN~2015ce   & $-1.49\pm0.25$                  & $8.88\pm0.08$                 & $8.29\pm0.13$          & \ldots                & $-0.36\pm0.16$                  & $8.61\pm0.05$                 & $8.46\pm0.08$          & \ldots                \\ 
\hline                                                                                                                                                                                                                                         
     \end{tabular}
\label{tab:zgrad}
 \begin{tablenotes}
 \item[$\dagger$] The N2 relation of \citetalias{marino13} was used instead of O3N2.
 \item[a] Only two \Hii{} regions were used to determine the gradient. We adopt an uncertainty of $\pm$0.15 dex on $Z_\textrm{SN, grad}$.
 \end{tablenotes}
\end{threeparttable}
\end{center}
\end{table}
\end{landscape}

\section{Results}
\label{sec:results}

We present our main results of the SNe explosion sites and host nuclei in \cref{tab:results}. We indicate those observations where the host nucleus was not captured in the FOV or where the ionising source was deemed to be powered by an AGN (\cref{sec:ionising,fig:bptmap}). Individual line fluxes for the explosion sites (where they could be measured) are given in \cref{tab:fluxes}.

\subsection{Metallicity}
\label{sec:resz}

\subsubsection{Gradients and estimating explosion site metallicities}
\label{sec:zgrads}

In some cases we were not able to directly measure emission lines at the explosion site of the SN. As discussed in \cref{sec:notmethods}, in these cases we resort to estimating the metallicity based on the observed metallicity gradient of the galaxy as has been done previously for other studies of SN environments.

\begin{figure}
 \centering
\subfloat{\includegraphics[width=\columnwidth]{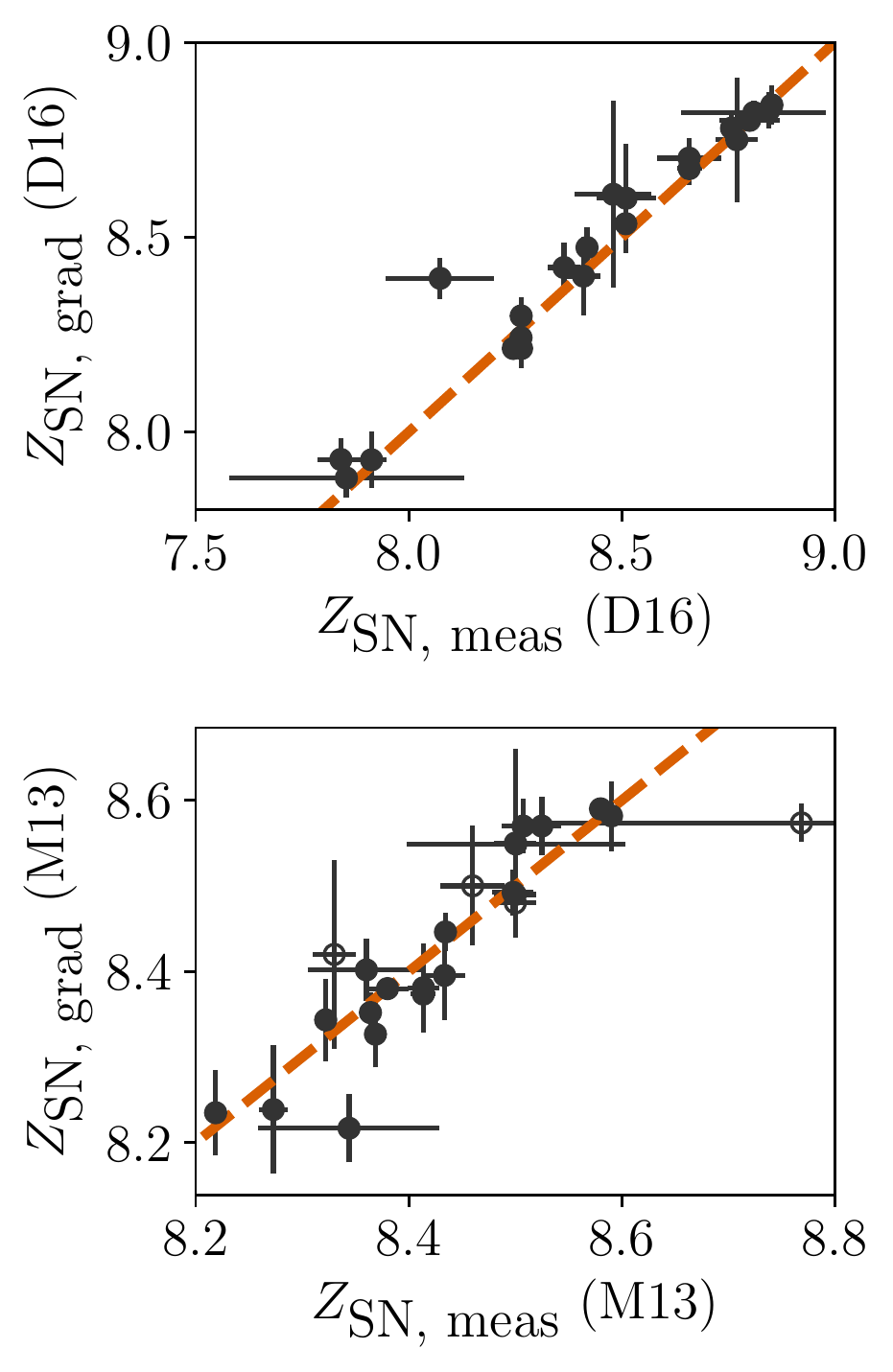}}
 \caption{A comparison between the metallicity determined directly at the location of SN explosion sites ($Z_\textrm{SN, meas}$) and that which would be estimated based on the metallicity gradient alone ($Z_\textrm{SN, grad}$). Metallicities are determined using the calibrations of \citetalias{dopita16} and \citetalias{marino13} in the top and bottom panels, respectively. Empty markers in the lower panel indicate the N2 relation was used (versus O3N2 for filled markers). Uncertainties on the measured metallicities are statistical only. The gradient uncertainties are either the uncertainty on the gradient linear fit, or the root mean square of \Hii{} region metallicity residuals about the gradient (see text). Orange dashed lines indicate the one to one relation for each.}
 \label{fig:Zgrad}
\end{figure}

Here we utilise the power of the MUSE data, in combination with the results of NOT to assess the reliability of this method in the extreme regimes of many and few \Hii{} regions from which to determine the gradient. In \cref{tab:zgrad} we present our determined metallicity gradients alongside the estimated (based on the gradient) and measured metallicities of the SN explosion sites, which are plotted in \cref{fig:Zgrad}.\footnote{We exclude NGC~1527, the host of SN~2008ge, as even if the ionised gas we detect is driven by SF (see \cref{sec:08ge}), it is very irregular in morphology, and perhaps not intrinsic to the luminous S0 host. Thus a metallicity gradient is not likely to be an appropriate description.} The estimates agree very well with the measured values -- the one outlier in the figure is SN~2005P, although this is still only at the $\sim2.3\sigma$ level and interestingly only in the \citetalias{dopita16} indicator. We crucially also see that even with a low number of regions from which to fit a metallicity gradient (in the case of NOT data), the measured and estimated values agree well, offering further support of results derived using this method. The associated uncertainties (determined from the residuals of the measured \Hii{} region metallicities about the gradient or based on the uncertainties of the linear fit; \cref{sec:musemethods,sec:notmethods}), are generally larger than the statistical uncertainties on the measured values, but are still mainly dominated by the systematic uncertainties on the calibrations. We note that these findings corroborate those of \citet{galbany16b} who investigated a number of local metallicity estimators in IFS data, finding that an interpolation of the metallicity gradient is a robust estimate.

For some observations we were not able to either measure the metallicity at the SN explosion site or determine a metallicity gradient. In the cases of SNe~1999ax, 2009ku, PS~15aic and PS~15csd we were only able to measure the host nucleus metallicities, and for SN~2008A we could only measure a single bright \Hii{} region slightly offset from the host nucleus\footnote{In agreement with our findings based on the NOT spectroscopy here, \citet{lyman13} show a \Ha{} image of the SN location with no underlying detected emission and mainly faint, diffuse emission throughout the host. \citet{mccully14b} also present {\em HST} broadband observations of the environment, showing it to be in the outskirts of its host, in a fairly low density environment.}. SN~2009ku and PS~15csd lie on top of their respective hosts which are of small apparent size and so we adopt the host nucleus metallicities as those of the SN (\cref{tab:results}). Our determined metallicity gradients cover a large range, making any estimates for SNe~1999ax, 2008A or PS~15csd based on an average gradient of limited use as the associated error would be very large (\citealt{galbany16b} also caution against this methodology). We are limited to supposing their metallicities are less than or equal to their respective host nucleus metallicities (for SN~2008A our sole \Hii{} region metallicity is at an offset of $r$/$R_{25}$ = 0.168 with a metallicity of $Z_\textrm{D16} = 9.10\pm0.07_{stat}$ dex, $Z_\textrm{M13, N2} = 8.60\pm0.01_{stat}$ dex). This does not introduce the requirement for especially low metallicities (compared to the rest of the sample) for these events.

Given the very good agreement between our direct explosion site metallicities and those estimated based on the hosts' gradients, we supplement our explosion site metallicity values with gradient estimates where appropriate, and present and discuss this enlarged sample in our results.

\subsubsection{Distribution}
\label{sec:reszdist}
\begin{figure}
 \centering
\subfloat{\includegraphics[width=\columnwidth]{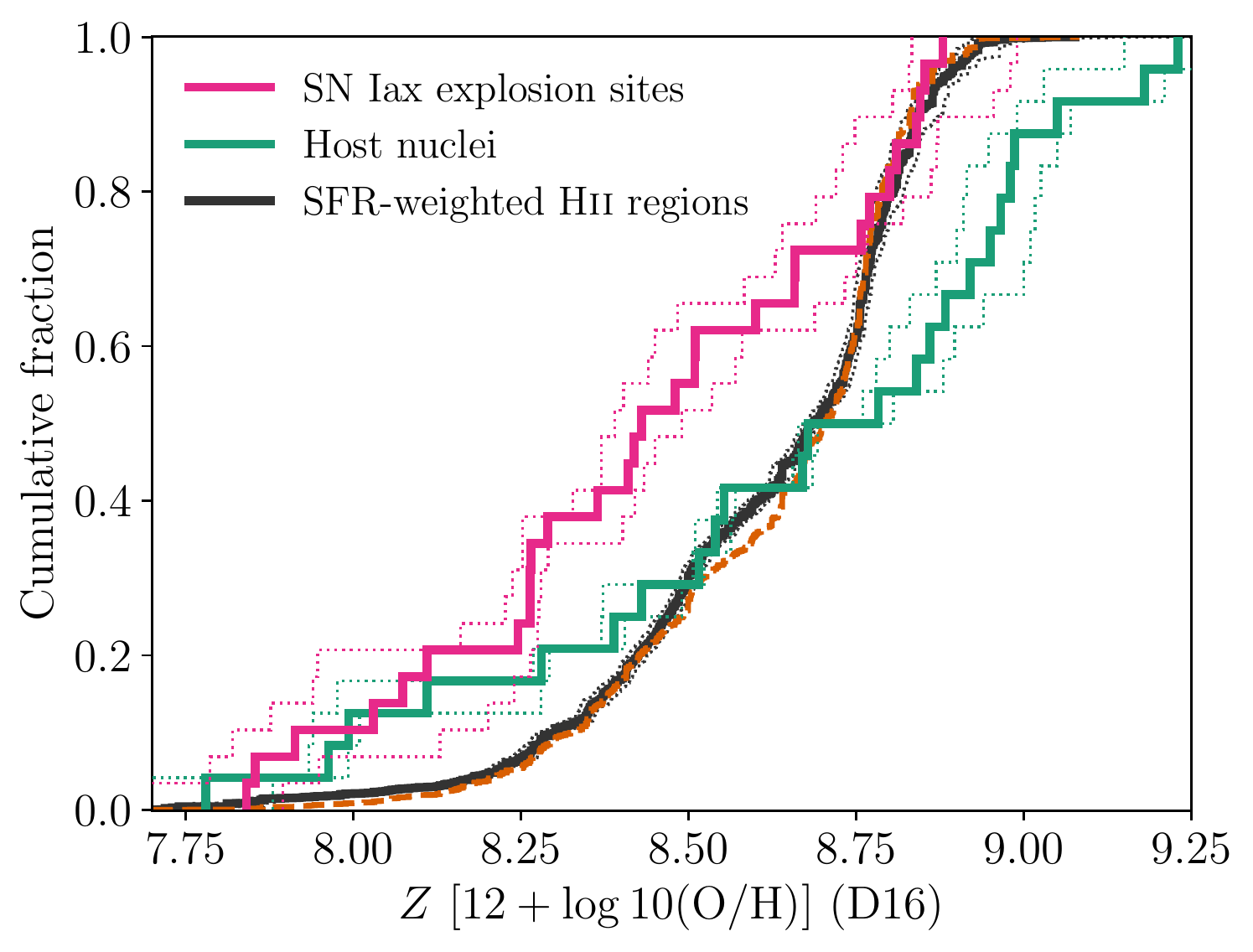}}\\
\subfloat{\includegraphics[width=\columnwidth]{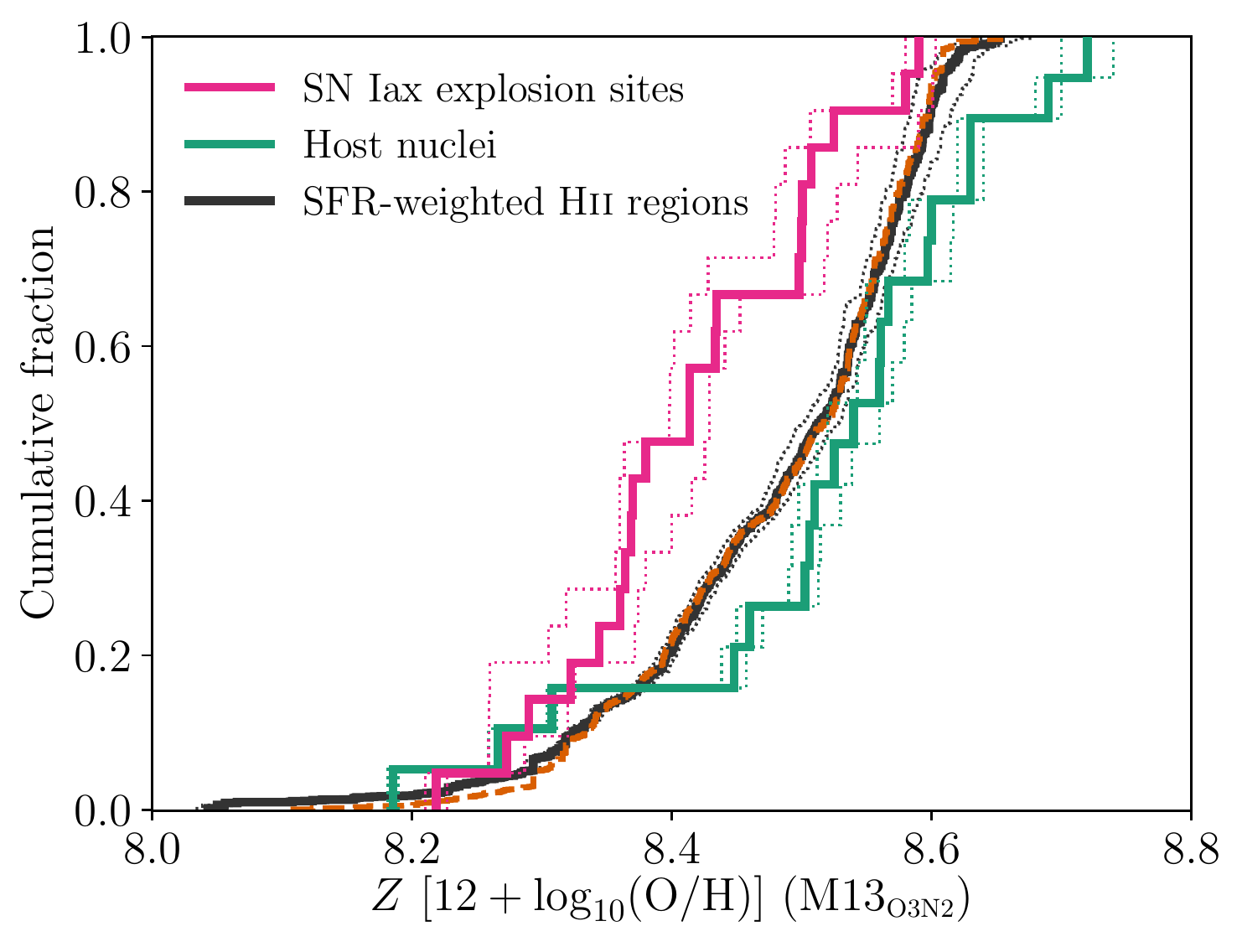}}
 \caption{The metallicity distributions of SNe~Iax explosion sites, their host nuclei, and all detected H{\sc ii} regions in the MUSE sample. Metallicities are determined using the calibrations of \citetalias{dopita16} and \citetalias{marino13} (O3N2) in the top and bottom panels, respectively. When appropriate, SN explosion site metallicity estimates based on the gradients of the hosts are included (see \cref{sec:zgrads}). Dotted lines are the distributions after adding/subtracting $1\sigma$ statistical uncertainties to/from all values and are somewhat extreme limits for the true distribution. The cumulative sum of SN explosion sites and host nuclei are unweighted (i.e. a step in these histograms equals one SN or host nucleus, as appropriate). The H{\sc ii} regions have been weighted by their \Ha{} luminosity such that this histogram shows the cumulative fraction of all detected SF with metallicity across the MUSE sample. The thin orange dashed line shows the sub sample of MUSE \Hii{} regions for which the observations covered the whole host and appears almost indistinguishable from the full \Hii{} region sample (see text).} 
 \label{fig:Zall}
\end{figure}

In \cref{fig:Zall} the metallicity distribution of SNe Iax explosion sites is shown alongside that of the host nuclei and all \Hii{} regions extracted in our MUSE sample\footnote{We do not include the emission from the host of SN~2008ge, see \cref{sec:08ge}.}. The cumulative weight of each \Hii{} region in this plot is given by its \Ha{} luminosity, i.e. SFR, in order to show the cumulative distribution of SFR with metallicity in SN~Iax hosts. Although we only have global metallicity determinations for the MUSE half of our sample, the main discriminating factor between this and the NOT sample is only declination of the source, and so we expect no bias in terms of host properties. Additionally, a few of our MUSE observations do not cover the entire hosts, but they do cover areas around the same location and offset as the SNe, and cover reasonable fractions of their respective hosts. As a check, we also plot in \cref{fig:Zall} the metallicity distribution of SF in our MUSE sample only including observations that cover approximately all the host galaxy (i.e., we exclude observations of SNe 2005P, 2010el, 2012Z, 2013gr and 2014dt; see \cref{fig:zstamps}). The distribution for this sub-sample of our MUSE galaxies is almost identical to the full MUSE sample (as was found in our other environmental measure also) and so the partially covered hosts do not introduce significant biases into the distributions. 

We find that the metallicity of SNe~Iax sites are at lower metallicities than where typical SF is occurring in the hosts (in each metallicity indicator shown the KS test gives $p \sim 0.01$ that the two are drawn from the same distribution). As expected, given the SN explosion sites are offset from their host nuclei and we find mostly negative radial metallicity gradients in our hosts, there is an offset between the host nuclei and explosion site metallicities.

\subsubsection{Metallicity ranking of explosion sites}
\label{sec:reszrank}
\begin{figure}
 \centering
\subfloat{\includegraphics[width=\columnwidth]{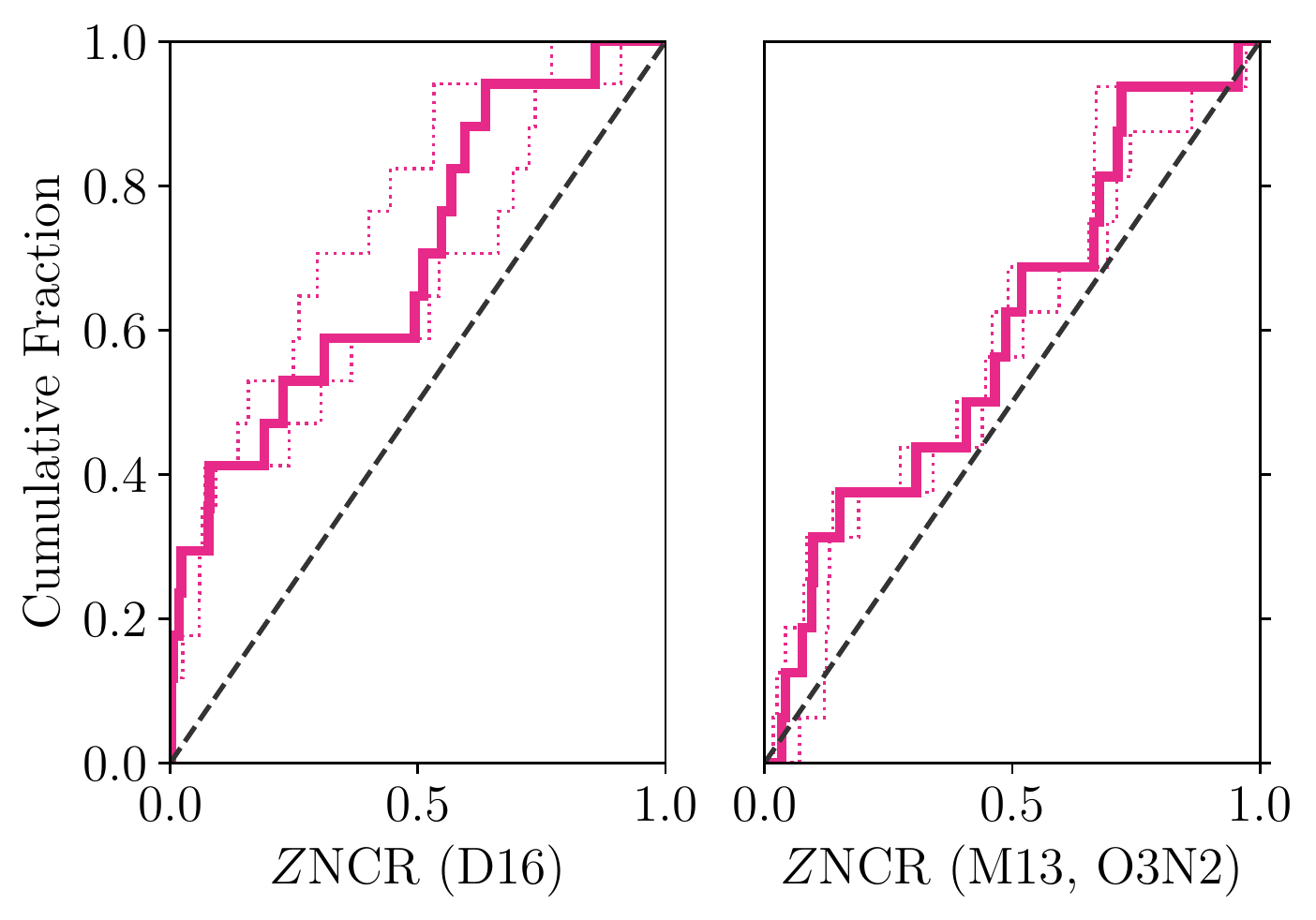}}
 \caption{The SFR-weighted metallicity NCR ($Z$NCR) values of SNe~Iax explosion sites in two metallicity indicators, as observed with MUSE. In the case of metallicity-unbiased production of the progenitors, the SNe should create a uniform distribution in $Z$NCR (thick dashed lines). Uncertainties on the distributions were found by creating many realisations of the cumulative sums (based on the uncertainties of the values) for each SN host and recalculating each explosion site's rank. Dotted lines represent the lower and upper distributions of the 95~per~cent confidence interval of the ranks.} 
 \label{fig:ZNCR}
\end{figure}

In order to further assess the indications of \cref{sec:reszdist}, we have calculated SFR-weighted metallicity NCR ($Z$NCR) values for the explosion sites of out MUSE sample of SNe~Iax. The $Z$NCR is calculated in a similar manner to the traditional NCR value of SNe \citep[\cref{sec:pixelstat}, see also][]{james06, anderson12}, but the cumulative distribution of the SFR-weighted metallicity in the galaxy is used to determine the rank of each SN. The $Z$NCR is thus the fraction of stars being formed at or below the metallicity of the explosion site in that host. This may be seen visually in the top right panel of \cref{fig:d16map}, where the SN in question has $Z$NCR = 0.003. In \cref{fig:ZNCR} we show the distributions of $Z$NCR for \citetalias{dopita16} and \citetalias{marino13}. In the case of metallicity-unbiased formation of the progenitors, a uniform distribution (indicated by dashed lines) should be recovered, i.e. the SNe have no dependence on where in the metallicity distribution of SF they form, and so sample it uniformly. We see for both distributions the SN explosion sites are systematically shifted to lower values than the uniform distribution, indicating the SNe are preferentially exploding in metal poor regions, and not unbiasedly tracing the SF in their hosts. For \citetalias{marino13}, although the distribution is systematically below the uniform distribution, the difference is more marginal than for \citetalias{dopita16}.

\subsection{Ages and SFR}
\label{sec:resultsages}

\begin{figure}
 \centering
\subfloat{\includegraphics[width=\columnwidth]{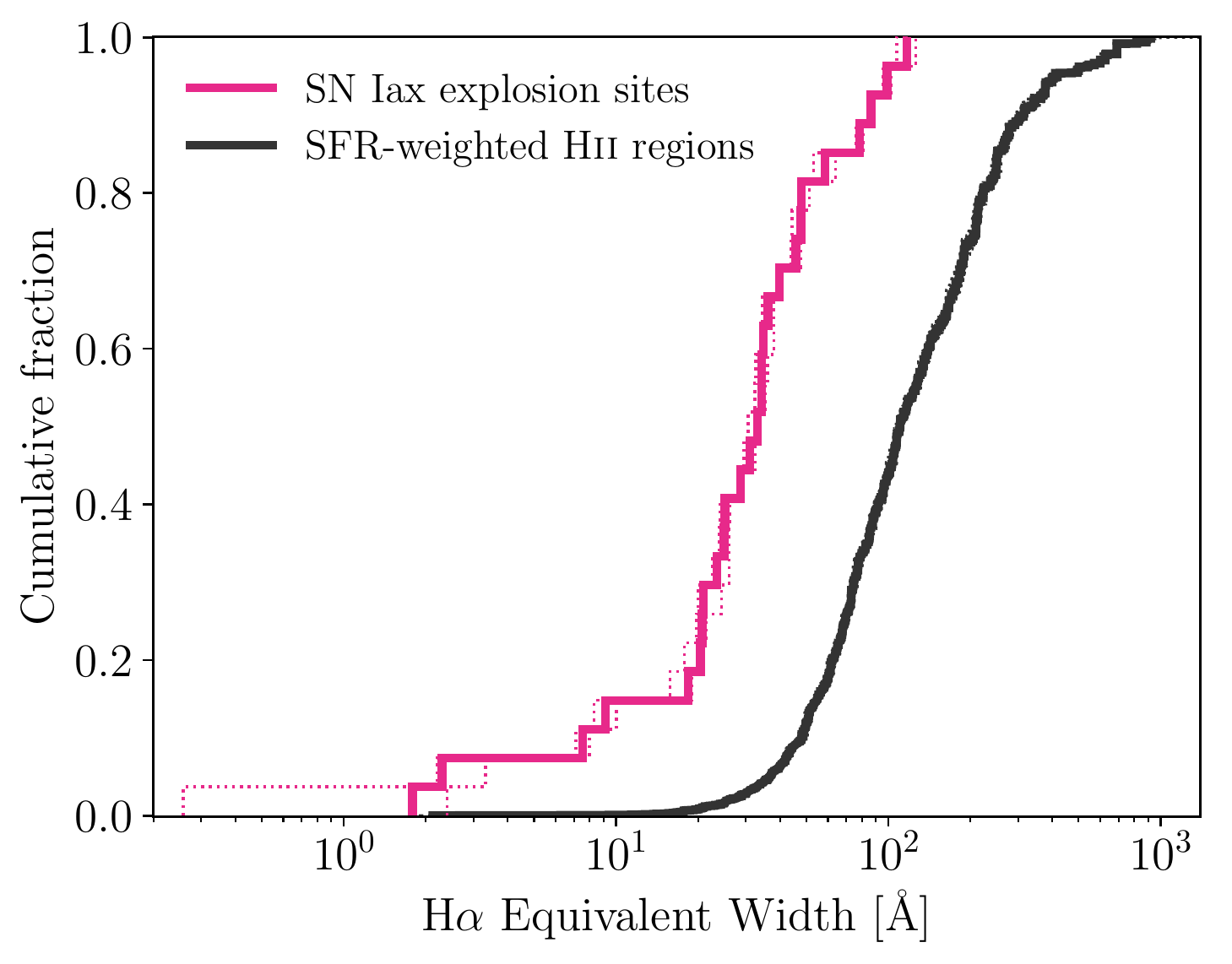}}
 \caption{As for \cref{fig:Zall} but here showing EW(\Ha{}) of SN~Iax explosion sites.} 
 \label{fig:HaEWall}
\end{figure}

\begin{figure}
 \centering
\subfloat{\includegraphics[width=\columnwidth]{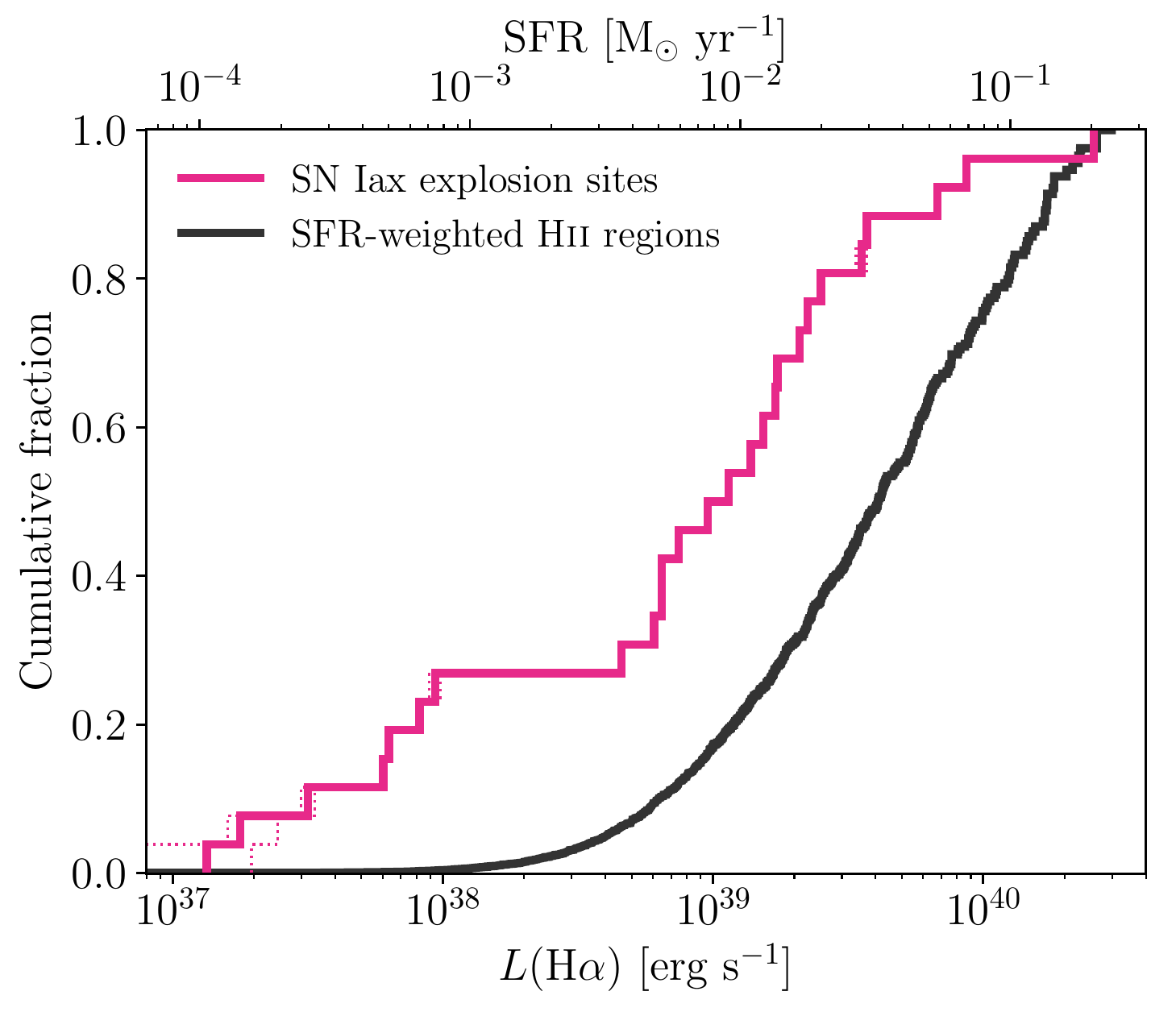}}
 \caption{As for \cref{fig:Zall} but here showing L(\Ha{}) of SN~Iax explosion sites. The top axis indicates the corresponding SFR \citep{kennicutt98}.} 
 \label{fig:Halumall}
\end{figure}

We show the EW(\Ha{}) and L(\Ha{}) measurements for our explosion site bins in \cref{fig:HaEWall,fig:Halumall}, respectively. As was done for the metallicities, we also show the cumulative distribution of SF in these measurements for our MUSE-observed hosts. The regions of SF at the locations of SNe~Iax are occurring at significantly lower EW and luminosities than the overall SF in the hosts (KS test $>3\sigma$ for each). Adopting a \Ha{} luminosity--SFR relation \citep{kennicutt98}, the \Ha{} luminosities imply that the explosions sites are typically of lower SFR than that of the population of SF regions in the hosts, ranging from $10^{-4}-10^{-2}$ \msun{}~yr$^{-1}$.

As discussed in \cref{sec:methages}, ages for the young stellar component of star forming regions can be estimated from EW measurements of emission lines but are subject to sources of uncertainty. The bulk of our explosion site EW(\Ha{}) measurements are $\sim$tens, up to $\sim 100$~\AA{}. We show in \cref{fig:bpassew} (see \cref{sec:methages}) the evolution of EW(\Ha{}) with age for {\sc bpass} models for an instantaneous SF episode at a range of metallicities. As the consensus is that the vast majority of (particularly massive) stars are in some form of binaries \citep[e.g.][]{sana12} we compare results with the binary population models.

Our observed range of EW(\Ha{}) suggests young SP components of several~$\times 10^7$ to $10^8$~yr at the locations of SNe Iax. We again note the inherent uncertainties in selecting a single fiducial model for the nebular gas properties, however these values indicate that there are moderately young SP components at the location of the majority of SNe~Iax explosion sites.\footnote{Although we were not able to measure emission lines at the locations of some NOT-observed examples, these data were comparatively shallower than the MUSE data, where signatures of ongoing SF were found for all but one explosion site (SN~2008ge). This is demonstrated by our duplicate observations of SN~2015H: in the NOT data we were not able to extract an emission line spectrum at the explosion site but we could in the MUSE data.} Notably, although our EW values are actually somewhat lower limits of the true young SP EW (as there will be an existing underlying, older SP that contributes to the continuum but not emission lines), they are not exceptionally high. This would seem to disfavour very young, and therefore very massive, SPs at their locations with ages of several Myr, since initially very high EW values tend to drop off quickly and largely independent of model differences and reasonable gas parameters.

As an additional check for the presence of very young SPs at the location of SNe~Iax we also attempted fitting for He {\sc i} $\lambda 4922$ in our emission fitting routine (\cref{sec:emisfitting}) for the MUSE sample.\footnote{Our comparatively shallower NOT data with lower spectral resolution is not conducive to providing meaningful detections of this faint line, which is close to other, much stronger features.} He {\sc i} $\lambda 4922$ in emission is present only for ages up to a few Myr \citep[e.g.][]{gonzalezdelgado99}. We did not detect this line at any SN~Iax explosion site. Within our MUSE data we found detections of He {\sc i} $\lambda 4922$ within 3~kpc of the SN for SNe~2008ha, 2009J, 2010ae, 2012Z and 2013gr. Assuming a 5~Myr age for the young SP (also consistent with EW(\Ha{}) $\gtrsim 100$ found in these regions), the average velocities of the SN~Iax progenitors would have to have been in excess of 234, 128, 269, 580, 159 \kms{}, respectively, in order to have originated from these regions. 

Two events with higher EW(\Ha{}) measurements at their explosion sites are debated members of the SN~Iax sample as they displayed helium and perhaps hydrogen in their spectra. The removal of these events (SNe~2004cs and 2007J) would lean the typical ages of the young SPs at the locations of the rest of the sample slightly higher (\cref{sec:hetero}).

\subsection{Offsets}
\label{sec:resultsoffsets}

\begin{figure}
 \centering
\subfloat{\includegraphics[width=\columnwidth]{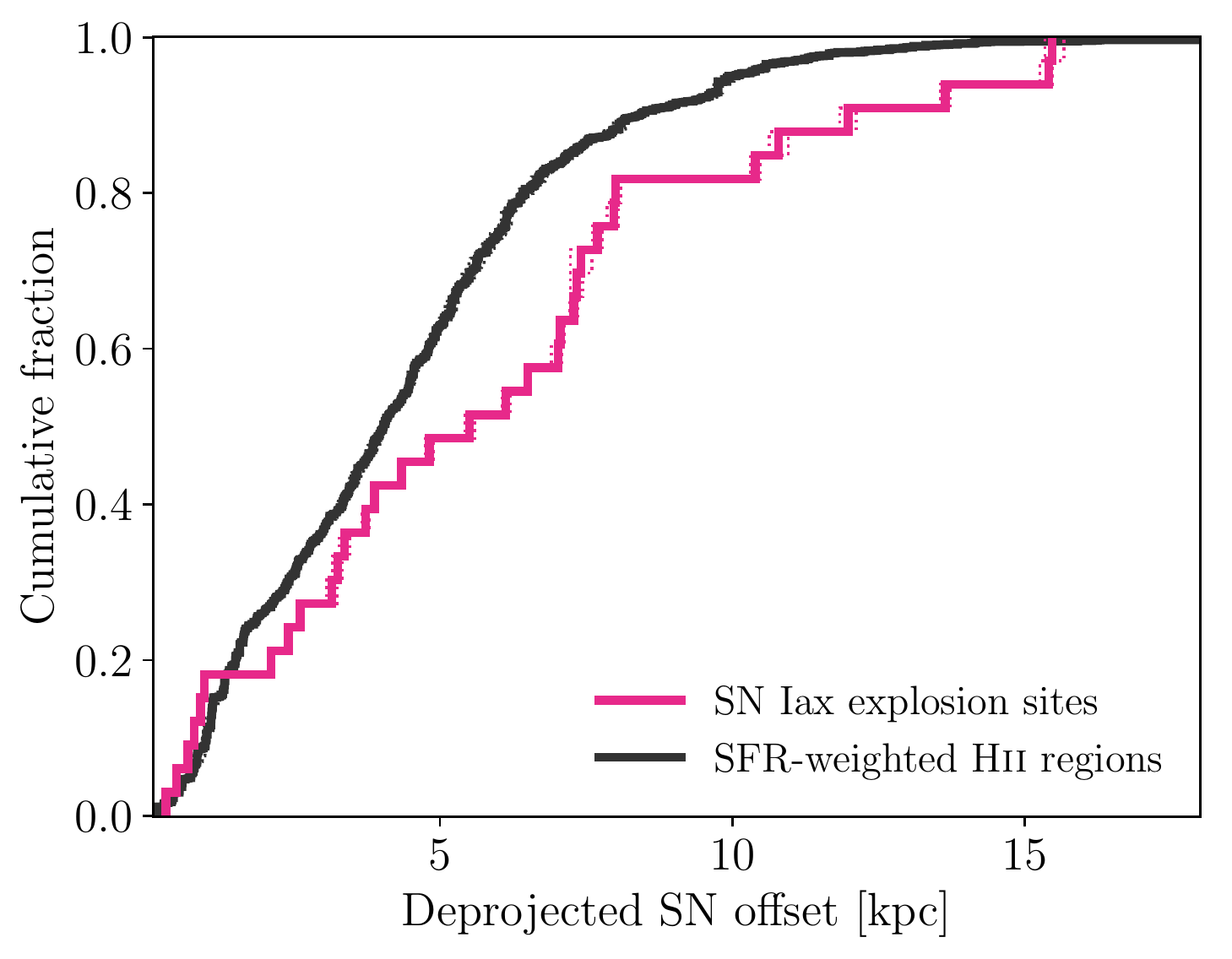}}\\
\subfloat{\includegraphics[width=\columnwidth]{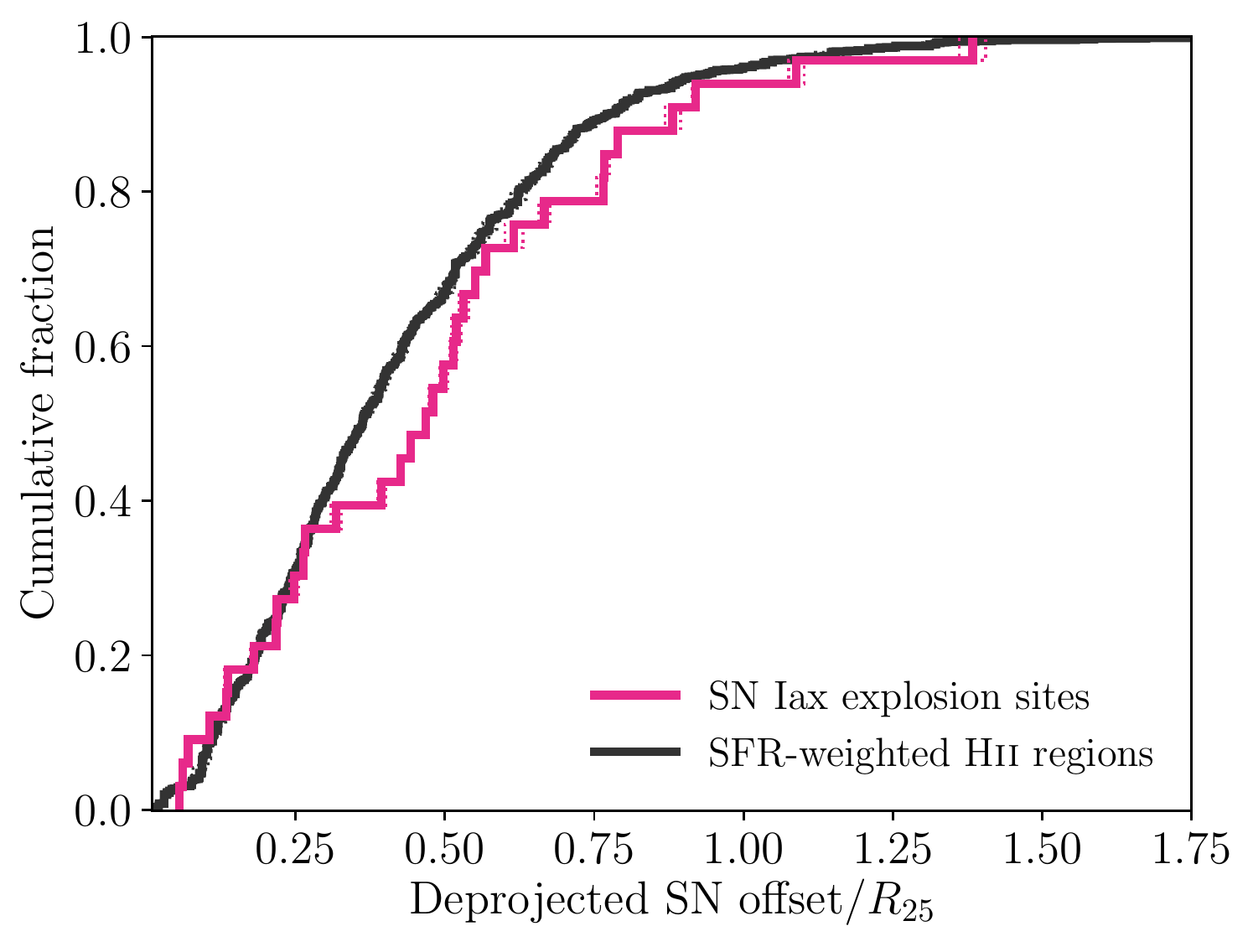}}
 \caption{As for \cref{fig:Zall} but here showing the deprojected galactocentric offsets of SN~Iax explosion sites. The offsets are given in kpc and normalised to the host galaxies' $R_{25}$ values in the top and bottom panels, respectively.} 
 \label{fig:Offsetall}
\end{figure}

Offsets of SNe~Iax and \Hii{} regions are shown for kpc and host-normalised values in \cref{fig:Offsetall}. SNe~Iax appear to trace a similar offset distribution as the overall SF of their hosts when accounting for the varying sizes of the host galaxies, although shifted systematically to slightly larger offsets. Since the outer regions of late type galaxies are likely to be more metal poor and less intensely star forming, this may be a contributing factor to the difference we observe in our other measurements.

\section{Discussion}
\label{sec:discussion}

\subsection{Correlation between environment and SN properties}
 
Tying the diversity of progenitor systems to the observed properties of SNe is an open issue amongst all SN types. It may be expected that some imprint of the nature of the progenitor (which can be determined via direct detection or, as here, inferred through environmental analysis) is evident in the SN light curve and spectra.

\begin{figure*}
 \centering
\subfloat{\includegraphics[width=\linewidth]{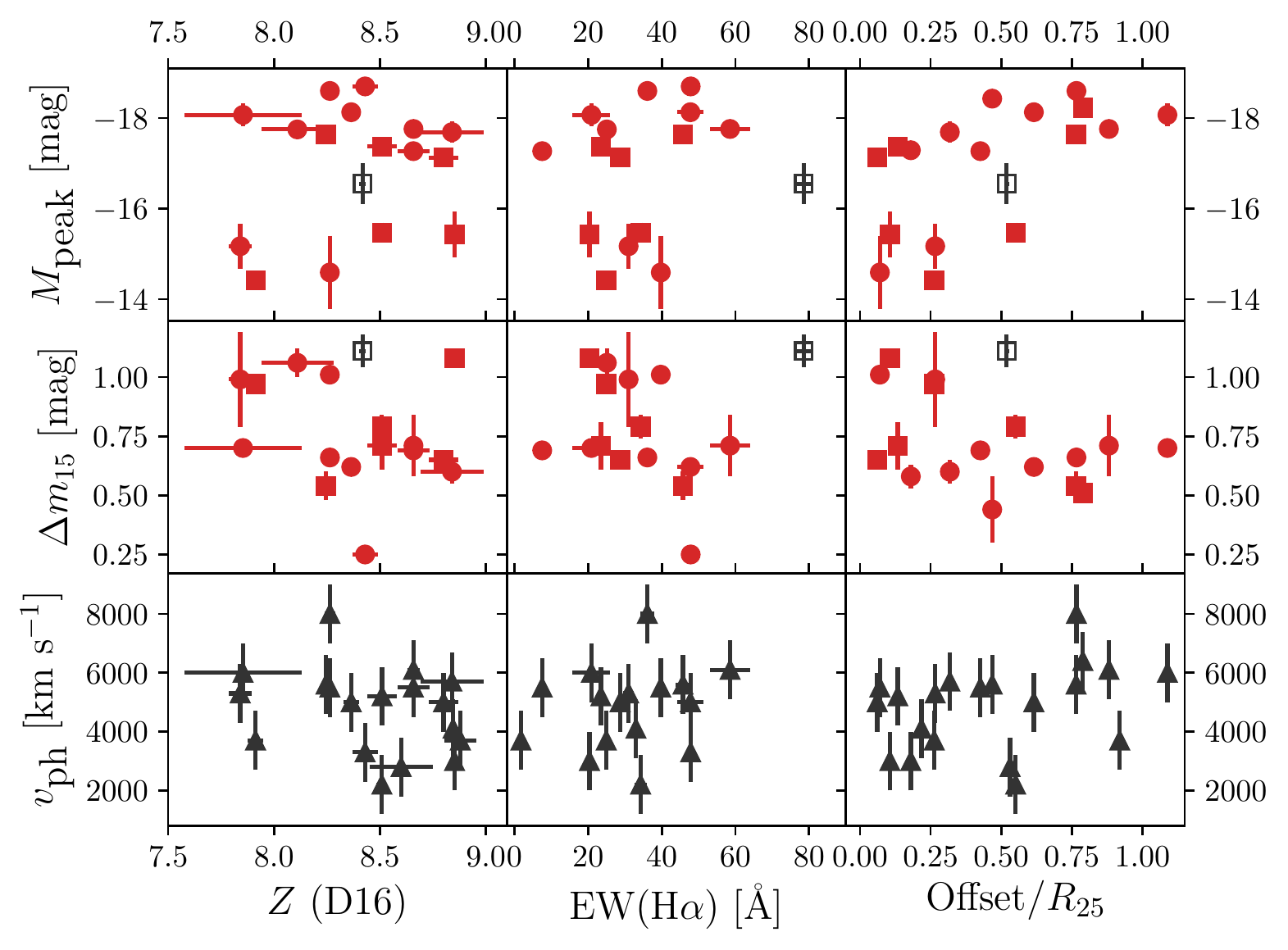}}
 \caption{The properties of SNe~Iax and their environments. $M_\textrm{peak}$ and $\Delta m_{15}$ values are given in the $R$- (red squares), $r$-band (red circles) or unfiltered (for SN~2004cs, empty squares). $v_\textrm{ph}$ values are estimates of the photospheric velocity.} 
 \label{fig:snprops}
\end{figure*}

To investigate this for SNe~Iax we plot SN properties, where they have been measured in the literature, against our environmental measures in \cref{fig:snprops}. The light curve peak ($M_\textrm{peak}$) and decline rate ($\Delta m_{15}$) in $R$- or $r$-band, as well as an estimate of the photospheric velocity around peak light ($v_\textrm{ph}$) are shown versus the \citetalias{dopita16} metallicity, EW(\Ha{}) and the host-normalised offsets. Values and references for SN properties are provided in \cref{tab:litsn}. In order to expand our comparison sample of SN properties we include preliminary analyses for SNe~2010el, 2013gr and 2014ey (Stritzinger et al., in preparation)\footnote{For the three SNe we find $M_\textrm{peak,r} \simeq [-15.4, -15.2, -18.1]$~mag, $\Delta m_{15,r} \simeq [1.08, 0.99, 0.62]$~mag and  $v_\textrm{ph} \simeq [3000, 5300, 5000]$~\kms{}, respectively. For SN~2010el we assign an uncertainty of 0.5~mag on the peak owing to the quite uncertain distance to NGC~1566.} -- these objects are to be the subject of more detailed studies in preparation, but our values here are representative.

\begin{table*}
\centering
\begin{threeparttable}
\caption{Values for literature SN properties.}
\begin{tabular}{lccccl}
  \hline
  SN name   & Filter & $\Delta m_{15}$ & $M_\textrm{peak}$ & $v_\textrm{ph}$ & Reference          \\
  \hline  
  2002cx     & $R$     & 0.54$\pm$0.06 &  -17.64$\pm$0.15  &  5600             & \citet{foley13}       \\
  2003gq     & $R$     & 0.71$\pm$0.10 &  -17.37$\pm$0.15  &  5200             & \citet{foley13}       \\
  2004cs     & $R$     & 1.11$\pm$0.07 &  -16.55$\pm$0.45  &  \ldots           & \citet{magee16}       \\ 
  2005cc     & $R$     & 0.65$\pm$0.01 &  -17.13$\pm$0.15  &  5000             & \citet{foley13}       \\
  2005hk     & $r$     & 0.70$\pm$0.02 &  -18.07$\pm$0.25  &  6000             & \citet{stritzinger15} \\ 
  2007qd     & \ldots  & \ldots        &  \ldots           &  2800             & \citet{foley13}       \\
  2008A      & $R$     & 0.51$\pm$0.01 &  -18.23$\pm$0.15  &  6400             & \citet{foley13}       \\ 
  2008ae     & $r$     & 0.71$\pm$0.13 &  -17.76$\pm$0.16  &  6100             & \citet{foley13}       \\ 
  2008ha     & $R$     & 0.97$\pm$0.02 &  -14.41$\pm$0.15  &  3700             & \citet{foley10a}      \\ 
  2009J      & $R$     & 0.79$\pm$0.05 &  -15.47$\pm$0.22  &  2200             & \citet{foley13}       \\ 
  2009ku     & $r$     & 0.25$\pm$0.03 &  -18.70$\pm$0.15  &  3300             & \citet{foley13}       \\ 
  2010ae     & $r$     & 1.01$\pm$0.03 &  -14.59$\pm$0.81  &  5500             & \citet{stritzinger14} \\
  2010el     & $R$     & 1.08$\pm$0.03 &  -15.43$\pm$0.50  &  3000             & This work             \\ 
  2011ay     & $r$     & 0.44$\pm$0.14 &  -18.43$\pm$0.19  &  5600             & \citet{foley13}       \\
  2012Z      & $r$     & 0.66$\pm$0.02 &  -18.60$\pm$0.09  &  8000             & \citet{stritzinger15} \\
  PS1-12bwh  & $r$     & 0.60$\pm$0.05 &  -17.69$\pm$0.24  &  5700             & \citet{magee17}       \\
  2013gr     & $r$     & 0.99$\pm$0.20 &  -15.17$\pm$0.50  &  5300             & This work             \\ 
  2014ck     & $r$     & 0.58$\pm$0.05 &  -17.29$\pm$0.15  &  3000             & \citet{tomasella16}   \\
  2014dt     & \ldots  & \ldots        &  \ldots           &  4100\tnote{\dag} & \citet{foley15}       \\
  2014ey     & $r$     & 0.62$\pm$0.02 &  -18.13$\pm$0.10  &  5000             & This work             \\
  2015H      & $r$     & 0.69$\pm$0.04 &  -17.27$\pm$0.07  &  5500             & \citet{magee16}       \\ 
  PS~15csd   & $r$     & 1.06$\pm$0.06 &  -17.75$\pm$0.06  &  \ldots           & \citet{magee16}       \\
  \hline
\end{tabular}
\label{tab:litsn}
 \begin{tablenotes}
 \item[\dag] Based on a similarity to SN~2002cx post-peak.
 \end{tablenotes}
\end{threeparttable}
\end{table*}

From these plots we see that brighter ($M_R \lesssim -16$~mag) members appear to cover almost the full range of the metallicity, EW and offset distributions as found for the full sample, with large spreads in each parameter. Although we have fewer fainter members with peak absolute peak magnitude determinations, we find none at large galactocentric offsets. This trend with galactocentric offset cannot be attributed to observational biases as these work in the opposite direction (very faint transients are more difficult to detect in the brighter central regions of galaxies) and so may actually be more pronounced than is shown. 
The faint members appear to cluster at a very small range in EW(\Ha{}) of $\sim$ 25--40 \AA{}, perhaps indicating more strict age constraints for these members, assuming they arise from the local young stellar population. Their metallicities appear diverse although most appear quite metal-poor.\footnote{For another member, SN~2007qd, \citet{mcclelland10} find a faint peak magnitude ($M_R \simeq -15.8$~mag), although this is uncertain since the light curve does not have full coverage \citep{foley13}. We note for this event we only have a poorly constrained (2 regions) metallicity gradient from which to estimate the \citetalias{dopita16} abundance as 8.60.}

\begin{figure}
 \centering
\subfloat{\includegraphics[width=\columnwidth]{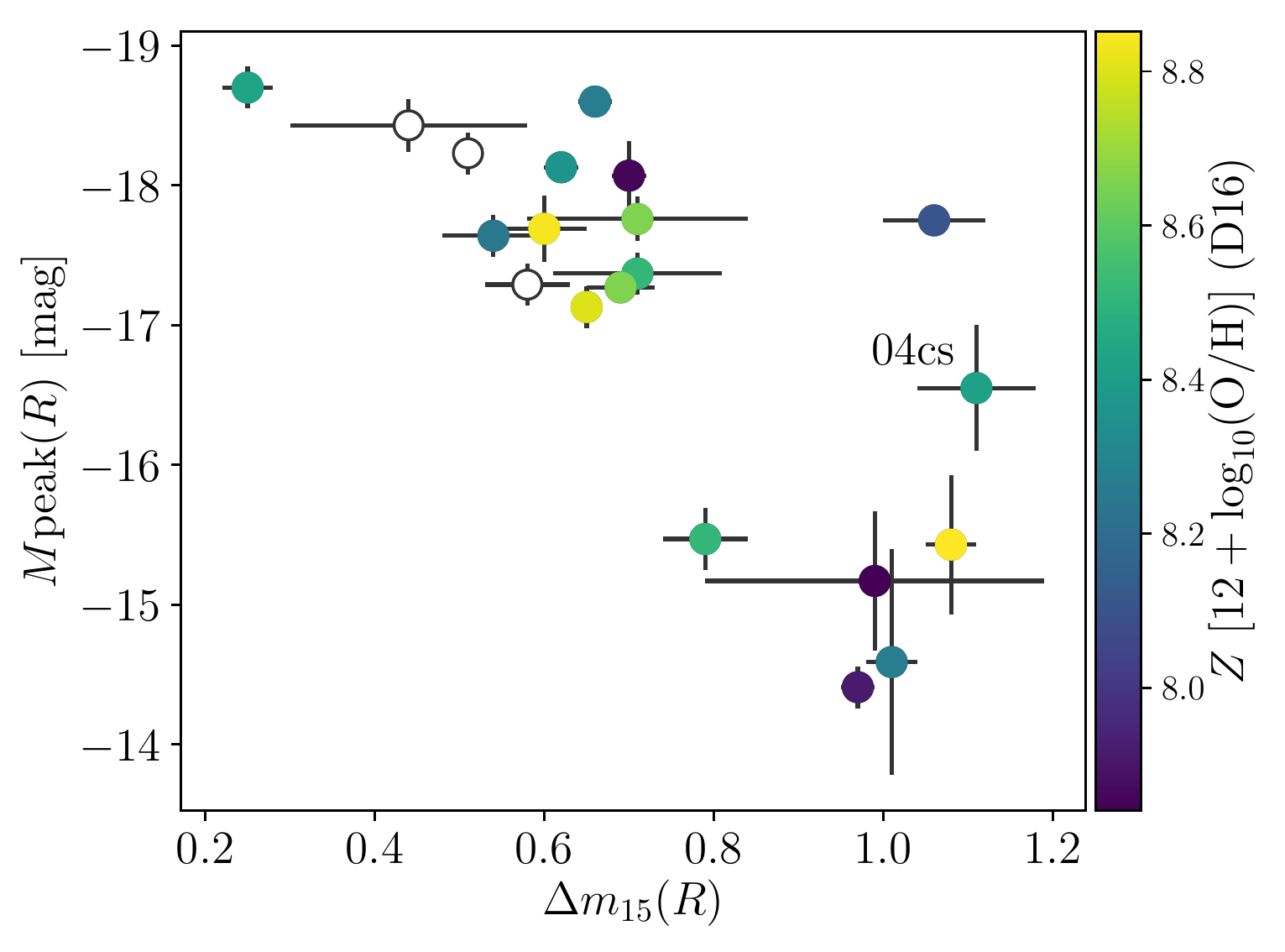}}
 \caption{$R$- and $r$-band $M_\textrm{peak}$ versus $\Delta m_{15}$ for SNe~Iax \citep[e.g.][]{narayan11, foley13, white15, tomasella16} in our sample, now colour coded by their environmentally-derived metallicities in the indicator of \citetalias{dopita16} where possible. SN~2004cs is highlighted as the classification of this event is debated (\cref{sec:hetero}).} 
 \label{fig:mm15}
\end{figure}

Evidence has been presented for a relation between $M_R$ and $\Delta m_{15}$ for SNe~Iax, but any relation is certainly less tight than that seen for normal SNe~Ia and with notable outliers \citep[e.g.][]{mcclelland10,narayan11,foley13}. The slowly fading ($\Delta m_{15} \lesssim 0.75$~mag) members of our sample occupy a wide range of environments, with the faster declining members generally occupying more restricted ranges, analogous to the trends seen for the bright and faint members. In particular we observe distinct clustering in metallicity with $\Delta m_{15}$ -- slowly declining members are systematically more metal-rich than the faster decliners (barring the fast declining, metal-rich SN~2010el).
A composite figure showing the $M_R$ and $\Delta m_{15}$ parameter space for SNe~Iax now coded by the explosion site metallicity is shown in \cref{fig:mm15}.
SN~2004cs (based on unfiltered imaging) was a fast declining member although somewhat brighter than the other fast-decliners in this plot, and shows a significantly larger EW(\Ha{}) than the other fast decliners (or any of the sample for which we have light-curve information). This was one of two proposed SN~Iax members that showed He features, and is further discussed in \cref{sec:hetero}.

We do not observe any strong clustering or correlated behaviour between our environmental measures and the estimates of photospheric velocity near peak. We caution here, however, that these values have been determined using a variety of methods (spectral synthesis, line measurements based on differing elements) at slightly varying epochs around maximum light, and there thus may be some systematics within the sample. We assign 1000~\kms{} uncertainties when plotting, which are likely to be overestimates for individual measurements but account better for potential systematics arising from differing methods.

\subsection{SNe~Iax environments in the context of other transients}
\label{sec:sncompare}

Despite sharing some similarities to SNe~Ia in the general spectral classification sense, the host galaxies of SN~Iax are almost exclusively (barring SN~2008ge) late type and star-forming \citep[e.g.][]{perets10,foley13} and the transients appear to be associated with regions of ongoing SF \citep[this work;][]{lyman13}. This is the case for CCSNe and thus a comparison between our results here and those of other SN types in the literature can inform on similarities or differences in the progenitor environments.

The construction of our comparison SN samples do not constitute unbiased, blindly-targeted events by any means. However, the same holds for the SN~Iax sample, which were discovered over a variety of surveys. Our comparison is thus limited to an initial, indicative comparison until such time as a reasonable sample of homogeneously discovered SNe~Iax exists.

\subsubsection{Metallicity}

As we are working with literature samples, there are almost no abundance measurements in the \citetalias{dopita16} scale owing to its recent inclusion in the literature. We therefore collect available values based on the O3N2 indicator (these were generally presented in the calibrations of \citealt{pettini04} and have been translated to those of \citetalias{marino13}). We used only values based at or nearby the SN explosion site, or based on determined gradient values (i.e. we exclude those where the metallicity was determined based on the host nucleus but the SN was at a significant offset and not covered by the slit/fibre).

We take values from the analyses of \citet{anderson10,modjaz11,leloudas11,sanders12,kuncarayakti13a,kuncarayakti13b,stoll13} and \citet{galbany16}.
In the case of SNe~II (including both II-L and II-P) we take values from \citet[][which includes the results of \citealt{anderson10}]{anderson16} and \citet{galbany16} directly in \citetalias{marino13}, and values from \citet{stoll13}. On the values of \citet{anderson16} we impose a distance cut of 3~kpc on the \Hii{} region---SN distance.
We present SNe Ib and Ic separately, however uncertain classifications mean there is likely to be some cross-contamination. Where appropriate we use the updated classifications of \citet{modjaz14, shivvers17} and count still uncertain `Ib/c' designations with half weight in each cumulative distribution.

\begin{figure}
 \centering
\subfloat{\includegraphics[width=\columnwidth]{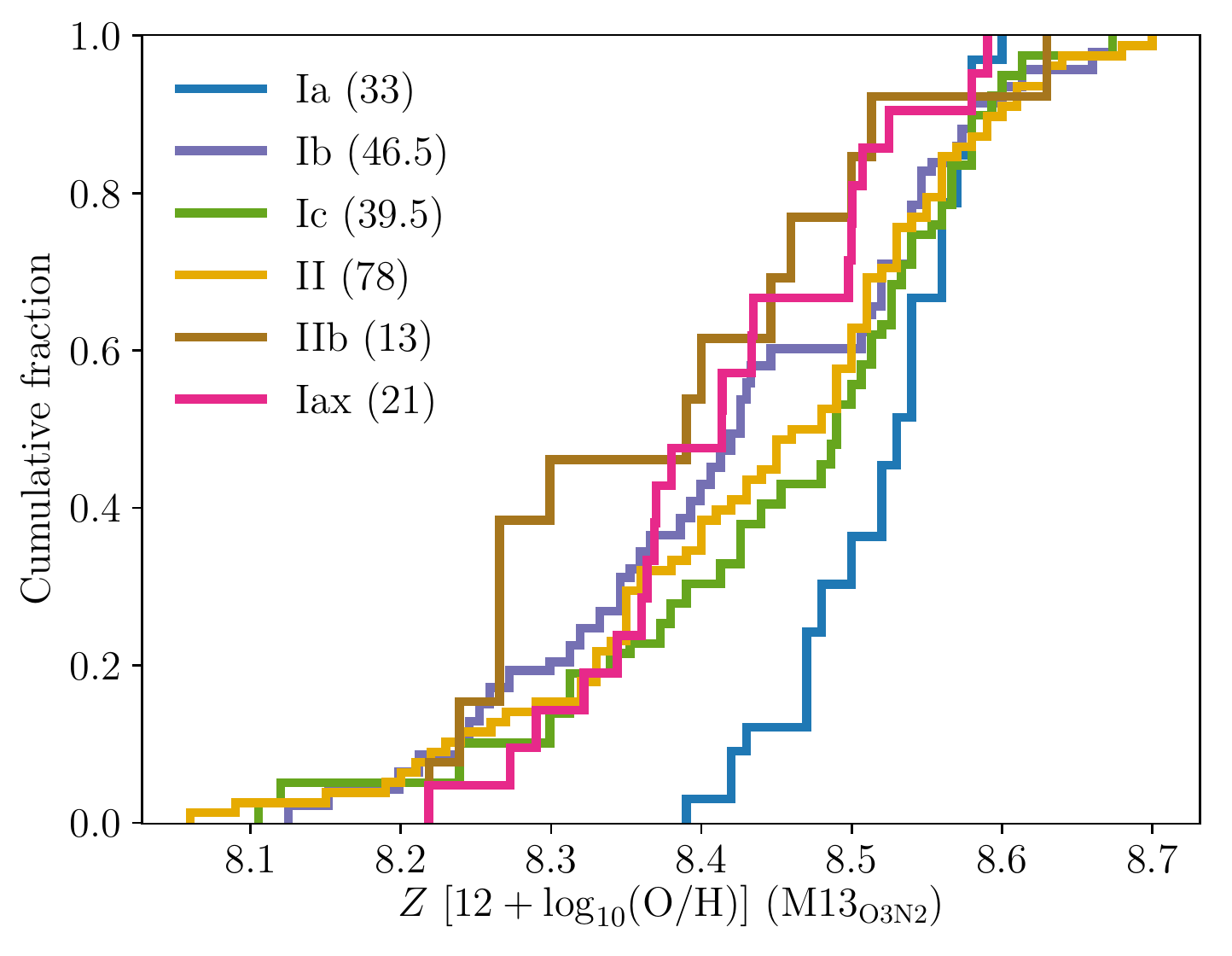}}
 \caption{The explosion site metallicity distributions of SNe~Iax compared to literature samples of other principal SN types. Metallicities are based on the O3N2 indicator of \citetalias{marino13}.}
 \label{fig:Zcompare}
\end{figure}

We plot the explosion-site metallicity distributions for SNe~Iax and our comparison samples in \cref{fig:Zcompare}. There is no significant difference between SNe~Iax explosion-site metallicities and those of SNe Ib, Ic, II or IIb and the median value for SNe~Iax (8.41 dex) is close to that of the others (Ib: 8.42 dex, Ic: 8.49 dex, II: 8.47 dex, IIb: 8.39 dex). The distribution for SNe~Iax covers broadly the same range as these other SN types, perhaps not extending as metal-rich or -poor as SNe~Ib, Ic or II, however there are only a very small number of events in these samples on the extreme edges of their distributions. Compared to the SNe~Ia distribution \citep[local metallicity measurements explosion sites from IFS data taken from][]{galbany16}, median = 8.53, we see SNe~Iax are metal-poor (KS test Ia vs Iax metallicities $p = 5\times10^{-4}$).

Although the overall SN~Iax distribution appears to follow broadly the distributions of other well-known SN types, barring SNe~Ia, it is overall quite different from the distribution of low redshift LGRB explosion sites, which are in the range $Z_\textrm{O3N2} \simeq 8.0-8.4$ dex \citep[e.g.][]{modjaz08, sanders12}. Similar very low metallicities have also been found for the the vast majority of super-luminous supernova hosts (e.g. \citealt{lunnan14,leloudas15}).

\subsubsection{Offsets}

$R_{25}$-normalised offset distributions for SNe Ib, Ic, II and IIb were taken from \citet{taddia13}. SNe~Ia offsets were taken from the catalogue of \citet{hakobyan16} and calculated following the prescription described therein. The SNe~Ia host-normalisation are based on $g$-band $R_{25}$ measurements as opposed to $B$-band measurements used for the other samples. This may introduce a systematic offset in the values, however, given the considerable filter overlap and the modest variation of apparent galaxy sizes around these wavelength ranges \citep{vulcani14}, the effect is likely to be small. SNe~Ia in the catalogue are those that exploded in disk galaxies (types S0-Sm).

\begin{figure}
 \centering
\subfloat{\includegraphics[width=\columnwidth]{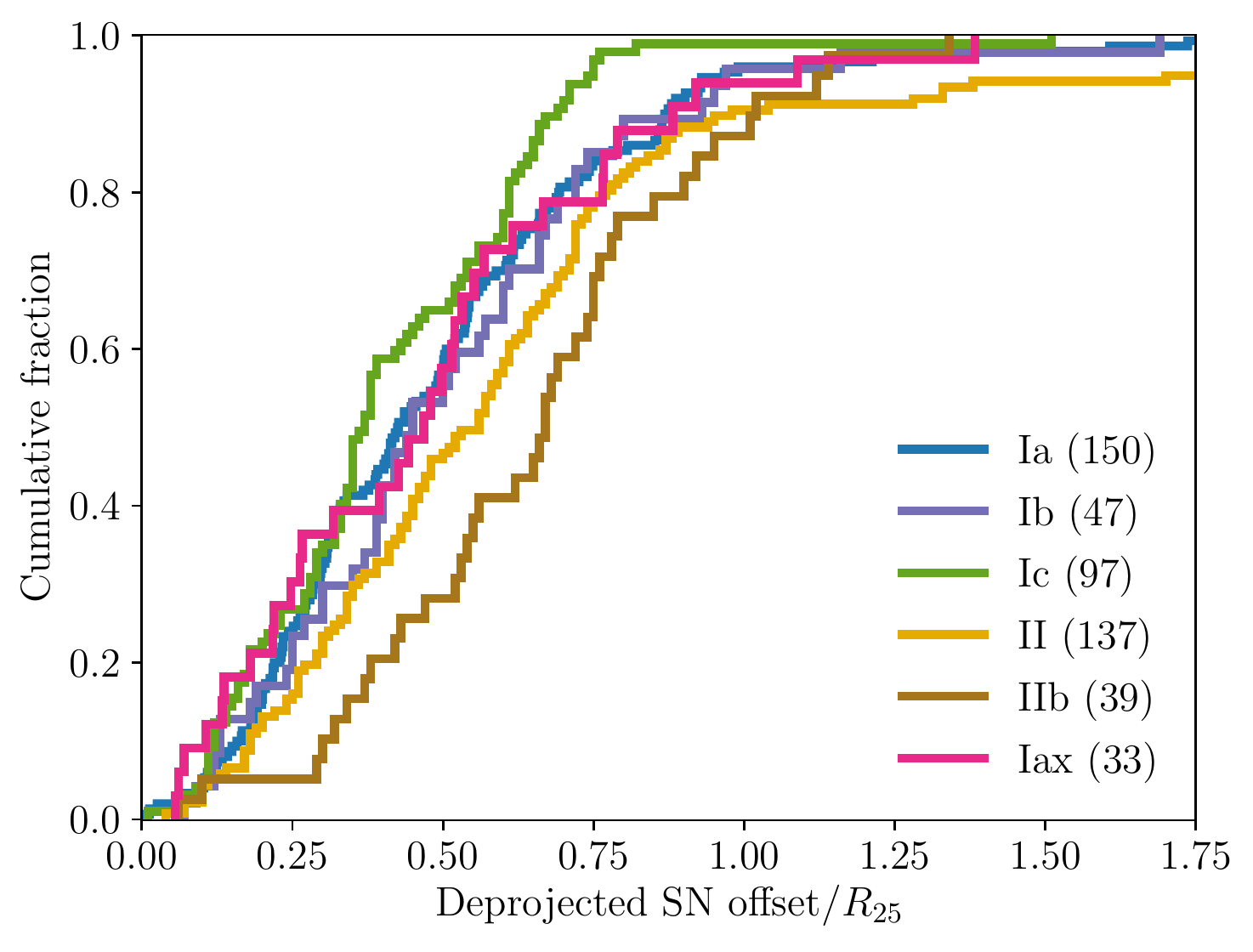}}
 \caption{The explosion site galactocentric offset distributions of SNe~Iax compared to literature samples of other well-known SN types. Offsets have been normalised by their hosts' respective $R_{25}$ values.} 
 \label{fig:offsetcompare}
\end{figure}

Similarly to the metallicities, we see in \cref{fig:offsetcompare} that the offsets of SNe~Iax closely match other SNe types, in particular SNe~Ia and Ib. Offsets do not appear to be a strong factor in distinguishing SNe~Iax, whereas their host galaxy type distribution proves more discriminatory against that seen for SNe~Ia \citep{perets10,lyman13}.

\subsubsection{Pixel Statistics}
\label{sec:pixelstatres}

\begin{figure}
 \centering
\subfloat{\includegraphics[width=\columnwidth]{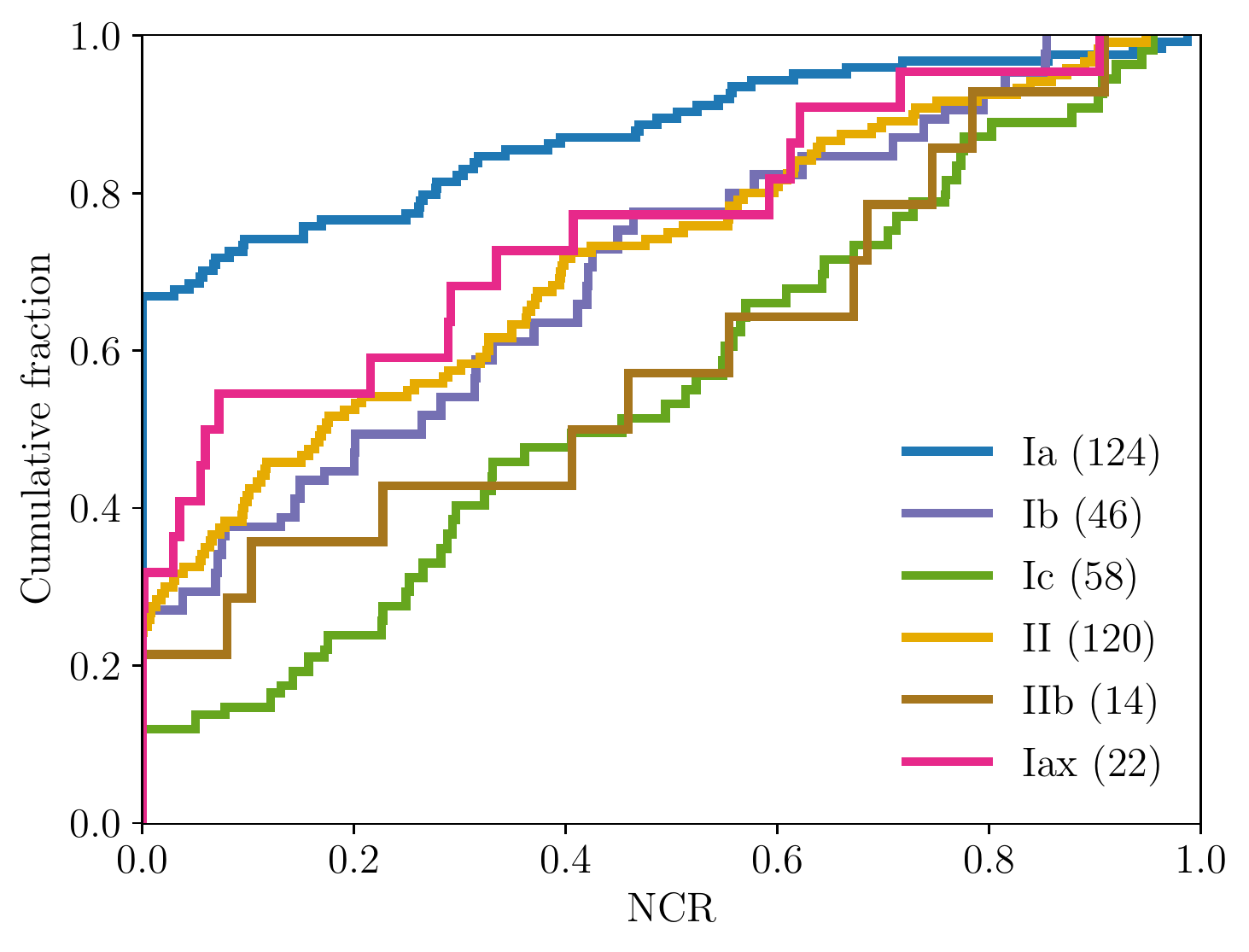}}
 \caption{The explosion site Normalised Cumulative Rank (NCR) distributions of SNe~Iax compared to literature samples of other well-known SN types. NCR values were calculated using \Ha{} maps from our MUSE data cubes and supplemented with values presented in \citet{lyman13}.} 
 \label{fig:ncr}
\end{figure}

An analysis of the NCR (\cref{sec:pixelstat}) distribution of SNe~Iax was presented by \citet{lyman13}, who found the association of SNe~Iax to SF in their hosts was at a similar level to that of SNe~IIP and inferred similar progenitor ages (tens of Myr). As we are able to create \Ha{} maps from our MUSE data cubes we present an updated comparison figure with this extended combined sample of SNe~Iax in \cref{fig:ncr}. Other SN sample data are taken from \citet{anderson08, anderson12, anderson15} and, following \citet{lyman13}, we have corrected the SN~Ia sample to account for the fact that only late-type galaxy hosts were used but $\sim$27~per~cent of SNe~Ia explode in early-type hosts \citep{li11}, with no appreciable ongoing SF. For SNe~2005P, 2008ha, 2009J and 2012Z we use values from \citet{lyman13} as \Ha{} imaging presented there covers the full spatial extent of the hosts. The NCR values for these determined in \citet{lyman13} and here are [0.06, 0.41, 0.00, 0.00] and [0.01, 0.33, 0.06, 0.03], respectively.

Following \citet{lyman13}, we confirm for a larger sample that SNe~Iax display a level of association to the SF of their hosts most similar to that of SNe II. Their distribution is formally discrepant, based on the KS test, with the SNe~Ia and Ic distributions at $3\sigma$ (considering only SNe~Ia in late-type hosts and excluding SN~2008ge from the SN~Iax sample, the SN~Ia vs Iax discrepancy is $p = 0.05$).\footnote{Anderson-Darling tests provided very similar levels of significance, with slightly higher significance for the uncorrected SNe~Ia vs Iax ($p \simeq 0.006$).} Although there are few \Ha{} NCR values for LGRBs (due to their typically much larger distances), studies have shown these to be very strongly associated to the brightest SF regions of their hosts, at a level exceeding SNe~Ic \citep{fruchter06, svensson10,lyman17}. Assuming the same would hold for SF as traced by \Ha{}, this infers drastically different environments of SNe~Iax compared to LGRBs.

We note that although we detected \Ha{} at the locations of all our SNe~Iax in the MUSE sample, we have some where NCR = 0.00. In order to facilitate comparison to literature samples, which were performed using narrow band imaging, we construct \Ha{} narrow band images (filter width 30\AA{}) and subtract the neighbouring continuum within our data cubes (\cref{sec:hiibinning}). As such, very faint sources of emission may become dominated by shot noise or inaccuracies in continuum subtraction during the \Ha{} map construction, leaving the pixel within the noise floor of the image \citep[defined as NCR=0, see][for more thorough discussion]{james06,anderson08}. Additionally, the spatial extend of our spaxel bins is larger than the $3 \times 3$ binning employed for the NCR calculations in order to replicate the method used in the literature.

\subsection{Ionised gas in the early-type host of SN~2008ge}
\label{sec:08ge}

Although there are likely to be contributions from different progenitor channels to such a diverse class of objects (see \cref{sec:hetero}), one SN in the Iax sample has been marked out by its unique environment and host. SN~2008ge was hosted by NGC~1527, an early, weakly-barred S0 galaxy \citep{deVauc91}, in contrast to the strongly star-forming, late-type hosts of other known Iax \citep{perets10, lyman13}. Motivated by this, \citet{foley10b} investigated the host galaxy and explosion site for signs of a young SP. They concluded from a smooth galaxy profile, non-detection of far-infrared or H{\sc I} 21cm emission in the host, and the lack of narrow nebular emission lines in their host galaxy spectrum that there is no significant ongoing SF (and thus young SP) in NGC~1527.

Our MUSE observations provide deep optical spectra, spatially resolved, across a reasonable fraction of the central regions of the host galaxy. After an initial pass with our emission line fitting as presented in \cref{sec:musemethods}, we reran the emission line fitting this time fitting only for \Ha{} and \Nii{}~$\lambda$6548,6583 in order to determine a subset of the bins with evidence for ionised emission lines. We then attempted to fit \Hb{} and \Oiii~$\lambda$5007 for these bins. In accordance with \citet{foley10b}, we found no significant emission lines at the location or in the close vicinity of the SN explosion site (either within the Voronoi bin directly underlying the explosion site or by extracting circular apertures of varying sizes). However, within our MUSE data we detected an ionised gas stream arcing from the NE of the nucleus with evidence for a further region of emission to the SE of the host nucleus. As NGC~1527 is particularly nearby ($\sim$ 17 Mpc) the MUSE observations only cover the central $\sim$2--3 kpc around the nucleus. We have summed the \Ha{} flux captured by the MUSE FOV and determine $\log_{10}$(L(\Ha{}) = 37.6~erg~s$^{-1}$ and note that this is not corrected for the effects of reddening as we were unable to detect \Hb{} in many of the bins. For the case that this emission is driven by ongoing SF, the relation of \citet{kennicutt98} suggests the detected emission equates to $\log_{10}$(SFR)~=~-3.48 \msun{}~yr$^{-1}$. This is consistent with the limit that \citet{foley10b} provide for NGC~1527 of $\log_{10}$(SFR)$ < -2.14$~\msun{}~yr$^{-1}$. The metallicity (for SF-driven ionisation) appears to be around solar to slightly sub-solar based on the N2 indicator \citepalias{marino13}. The EW(\Ha{}) of the bins are $\sim 0.8$\AA{}.

\begin{figure*}
 \centering
\subfloat{\includegraphics[width=\linewidth]{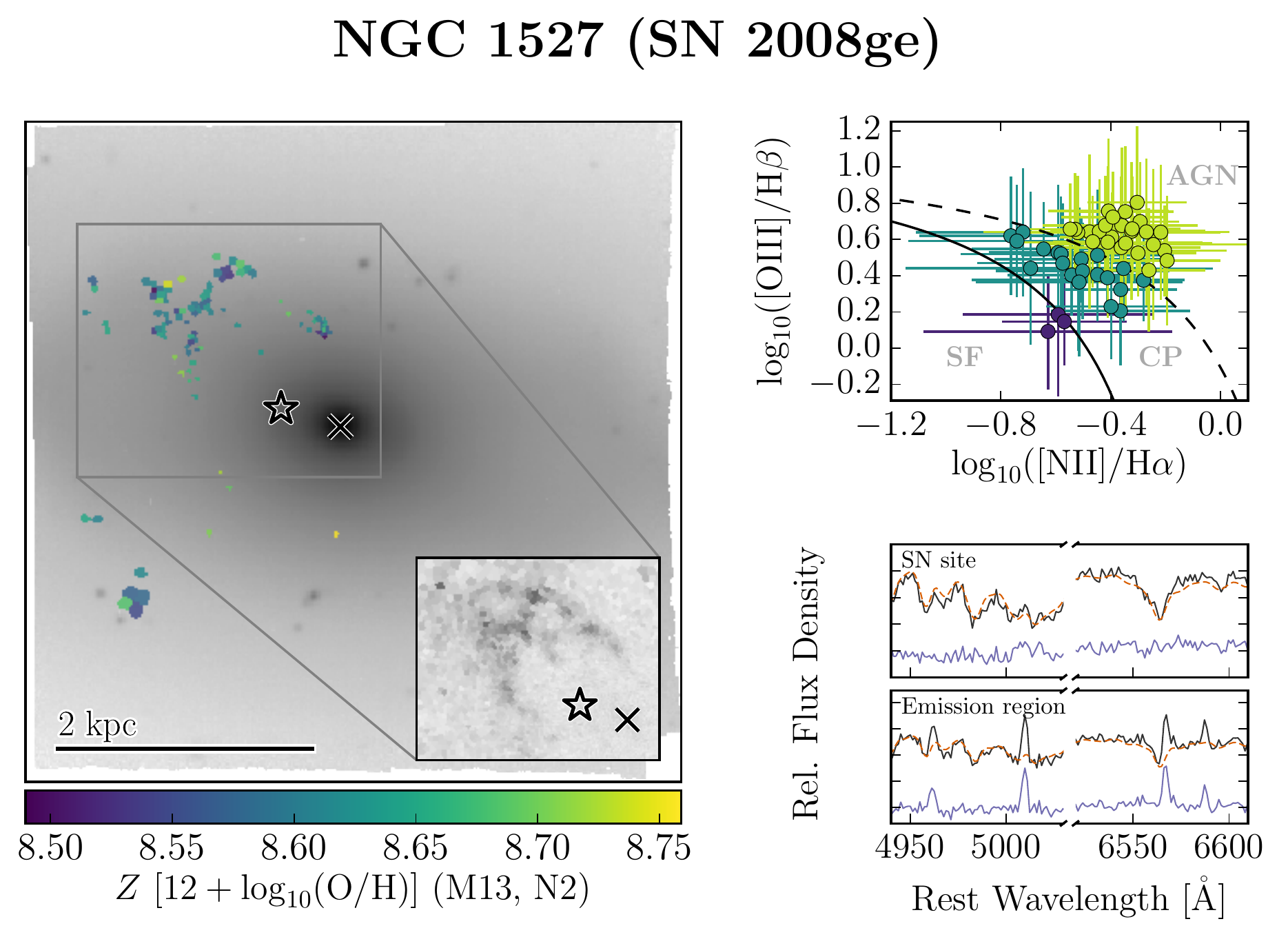}}
 \caption{The central ionised gas component of NGC~1527 (the host of SN~2008ge). {\em Left:} A heatmap of the metallicity (using the N2 relation of \citetalias{marino13}) for each bin where both \Ha{} and \Nii{}~$\lambda$6583 has SNR~$> 2$, overlaid on an inverted $R$-band image of the galaxy; North is up, East is left. The adopted location of the nucleus and the SN explosion site are indicated by the black $\times$ and `star' symbols, respectively. {\em Left, inset:} An inverted, continuum-subtracted \Oiii{} image of the central region of NGC~1527. Each Voronoi bin has had its best fitting stellar continuum model (\cref{sec:contfitting}) subtracted. The residual ionised gas component is evident {\em Top right:} A BPT diagram for those bins with detected emission. The format is the same as shown in \cref{fig:bptmap}. {\em Bottom right:} The observed spectrum (black) with the {\sc starlight} fit to the continuum (orange, dashed) and the residual emission (purple) around \Ha{} and \Nii{}. This is shown for the SN explosion site, where we do not detect any significant emission lines, and in one of the bins NE of the galaxy nucleus, from which we were able to measure a metallicity. Wavelengths are in the rest-frame of the nucleus of NGC~1527.}
 \label{fig:08ge}
\end{figure*}

Morphologically the main stream appears somewhat coherent, in a thin parabolic stream about the nucleus. The emission looks different to the more extended, and generally more symmetric, spiral arms seen in some early-type galaxies \citep[ETG; see][and references therein]{gomes16a}. The nature of the ionising sources in ETGs is debated and there appears to be contributions from a variety of phenomena, with different mechanisms dominating in different galaxies \citep[e.g.]{goudfrooij99,sarzi06,sarzi10}. Unfortunately, we are limited in our analysis for the ionising source based on emission line flux ratios  as we are close to our detection limits and thus flux ratios are not well constrained. Within \cref{fig:08ge} we plot the BPT diagram for bins where all lines were detected at SNR > 1. Although the position of each bin is rather uncertain within this diagram, there appears to be a general clustering of the bins around or above the maximal star-formation relation of \citet{kewley01}, indicative that ongoing SF may not be the dominant process driving this emission. (Those bins that are more consistent with being ionised solely due young, massive stars are those SE of the nucleus, not in the main arc). With a comparison to the study of \citet{sarzi10}, the stronger line ratios and morphology may be more reminiscent of emission due to shocks in the galaxy, although these are not expected to be dominating sources of ionisation in ETGs. Alternatively it may be due to diffuse ionising SPs, such as post-asymptotic giant branch (pAGB) stars. 

The ionising flux from pAGB stars can lead to line ratios similar to those seen for low-ionization emission-line regions  \citep[LIERs][]{binette94,sarzi06}, and the bulk of the bins in \cref{fig:08ge} are located close to LIER areas of the parameter space \citep[e.g. $\log_{10}\frac{\Nii{} \lambda6583}{\Ha} \gtrsim -0.4$ and EW(\Ha{})~$< 6$\AA{},][]{cidfernandes11}. One would, however, expect a pAGB population to be more or less pervasive across the galaxy and follow the distribution of stellar mass. This results in the correlation between ionised line flux and stellar continuum flux seen in ETGs with pAGB ionising sources \citep{sarzi10}. A more structured ionised region (as seen here) could therefore be due to variations in gas column density within the galaxy.
The ionising output and subsequent EW(\Ha{}) predictions from various models for pAGB star contributions predict roughly constant low values from $\sim 10^8$ to $10^{9.5}$~years \citep{cidfernandes11}, meaning such values do not offer strong constraints on the relative age of the underlying population.

\subsection{Heterogeneity of the class and contaminants}
\label{sec:hetero}

As is the case for many astrophysical transients, in particular for peculiar and relatively rare objects, membership to a particular class (or even designating a single class) can be contentious. Often the underlying continua of the events properties are at odds with defining distinct regions of parameter space to assign one class or another. This is exemplified in the case of SNe Iax where multiple progenitor channels and explosion mechanisms may be contributing to empirically similar transients.

Two examples in the sample of \citet{foley13} showed evidence for helium in their spectra: SNe 2004cs and 2007J. Helium can remain hidden in SN spectra as it is difficult to ionise and so it is possible that helium is present in the ejecta of other members, however at the time of writing these two members remain the only posited SNe~Iax to show detections. One potential confusion, as highlighted by \citet{foley13}, is with SN~2005E-like events \citep[][also known as Ca-rich transients/SNe]{perets10} -- these similarly faint-and-fast events display helium features in their spectra before quickly evolving to unusually Ca-dominated nebular spectra \citep{kasliwal12}. Instead \citet{white15} favour the original classifications of SNe~2004cs and 2007J as CCSNe. In addition to helium they also find evidence for hydrogen in their spectra, prompting classifications of SNe~IIb. However, a reanalysis by \citet{foley16} argued against the presence of hydrogen in SN~2007J, citing an identification as [Fe{\sc ii}] of the same feature in SN~2002cx. Furthermore, they find inconsistencies between the light curve of SN~2004cs and other known SNe~IIb. 
As we are discussing only 2 objects, statistical inferences from environmental measures are weak. This is further exacerbated by the fact that our measures overlap significantly for SNe~Iax and SNe~IIb (\cref{sec:sncompare}). For completeness we note EW values at the helium-SNe~Iax explosion sites are the among the highest (1st and 4th) of our sample, pointing to the presence of younger SPs at their explosion sites than typical SNe~Iax. Their offsets ($\simeq 0.5 \times R_{25}$) and metallicities (O3N2~$\simeq 8.4$) are typical of the samples of both SNe~Iax and IIb. Their exclusion would revise our estimates of the typical ages of the young SP at SN~Iax explosion sites slightly higher, although the quantitative effect is likely to be dwarfed by the inherent uncertainties present in determining ages from nebular gas spectra \citep[][\cref{sec:methages}]{stanway14b}, and would not significantly affect our discussion elsewhere.

SN~1999ax was classified as a ``somewhat peculiar'' SN~Ia by \citet{galyam08} at $z \simeq 0.05$. On the basis of SDSS spectra of a nearby potential host showing it to be at $z = 0.023$, \citet{foley13} re-analysed its spectrum and classified it as a SN~Iax based on similarity to SN~2002cx post-peak. We note our NOT spectroscopy of the explosion site also covered another nearby object in SDSS that is at a similar (projected) offset from SN~1999ax as the galaxy at $z = 0.023$ (with the SN being located between the two). We find this galaxy is at $z = 0.059$ based on the \Ha{} line in the NOT spectrum. As the SN is located directly between the two potential hosts it is difficult to distinguish the likely host from environments alone. The velocities of the features in the spectrum appear low compared to normal SN~Ia, and a consideration of the SN at $z \simeq 0.05$ still poses peculiarities to a typical SN~Ia classification, as such we retain the SN~Iax classification making the lower redshift galaxy the probable host. 

\subsection{Implications for progenitor models}
\label{sec:progmodels}

Pure deflagrations \citep[e.g.][]{branch04,jha06,phillips07,jordan12,kromer13,fink14,magee16}, pulsational-delayed detonations \citep{stritzinger15} and helium-ignition \citep{wang14} of WDs have been presented as progenitor models to explain SNe~Iax. Although too weak to account for normal SNe~Ia, they broadly agree with the low $^{56}$Ni masses and kinetic energies displayed by SNe~Iax. There are currently limited predictions as to the progenitor environments for these based on population studies, 
with the main constraints in the form of delay-time distributions and age constraints.

Any double-degenerate channel (either merger or accretion between two WDs), would require circumstances to explain the young environments since such progenitor systems would be expected to be prevalent in old SPs and early-type galaxies, inconsistent with the locations of SNe~Iax as a whole. As such, single-degenerate formation channels are generally favoured for SN~Iax.

\citet{meng14} found that delay times as low as 30~Myr were possible for massive hybrid CONe WDs accreting material until the Chandrasekhar mass, and would likely produce lower-luminosity events (cf. normal SNe~Ia) making them an attractive possibility for SNe~Iax. 
\citet{wang13} explore detonation of CO WDs via ignition in a helium envelope that is accreted from a He companion. For the case of a non-degenerate companion their delay times to explosion are $\sim 10^8$~yrs, consistent with the ages of young SPs at the location of most SNe~Iax. If the companion is a He WD the delay time can be significantly extended (up to the Hubble time), potentially providing an explanation for SNe~Iax in old populations (e.g., SN~2008ge) or those with no detected signs of SF at the explosion site and in the outskirts of their hosts (e.g., SN~2008A). We note that these are for solar metallicity populations, which is metal-rich compared to the SN~Iax population (\cref{fig:Zall}). Similarly, \citet{liuzw15a} find that the WD + He star channel best reproduces the young ages of SN~Iax explosion sites, but note an extended delay time exists for main sequence or red giant companions. 
Our age constraints on SNe~Iax seem to be in good agreement with predictions of young explosion scenarios of single-degenerate thermonuclear progenitor models.

\subsubsection{Faint and fast `SN~2008ha-like events'}
\label{sec:08halike}

The faintest, lowest-energy members of SNe~Iax could represent a distinct population of events, with the most famous example being SN~2008ha \citep[see][for further discussion of models in light of the physical properties of the SNe]{foley09}. These very faint examples are difficult to produce with models proposed for other SNe~Iax \citep{fink14}, perhaps indicating a single model is not able to explain the full diversity of all events labelled as such. For the purposes of this section we simply refer to faint ($M_R > -16$~mag) and (/or) fast evolving ($\Delta m_{15} \gtrsim 1$~mag) members of the sample, ignoring a detailed spectroscopic distinction.

\citet{valenti09} suggested SN~2008ha and, by extension, other SNe~Iax were of core-collapse origin. Our environmental analyses indicate that the fainter members arise from low metallicity regions, quite centrally located on their host galaxies (\cref{fig:snprops}). From an environmental viewpoint, their distributions are largely consistent with those seen for well-known CCSN types. The question of whether they are due to lower mass, $\sim 8$~\msun{}, stars stripped of their hydrogen envelopes or very massive WR stars, probable progenitors of LGRBs, experiencing significant fall-back, was posed by \citet{valenti09}. Our analysis of the ages of the youngest SPs here suggests ages of order tens of Myr for their explosion sites (\cref{sec:resultsages}). This is in agreement with resolved SP studies of explosion sites of nearby SNe~Iax \citep{foley14,mccully14} and the comparative association of SNe~Iax to \Ha{} emission in their host galaxies at the level of SNe~II \citep[\cref{sec:pixelstatres}, ][]{lyman13}. For WR stars, however, we may expect much younger ages, $\lesssim 10$~Myr (for stars with initial masses of $\gtrsim 25$~\msun{}). Consequently we would expect larger EW values at the explosion sites \citep[e.g. see the low redshift LGRB environment studies of][]{thoene08,christensen08,kruehler17,izzo17}, and an association to ongoing star formation of their hosts at a level much higher than is seen for SNe~Iax (\cref{sec:pixelstatres}, see also \citealt{kangas17}).

A search for He {\sc i} $\lambda 4922$ in emission (indicative of stellar populations of only a few Myr) in the hosts of our MUSE sample found 5 SNe~Iax with such young regions within 3~kpc. This include 4 of the 5 events with peak luminosities known to be $\geq -16$~mag plus SN~2012Z (the missing low-luminosity event is SN~2010el). The host galaxies of these fainter members are typically more irregular and exhibit signatures of extremely young stellar populations within them, with EW(\Ha{}) reaching hundreds of \AA{}. In order to establish a causal link between the SNe and their nearest very young regions, progenitor velocities of hundreds of \kms{} must be invoked (\cref{sec:resultsages}). To produce such high mass and velocity runaways requires very rare dynamical ejections or unfeasibly large ejection velocities after a binary companion goes SN \citep[][see also discussion in \citealt{kruehler17}]{eldridge11}.
Our results would therefore disfavour the very massive progenitor scenario (for the SN~Iax sample as a whole and the SN~2008ha subset). \citet{moriya10} concur, finding that their 25 and 40~\msun{} fall-back models cannot reproduce the observations of SN~2008ha, but find better agreement with a 13~\msun{} model.

Deflagrations of hybrid CONe have been proposed to explain the weak SN~2008ha by \citet{kromer15}. Although occurring in Chandrasekhar-mass WDs, the deflagration is not propagated into the outer ONe layer, which results instead in lower energy release and consequently low ejecta masses (cf. CO WD explosions), leaving behind a bound remnant. The ejecta mass in the model is below estimates for SN~2008ha, resulting in too quickly rising and fading light curves, and also affecting the spectral comparison. The delay time to explosion for CONe WDs with helium donor stars has been estimated at $\sim$few~$\times 10^7$ to $10^8$~yrs by \citet{wang14, kromer15}, albeit at solar metallicity which may not be representative of the typical environments of SN~2008ha-like events. This is in agreement with this study and \citet{foley14}.

\citet{moriya16} present an initial study of the light curves and rates for binary evolution channels of stripped-envelope electron-capture (SE-EC) SNe, which may have relation to the low-luminosity CCSN scenario of \citet{valenti09}. They find broad agreement with SN~2008ha-like events in terms of peak magnitude and their rapidly-evolving light curves, and that the rates of SE-ECSNe are significantly enhanced at low-metallicities. For our 5 events peaking at $M_R > -16$~mag we determined a wide spread of metallicities \citep[see also][]{foley09, stritzinger14}, including very low metallicity events and the quite metal-rich SN~2010el. 
In the case of the rapidly declining events (note that these are not all overlapping with the faint events, e.g., PS~15csd; \citealt{magee16}), we similarly find low metallicities for all but SN~2010el. SE-ECSNe progenitors are expected to come from the lower end of the mass range of massive SNe progenitors $\sim 7-8$~\msun{} and we therefore expect to see evidence for young, but not exceptionally young SPs at their explosion sites -- this is borne out in the relatively modest EW(\Ha{}) values we find of 25--40~\AA{}. We therefore confirm here the preference for faint and/or fast subsets of SNe~Iax (SN~2008ha-like events) to be preferentially in lower-metallicity environments with signatures of relatively young SPs at their explosion sites, consistent with current predictions for SE-ECSNe by \citet[][see also \citealt{pumo09}]{moriya16}. In this scenario the proposed donor or remnant of SN~2008ha \citep{foley14} would be the companion star that may have stripped the envelope of the SN progenitor. The spectral predictions for SE-ECSNe that are needed to provide a full comparison to SN~2008ha-like events are currently lacking.

\section{Conclusions}

We have presented a spectroscopic survey of the explosion sites and host galaxies of SNe~Iax and fitted emission line regions in our VLT/MUSE IFS and NOT/ALFOSC long-slit observations. 


We determined deprojected metallicity gradients for all hosts where possible and found that estimates of the explosion sites' metallicities based on these gradients agreed well with our direct measurements. With this our explosion site metallicity sample was supplemented with gradient estimates where we could not measure the location directly. The majority of SN~Iax explosion sites were found to be sub-solar ($12 + \log_{10} (\text{O/H}) \simeq 7.75-8.85$ dex in the scale of \citetalias{dopita16}), and generally metal poor compared to the distribution of SF in their hosts (\cref{sec:resz}). 

SNe~Iax explosion sites appeared to be less intensely star-forming and somewhat older than the typical SF region of the hosts ({\cref{sec:resultsages}), although they do follow a similar host-normalised offset distribution (\cref{sec:resultsoffsets}). Through comparison with a fiducial gas model and the results of the SP synthesis code {\sc bpass} we estimate the ages of the young SPs at the explosion sites of SNe~Iax to be several $\times10^7$ to $10^8$ years old (although note the caveats on assigning quantitative values in \cref{sec:methages}).
The relatively young ages at the explosion sites are confirmed through a similarity in their association to the ongoing SF of their host galaxies to that seen for SNe~II (\cref{sec:pixelstatres}), which are expected to have typical ages of tens of Myr, extending up to $\simeq 10^8$ years \citep{zapartas17}.

Assessing our environmental measures in terms of SN properties, we find that the brighter, more slowly fading objects define our range of metallicities, EW(\Ha{}) and offsets by covering the entire ranges (\cref{sec:sncompare}). Faint and faster events appear to occupy more restricted ranges of EW values and host offsets although we are limited to only a few objects for which we can investigate. Their metallicities extend across a wide range including some of the most metal-poor and -rich in the whole sample. We find no correlations between the velocity of the SNe and our environmental parameters. The only two designated SNe~Iax to display helium (SNe~2004cs, 2007J) have among the highest EW(\Ha{}) values of the sample, indicate the presence of comparatively younger SPs. Their membership as SNe~Iax has been debated \citep{white15, foley16}. With just two events we are limited in searching for statistical difference between these and the rest of the SNe~Iax and their environments appear typical of both SNe~Iax and IIb (the suggested typing of \citealt{white15}).

When compared to other SN types, SNe~Iax as a whole display a similar metallicity distribution to that seen for SNe Ib/c and II, with similar median values and covering a broadly similar range in metallicity. The host-normalised offset distribution of SNe~Iax follows closely that seen for SNe~Ia in disk galaxies and SNe~Ib. Using pixel statistics, SNe~Iax trace the ongoing SF of their hosts at a level most similar to SN~II and significantly more so than SNe~Ia, but less than SNe~Ic (and LGRBs).

For the S0 host of SN~2008ge, NGC~1527, we discover a stream and clumps of ionised gas not seen in previous studies. A limited analysis of emission line ratios would disfavour a SF-driven ionising flux, with an evolved SP or shocks more likely to be powering it. We find no evidence of ionised gas at, or nearby, the location of SN~2008ge and confirm its environment is old compared to the rest of SNe~Iax. The ionised component suggests there may have been SF in NGC~1527 relatively recently and thus $\sim$Gyrs old progenitors are not required.

The young explosion sites we confirm here remain in good agreement with the expectations for progenitors consisting of CO/CONe WD + He stars and moderately massive CCSNe with fall-back or SE-ECSNe. The lack of features associated with very young environments disfavour the presence of very massive stars at the explosion site. We would therefore consider the model of massive WR stars suffering fall-back on to a black hole upon collapse as unlikely progenitors (without having to invoke implausibly large runaway velocities), ruling out faint SNe~Iax as a way to explain local, apparently SN-less LGRBs.

\section*{Acknowledgements}

We thank Ryan Foley for providing the late time spectra of SN~2014dt. David Balam is thanked for providing the spectrum of PSN~J22412689+3917220 (now SN~2015ce) for its registration on the Transient Name Server. 
Eric Hsiao and the iPTF collaboration are thanked for their assistance registering PTF~14ans (now SN~2014ey) onto the Transient Name Server. 
Lise Christensen is thanked for comments on a draft of the manuscript.
JDL gratefully acknowledges support from the UK Science and Technology Facilities Council (grant IDs ST/L000733/1 and ST/P000495/1).
MDS gratefully acknowledges support provided by the Danish Agency for Science and Technology and Innovation realized through a Sapere Aude Level 2 grant, the VILLUM FONDEN  (research grant 3261),  and  the Instrument Centre for Danish Astrophysics (IDA).
Based on observations collected at the European Organisation for Astronomical Research in the Southern Hemisphere under ESO programmes 095.D0091, 096.D-0263 and 099.D-0022. 
Based on observations made with the Nordic Optical Telescope (proposal numbers 52-004 and P3-005; PI: Stritzinger), operated by the Nordic Optical Telescope Scientific Association at the Observatorio del Roque de los Muchachos, La Palma, Spain, of the Instituto de Astrofisica de Canarias.
This research has made use of the NASA/IPAC Extragalactic Database (NED) which is operated by the Jet Propulsion Laboratory, California Institute of Technology, under contract with the National Aeronautics and Space Administration. We acknowledge the usage of the HyperLeda database (http://leda.univ-lyon1.fr)
This research made use of Astropy, a community-developed core Python package for Astronomy \citep{astropy13}.

\bibliographystyle{mnras}
\bibliography{references}

\appendix

\section{The late-time spectra of SNe 2015H and 2014dt}
\label{app:14dt15H}

\begin{figure}
 \centering
\subfloat{\includegraphics[width=\columnwidth]{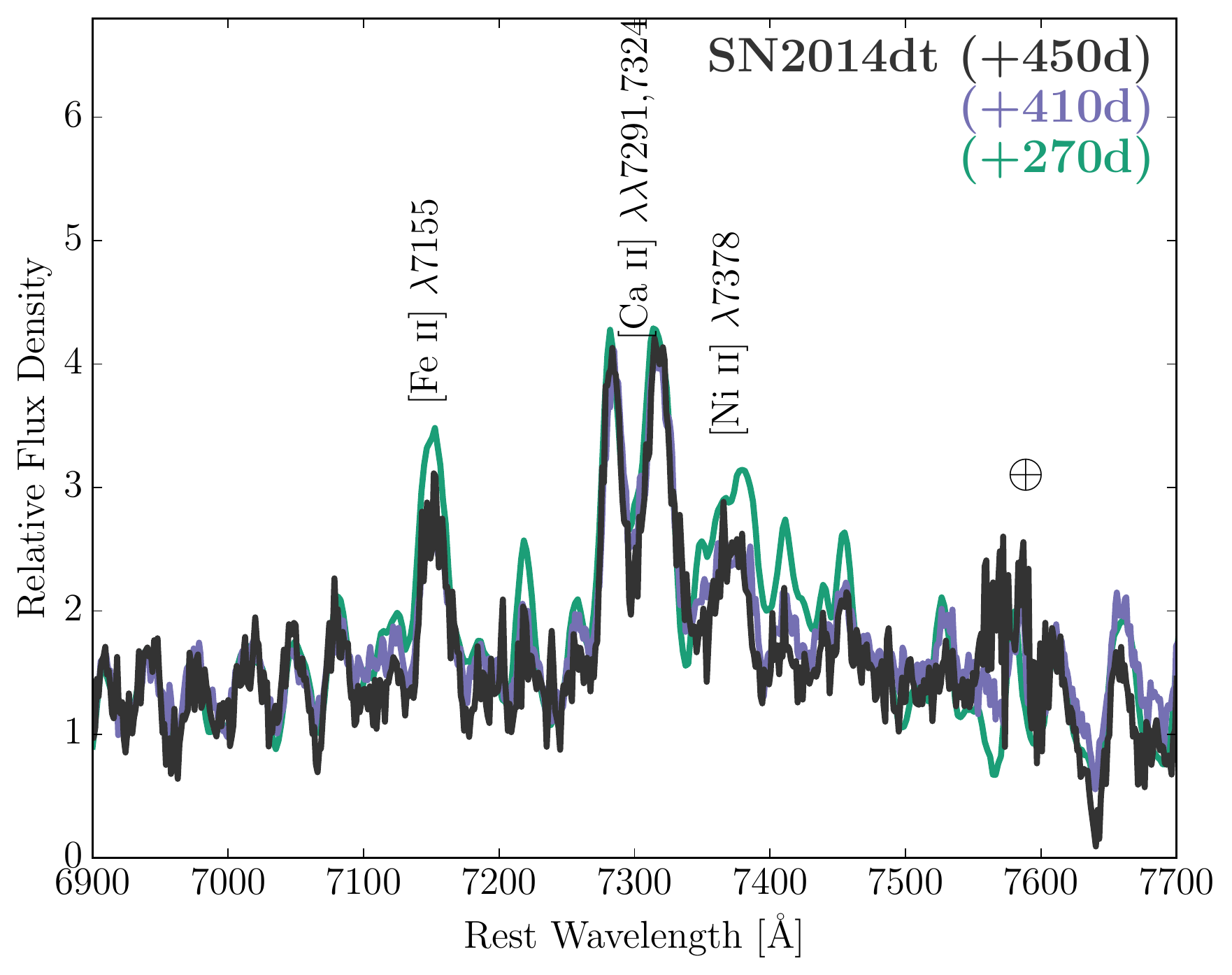}}
 \caption{The late-time spectrum of SN~2014dt extracted from the MUSE data cube at an epoch of $\sim$+450 days after maximum, centred around prominent forbidden lines. Also shown are the +270 and +410 days spectra of \citet{foley16}, we confirm a general lack of evolution in the strength or widths of the forbidden lines between the +270d spectra, although the [Ni {\sc ii}]~$\lambda$7378 flux is slightly depressed. Our spectrum shows very little evolution from the +410d spectrum, with only perhaps a slight weakening of the $\sim$7670\AA{} feature.}
 \label{fig:14dtlate}
\end{figure}

Our MUSE observations included a significant time lag in order to analyse the underlying galaxy light at the explosions sites. Nevertheless, for SNe 2014dt and 2015H we were able to detect prominent forbidden emission that is associated with SNe~Iax at late phases. Although a full analysis of the SN signal is beyond the scope of this work we present their spectra for completeness and compare to the comprehensive analysis of late-time SN~Iax spectra sample by \citet{foley16}. For this we extracted flux in spaxels centred on the SN emission within a radius equal to the seeing of the image.

Our SN~2014dt spectrum is at a phase of approximately +450 rest-frame days post maximum, which makes it the latest spectrum of a SN~Iax to our knowledge. In \cref{fig:14dtlate} we compare our spectrum in a region of strong forbidden line emission to those of \citet{foley16}, which were taken at +270 and +410d. As those authors noted based on their +410 days spectrum (to which our +450 days spectrum is almost identical), there is a general lack of evolution between these epochs in line strength or widths. Our SN~2014dt observations were taken under challenging sky conditions and the regions of interest are affected by residual sky emission not removed by the MUSE data reduction (\cref{sec:museobs}) -- for the sake of presentation of \cref{fig:14dtlate} we have manually subtracted the signal from a very faint region of the data cube.

The early spectra and light curve of SN~2015H were analysed by \citet{magee16} and our MUSE observation adds a late time spectrum at an epoch of +291 days past maximum. SNe~Iax display a wide range of morphologies in their spectra around these wavelengths at similar epochs, as shown in Figure 8 of \citet{foley16} and we note a very strong similarity between SN~2002cx and SN~2015H. As with SN~2014dt we present a region of strong forbidden line emission in \cref{fig:15Hlate} where we also plot the best-fitting 10 parameter model of \citet{foley16} in order to compare more quantitatively with SN~2002cx. As with SN~2002cx, we see that [Ni {\sc ii}]~$\lambda$7378 emission is dominated by its broad component (FWHM $\sim 8000$~\kms{}, cf. 7870~\kms{} for SN~2002cx), with the narrow components of [Fe {\sc ii}]~$\lambda$7155 and [Ca {\sc ii}]~$\lambda$7291,7324 being more prominent, albeit somewhat narrower (FWHM $\sim 800$~\kms{} cf. 1430~\kms{} for SN~2002cx).

We additionally note that we do not expect emission from these SNe that would affect our main environmental analysis, in the absence of circumstellar interaction, of which there appears to be no evidence. When running these SN spaxel bins as part of our main fitting procedure, {\sc starlight} masked over the SNe features (such as those shown in \cref{fig:14dtlate,fig:15Hlate}) as part of a sigma clipping routine. We found that manually masking large regions of the input spectra that may be affected by SN emission made very little difference to our emission line results.
Late in the manuscript's preparation we were able to check the effect of the SN emission for SN~2015H as MUSE re-observed the SN location. Although the observations were aborted due to scheduling restrictions, $2\times700$~s exposures were taken. We measured metallicities at the SN location on the first exposure (as the second was of poorer quality). Our values of $8.67\pm0.10$ and $8.56\pm0.03$ dex agree excellently with the values from our original observations of $8.66\pm0.07$ and $8.56\pm0.03$ dex for \citetalias{dopita16} and \citetalias{marino13} (N2), respectively.

With these results we highlight the power of MUSE as an extremely efficient optical spectrograph to obtain nebular phase spectra of SNe, whilst obtaining spectral information at the explosion site and across the host for free, providing legacy value to the data for a wealth of galactic studies not possible with traditional long slit spectroscopy.

\begin{figure}
 \centering
\subfloat{\includegraphics[width=\columnwidth]{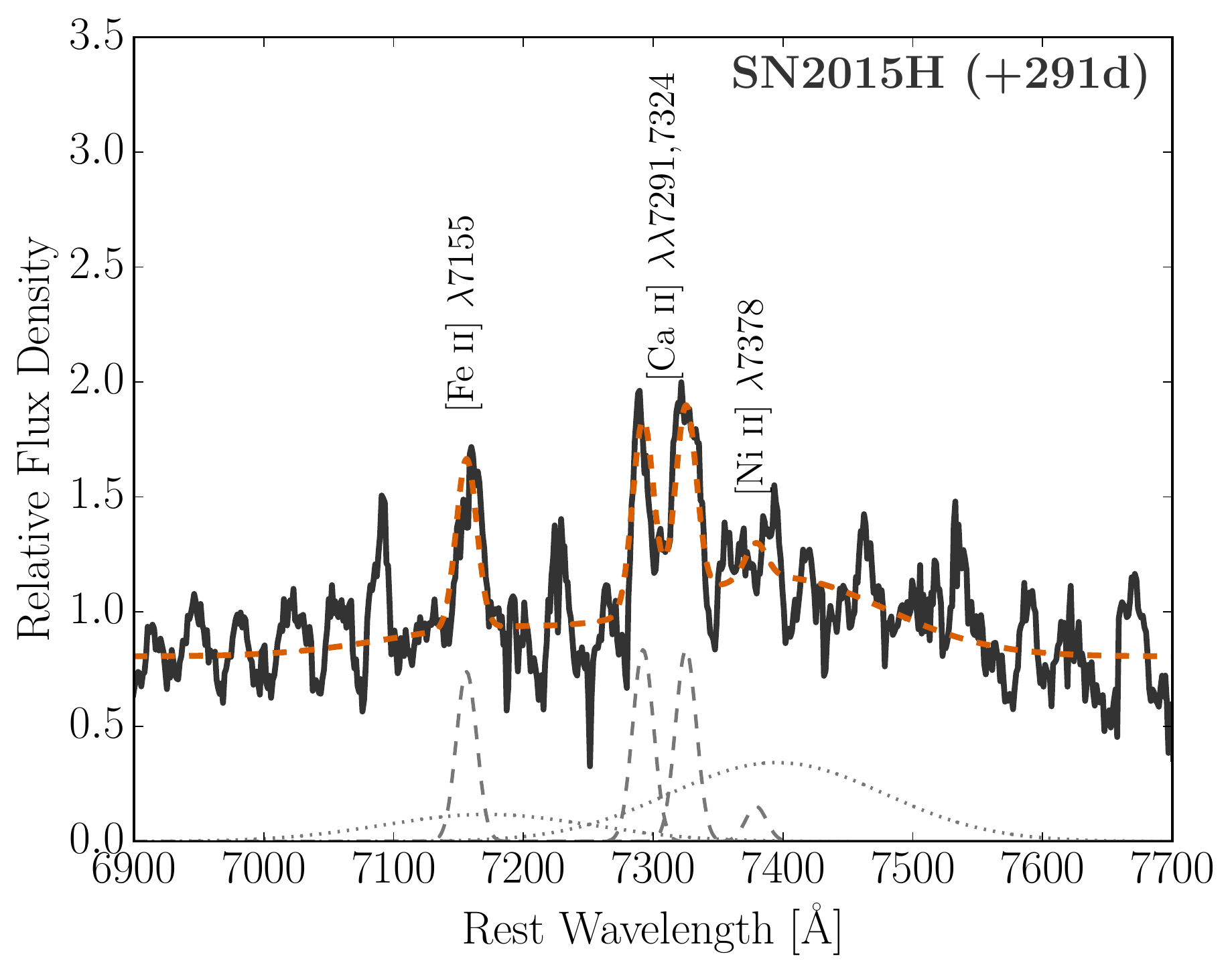}}
 \caption{As for \cref{fig:14dtlate} but here showing SN~2015H at an epoch of +291 days. The 10 component model, following the methodology of \citet{foley16} is shown. The morphology and line kinematics of the spectrum are similar to those of SN~2002cx at a similar epoch.}
 \label{fig:15Hlate}
\end{figure}

\section{Explosion site line fluxes}

In \cref{tab:fluxes} we provide individual flux measurements for the SN~Iax explosion sites. These values have been corrected for Galactic reddening but not local reddening, i.e., before our Balmer decrement correction.

\begin{center}
\begin{table*}
\begin{threeparttable}
\caption{Emission line fluxes at the explosion sites of SNe~Iax, corrected for Galactic extinction. Line strengths are given relative to the \Ha{} flux, $F($\Ha{}). For the NOT sample \Sii{} are given as the sum of the lines $\lambda 6716+\lambda 6731$.}
\begin{tabular}{lrccccccccc}
\hline                  
SN         & $F($H$\alpha)$  & H$\beta$        & \multicolumn{2}{c}{[O {\sc iii}]} & \multicolumn{2}{c}{[N {\sc ii}]} & \multicolumn{2}{c}{[S {\sc ii}]} \\                           
           & [$\times 10^{-16}$ erg~s$^{-1}$] &                 & $\lambda 4959$ & $\lambda 5007$                &  $\lambda 6548$ & $\lambda 6583$    &   $\lambda 6716$ & $\lambda 6731$  \\   
\hline     
\multicolumn{9}{c}{MUSE sample} \\
\hline
1991bj   &           21.59 &           0.278  &           0.100 &           0.309 &            0.064 &            0.184 &            0.224 &            0.159 \\                                                                                               
2002bp   &            0.15 &         \ldots   &         \ldots  &         \ldots  &          \ldots  &            1.136 &          \ldots  &          \ldots  \\                                                                                               
2002cx   &            9.05 &           0.301  &           0.113 &           0.355 &            0.066 &            0.188 &            0.231 &            0.177 \\                                                                                               
2004cs   &           27.00 &           0.290  &           0.078 &           0.218 &            0.083 &            0.254 &            0.231 &            0.172 \\                                                                                               
2005P    &            1.73 &           0.252  &         \ldots  &           0.264 &            0.044 &            0.163 &            0.301 &            0.202 \\                                                                                               
2005hk   &            0.51 &           0.290  &         \ldots  &           0.248 &            0.059 &            0.112 &            0.313 &            0.204 \\                                                                                               
2008ae   &            3.78 &           0.287  &           0.060 &           0.168 &            0.118 &            0.392 &            0.235 &            0.166 \\                                                                                               
2008ha   &           32.13 &           0.439  &           0.376 &           0.990 &            0.026 &            0.079 &            0.167 &            0.119 \\                                                                                               
2009J    &            7.72 &           0.284  &           0.124 &           0.339 &            0.102 &            0.323 &            0.258 &            0.185 \\                                                                                               
2010ae   &          378.23 &           0.280  &           0.169 &           0.493 &            0.058 &            0.177 &            0.216 &            0.150 \\                                                                                               
2010el   &           24.57 &           0.202  &           0.070 &           0.169 &            0.202 &            0.595 &            0.263 &            0.181 \\
2011ce   &          249.84 &           0.245  &           0.020 &           0.052 &            0.127 &            0.378 &            0.179 &            0.129 \\
2012Z    &            4.81 &           0.316  &           0.110 &           0.267 &            0.086 &            0.231 &            0.300 &            0.207 \\                                                                                               
2013gr   &            4.08 &           0.310  &           0.195 &           0.547 &            0.035 &            0.106 &            0.300 &            0.202 \\
2014dt   &           42.74 &           0.234  &           0.045 &           0.144 &            0.168 &            0.539 &            0.237 &            0.158 \\                                                                                               
2014ey   &            1.75 &           0.261  &           0.085 &           0.203 &            0.078 &            0.258 &            0.271 &            0.197 \\                                                                                               
2015H    &            2.95 &           0.144  &         \ldots  &           0.084 &            0.072 &            0.411 &            0.277 &            0.143 \\                                                                                               
\hline     
\multicolumn{9}{c}{NOT sample} \\
\hline
2003gq   &    38.99 &  0.158 & \ldots & \ldots & \ldots &  0.126 & \multicolumn{2}{c}{ 0.132} \\
2005cc   &    55.52 &  0.135 & \ldots & \ldots & \ldots &  0.302 & \multicolumn{2}{c}{ 0.204} \\
2006hn   &    15.69 &  0.309 & \ldots &  0.170 & \ldots &  0.398 & \multicolumn{2}{c}{ 0.282} \\
2007J    &    38.60 &  0.288 & \ldots &  0.288 & \ldots &  0.200 & \multicolumn{2}{c}{ 0.295} \\
2009ku\tnote{\dag}   &    17.82 &  0.282 & \ldots &  0.363 & \ldots &  0.219 & \multicolumn{2}{c}{ 0.316} \\
2013dh   &    37.82 &  0.324 & \ldots &  0.069 & \ldots &  0.372 & \multicolumn{2}{c}{ 0.282} \\
2013en   &    44.03 &  0.295 & \ldots & \ldots & \ldots &  0.251 & \multicolumn{2}{c}{ 0.331} \\
2014ek   &    60.88 &  0.209 & \ldots & \ldots & \ldots &  0.295 & \multicolumn{2}{c}{\ldots} \\
PS~15csd\tnote{\dag}    &     8.56 &  0.295 & \ldots &  0.525 & \ldots &  0.112 & \multicolumn{2}{c}{ 0.363} \\
\hline
\end{tabular}
\label{tab:fluxes}
  \begin{tablenotes}
  \item[\dag] SN overlaid on a small, unresolved host.
  \end{tablenotes}
 \end{threeparttable}
\end{table*}
\end{center}

\section{Metallicity maps}
\label{app:zstamps}

In \cref{fig:zstamps} we show stamps of \citetalias{dopita16} metallicity maps for the MUSE sample of SNe~Iax hosts, excluding SN~2008ge, which is discussed and shown in \cref{sec:08ge}).

\begin{figure*}
 \centering
\subfloat{\includegraphics[width=0.8\linewidth]{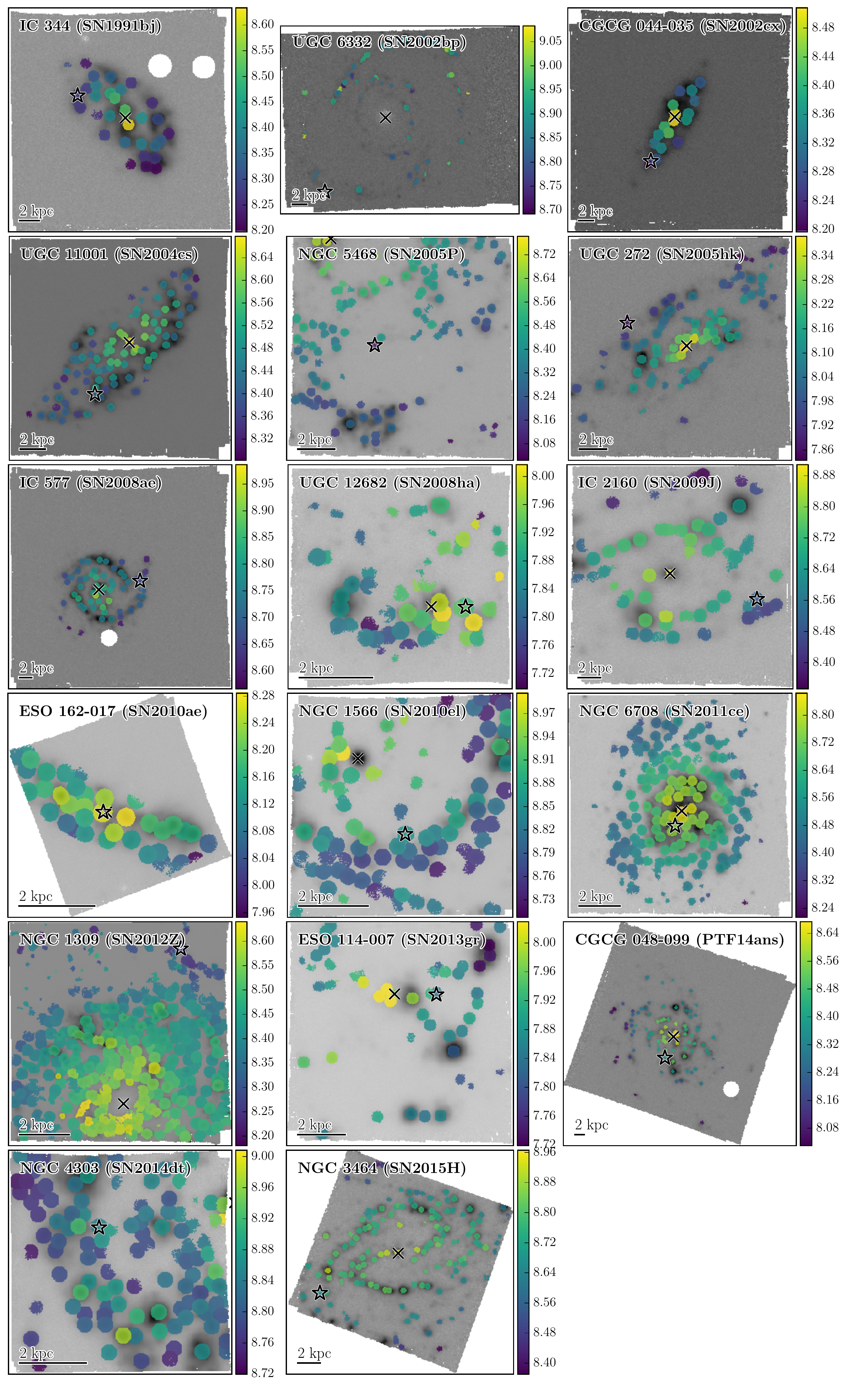}}
 \caption{The metallicity maps of SNe~Iax host galaxies using the indicator of \citetalias{dopita16} (see \cref{sec:methmetallicity}). Host nuclei and the SN explosion site positions are given by black $\times$ and `star' symbols, respectively. North is up, East is left. Note that PTF~14ans = SN~2014ey}
 \label{fig:zstamps}
\end{figure*}

\bsp	
\label{lastpage}
\end{document}